\documentclass[natbib]{svjour3}                     
\smartqed  
\usepackage{graphicx}

\newcommand{\aap}{{Astron. Astrophys.}}
\newcommand{\apj}{{Astrophys. J.}}

\newcommand{\solphys}{{Solar Phys.}}
\newcommand{\apjl}{{Astrophys. J. Lett.}}
\newcommand{\jgr}{{J. Geophys. Res.}}
\begin{document}

\title{Excitation of standing kink oscillations in coronal loops
}

\titlerunning{Excitation of kink oscillations in coronal loops}        

\author{Jaume Terradas 
}


\institute{J. Terradas \at
              Centre for Plasma Astrophysics and Leuven Mathematical Modeling and
Computational Science Centre, Katholieke Universiteit Leuven, Celestijnenlaan
200B, B-3001 Leuven, Belgium
              Tel.: +32-16-327690\\
              Fax: +32-16-327998\\
              \email{jaume@wis.kuleuven.be}           
}

\date{Received: date / Accepted: date}

\maketitle

\begin{abstract}

In this work we review the efforts that have been done to study the excitation of
the standing fast kink body mode in coronal loops. We mainly focus on the
time-dependent problem, which is appropriate to describe flare or CME induced
kink oscillations. The analytical and numerical studies in slab and cylindrical
loop geometries are reviewed. We discuss the results from very simple
one-dimensional models to more realistic (but still simple) loop configurations.
We emphasise how the results of the initial value problem complement the
eigenmode calculations. The possible damping mechanisms of the kink oscillations
are also discussed.

\end{abstract}

\section{Introduction}
\label{intro}

Transversal coronal loop oscillations are regularly observed by the EUV telescope
on board $TRACE$. Nowadays, there is abundant information about the period,
damping time and amplitude of the oscillations \citep[see for example][and see
also the review of M. Aschwanden in this volume]{aschetal02,schrijetal02}. In most
of the cases these oscillations are triggered by a flare or a CME, and they have
been interpreted as fast standing magnetohydrodynamic (MHD) kink oscillations of
cylindrical tubes \citep[see][]{nakaof01,rudrob02,goossetal02}.

In general, most of the theoretical studies on loop oscillations assume that the
system is in the stationary state and that the loop dynamics is given by the
normal modes. The calculation of the eigenmodes is possible in rather simple
configurations but their properties are usually unknown in complex structures.
Although normal modes should be seen as the building blocks to interpret coronal
loop oscillations they do not represent the whole picture, but their study
provides a basis for understanding the dynamics of the system. To have a more
accurate description, the time-dependent problem needs to be
analysed. Ideally, we would like to know which normal modes are excited given an
initial perturbation.

Although there is an extensive literature about MHD oscillations in coronal loops
only those theoretical works where the time-dependent analysis has been considered
are reviewed here. Usually the time-dependent problems are classified in driven or
in initial-value problems. Driven problems are useful to model, among others, the
effect of the photospheric motions in coronal loops. In this case a continuous
motion is applied at the base of the loop exciting several kinds of modes that
propagate along the tube. On the other hand, initial-value problems are more
appropriate to describe a sudden release of energy, as in flares. Such energy
release generates a disturbance, a blast wave or an EIT wave, that propagates in
the corona and produces kink oscillations of nearby loops. In this work we will
concentrate on initial-value problems. Moreover, since most of the observations of
kink oscillations are interpreted as the fundamental standing kink mode, here we
will focus only on this mode, characterised by a single frequency and  a dominant
wavenumber along the loop. For propagating kink modes, characterised by a wide
superposition of different longitudinal wavenumbers and frequencies, the reader is
referred to the works of
\cite{murrob93a,murrob93,murrob94,huangetal99,selwamur04,ogromur06,ogromur07} in
slab models and to \citet{robetal83,robetal84a,selwaetal04,terrahometal03} in
cylindrical loops.

In this work we first describe the results of the time-dependent problem starting
with simple loop models and then we increase the complexity of the equilibrium. We
also centre on those studies where a direct comparison with eigenmodes can be done
since this helps to understand and interpret the results of the time-dependent
problems.

\section{Basic Loop Models}
\label{model}

Two basic equilibrium models have been studied in detail in the literature. They
are the simplest representation of coronal loops and are important to understand
the basic properties of MHD waves. These models are the slab and the cylindrical
flux tube. 

\subsection{Slab}
\label{slab}

The simplest loop model is a Cartesian slab configuration. In this model the
loop is represented by a plasma column of enhanced density, $\rho_{\rm i}$,
respect to the coronal environment, with density $\rho_{\rm e}$. The slab has a
half-width of $a$ and the magnetic field is vertical (${\bf B}=B_0 {\bf e}_z$). Since the inhomogeneity is in one direction, hereafter
assumed in $x$, it is possible to make Fourier analysis in the other ignorable
directions ($y$ and $z$). The reader is referred to \citet{edrob82} and
\citet{terretal05} (with special attention to the leaky modes) for the details
about the dispersion diagram in this model. When the wave propagation is in the
plane of the slab then the frequency of the fundamental fast kink body mode, in
the limit of thin tubes (when $k_z a \ll 1$), tends to the external Alfv\'en
frequency,
\begin{equation}\label{kinkeqslab} \omega=k_z\, v_{\rm Ae},
\end{equation} 
where $v_{\rm Ae}=B_0/\sqrt{\mu\rho_{\rm e}}$ is the external
Alfv\'en speed. However, when the perpendicular propagation is allowed ($k_y\neq 0$) the
body kink mode is transformed in a surface wave
\citep[see][]{zheletal96,arrterretal07} and its frequency tends to the kink
frequency (when $k_y\gg k_z$)

\begin{equation}\label{kinkeq} \omega_k=k_z \sqrt{\frac{\rho_{\rm i} v_{\rm
Ai}^2+\rho_{\rm e} v_{\rm Ae}^2}{\rho_{\rm i}+\rho_{\rm e}}}.
\end{equation}

\noindent In general this model is appropriate if the structure is invariant in
the perpendicular plane (i.e. in the $y-$direction).

\subsection{Cylinder} \label{cylinder}

A cylindrical flux tube is the natural extension of the previous model. The
reader is referred to \citet{edrob83,cally86,cally03} for the details of the
dispersion diagram and the classification of the different modes (trapped and
leaky). Here we will restrict to body waves and the $m=1$ mode (the oscillation
that displaces transversally the whole tube). In the limit of thin tube the
frequency of oscillation is the kink frequency, given by
equation~(\ref{kinkeq}). 

The main advantage of the cylindrical model is that it represents a
three-dimensional structure. It is interesting to note that the kink body modes
are very similar to the kink modes of the slab model when $k_y$ is large. The
frequency of oscillation is the same, the kink frequency, and the eigenfunction is
quite localised near the loop boundary.  In fact in some papers \citep[see for
example][]{hollyang88}, the results of the slab models have been extended to the
cylindrical geometry by using the equivalent of the azimuthal wavenumber in
Cartesian coordinates, i.e. assuming that $k_y=m/R$. 

On the other hand, one of the assumptions that is usually relaxed in the
cylindrical and in the slab model is the Alfv\' en speed profile across the
boundary. In the simple models it is assumed that the density (and the Alfv\'en
speed) has a discontinuous jump at the interface between the loop and the coronal
environment. However, if the Alfv\' en speed has an inhomogeneous layer, with a 
typical width of $l$, then the process of resonant absorption takes place (for
$m\ge 1$ in the cylindrical model or for $k_y\neq 0$ in the slab model).

\subsection{Curved configurations}\label{curvedconf}

One of the configurations that takes into account the curvature, neglected in
the previous models is the circular loop model. In this force-free model the
shape of the magnetic field lines is given by a purely poloidal magnetic field,

\begin{equation}\label{poloidal}
B_{\theta}=B_0 \frac{r_0}{r}.
\end{equation}

\noindent The density profile is usually chosen (in the zero-$\beta$
approximation) to have the same dependence as the magnetic field and with a
density enhancement, representing the loop, at a given radius. The advantages of
this model is that it is possible to make Fourier analysis along the
$\theta-$direction and the loop cross section is constant along the tube. One of
the disadvantages is that it has a singularity at $r=0$ and this might be
difficult to handle from the numerical point of view.

Another popular model is the potential arcade. In this model the current-free
magnetic field is given by
\begin{eqnarray}\label{arcadepot}
B_x=B_0 \sin\left( x/\Lambda_B \right)\, e^{-z/\Lambda_B},\\
B_z=-B_0 \cos\left(x/\Lambda_B \right)\, e^{-z/\Lambda_B},
\end{eqnarray}

\noindent where $\Lambda_B=L/\pi$ and $L$ is the arcade width. The background density is
usually taken to be of the form
\begin{eqnarray}\label{arcadepotdens}
\rho=\rho_0\,e^{-z/\Lambda}.
\end{eqnarray}

\noindent The parameter $\Lambda$ allows to have a decreasing or increasing
Alfv\'en speed with height ($v_{\rm A}$ is independent of the $x-$coordinate)
and the loop is modelled by a density enhancement along certain field lines. The
configuration is well behaved at the origin but the loop cross section is
not constant along the tube and no Fourier analysis is allowed along the field
lines. 

These two equilibrium models are two-dimensional but they can still be used to
study curved loops in three dimensions assuming that the field is invariant in the
perpendicular direction to the plane. Other sophisticated equilibrium models
involve three-dimensional dipoles or even photospheric magnetic field
extrapolations, usually force free, together with the inclusion of a certain
density profile along the field lines. However, these models are still in a very
primitive stage.

\section{Waves in Slabs}\label{slabstand}

\subsection{Basic properties of fast MHD waves in a uniform line-tied medium}

We start with the most simple configuration, a uniform plasma permeated by a
vertical and uniform magnetic field. In the $\beta$
zero limit the ideal linearised MHD equations reduce to the well-known wave
equation for the fast modes,
\begin{eqnarray}
\frac{\partial^2 \xi}{\partial{t^2}}=v^2_A\left(
\frac{\partial^2 \xi}{\partial{x^2}}+\frac{\partial^2 \xi}{\partial{z^2}}\right),
\label{wave}
\end{eqnarray}
where $\xi$ is the horizontal displacement and $v_{\rm A}=B_0/\sqrt{\mu\rho}$ is the
Alfv\'en speed. Since $\beta=0$ slow modes and Alfv\'en waves
(we assume that $k_y=0$) are absent. We Fourier analyse in the $z$-direction, i.e. we assume that the perturbations are of the
form $e^{-i k_z z}$. Hence, equation~(\ref{wave}) reduces to the Klein-Gordon
equation,
\begin{eqnarray}
\frac{\partial^2 \xi}{\partial{t^2}}=v^2_A
\frac{\partial^2 \xi}{\partial{x^2}}-\omega_{\rm c}^2\xi,\label{wavecut}
\end{eqnarray}

\noindent where $\omega_{\rm c}=v_{\rm A} k_z$ is the cut-off frequency. Thus, after introducing the
longitudinal wavenumber $k_z$ the two-dimensional wave equation reduces to a
one-dimensional equation, and by selecting the appropriate value of this
parameter the effect of line-tying can be easily incorporated. For the
fundamental oscillation with a node at each footpoint we have that $k_z=\pi/L$,
$L$ being the length of the system (or loop if we include a density enhancement)
in the vertical direction. In this model the time a wave needs to travel along
the entire loop to feel the effect of the photosphere is neglected, and the loop
instantly feels the effect of line-tying.

Normal mode solutions of equation~(\ref{wavecut}), i.e. with a dependence of the
form $e^{i (\omega t - k x)}$, yield the dispersion relation $\omega=\sqrt{k^2
v_{\rm A}^2+\omega_{\rm c}^2}$ which indicates that wave propagation is
dispersive. The propagation speed for waves with wavenumber $k$ is the group
velocity, 
\begin{equation} v_g=\frac{d\omega}{dk}=\frac{k\,v_{\rm A}^2}{\sqrt{k^2v_{\rm
A}^2+\omega_{\rm c}^2}}= \frac{\sqrt{\omega^2-\omega_{\rm c}^2}}{\omega}\,v_{\rm
A},\label{vgroup} \end{equation}

\noindent which ranges between zero, for waves with infinite wavelength, and
$v_{\rm A}$ for waves with infinitely small wavelength. Therefore, modes with
short wavelengths propagate faster than modes with larger wavelengths. In
addition, any mode with frequency below the cut-off frequency is evanescent and
unable to propagate in the medium. The dispersive behaviour of waves is simply
due to the fact that we assume a fixed wavelength along the loop (a single
$k_z$) to model the effect of fixed footpoints.

\begin{figure}[hh]\center{
\includegraphics[width=0.65\textwidth]{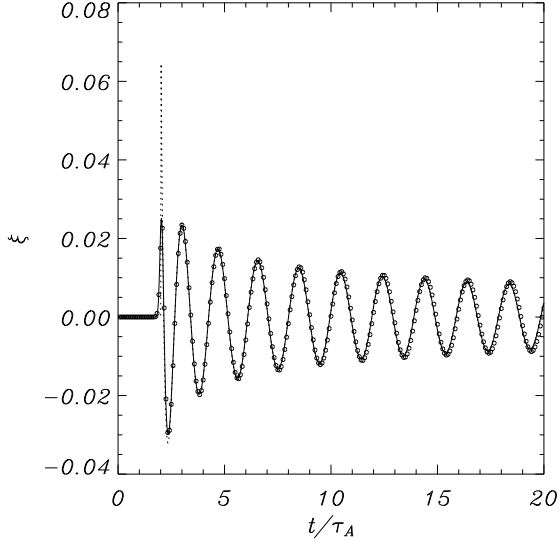}}
\caption{\small 
Displacement as a function of time at position $x=2L$ for different wave source
widths: the dashed line corresponds to a $\delta$-function and the continuous
line corresponds to a Gaussian profile with
$w=0.1L$. The numerical solution with a Gaussian profile with $w=0.1L$ for a
thin, low density slab model ($a=0.01L$ and $\rho_{\rm i}/\rho_{\rm e}=2$) is
displayed with circles. For comparison purposes all the profiles have been
normalised to the same amplitude. It is clear that different kinds of
impulsive excitation and the presence or not of the loop result in essentially
the same temporal behaviour for this low density contrast, thin loop.}
\label{gauss}
\end{figure}

Once we know the properties of fast MHD waves in the model we can study the
time-dependent problem by launching a perturbation in the system.  An analytical
solution of equation~(\ref{wavecut}) can be derived for the following initial
conditions \begin{equation} \xi(x,0)=0, \,\,\,\,\,\, \frac {\partial \xi}
{\partial t}(x,0)=v_0\delta(x), \label{initcond} \end{equation} $v_0$ being the
velocity of the initial perturbation. The displacement as a function of time in
terms of the Bessel function $J_0$ is given by \begin{equation}
\xi(x,t)=\frac{v_0}{4\omega_{\rm c}}J_0\left(\omega_{\rm c}\sqrt{t^2-(x/v_{\rm A})^2}\right)
{\cal H}(t-x/v_{\rm A}),\label{wakeeq} \end{equation} where $\cal H$ is the
Heaviside function. This solution \citep[see
for example also][]{raeroberts82}, is characterised by a wave front travelling at the
Alfv\'en speed in the direction normal to the magnetic field lines and by a wake
behind this pulse oscillating at the cut-off frequency $\omega_{\rm c}$ (see dashed
line in Fig.~\ref{gauss}). For $t\gg x/v_{\rm A}$ the asymptotic expression for
the displacement is \begin{equation}
\xi(x,t)=\frac{v_0}{\sqrt{8\pi}\omega_{\rm c}^{3/2}t^{1/2}}\cos(\omega_{\rm c} t
-\pi/4),\label{asymp} \end{equation}

\noindent which indicates that the displacement amplitude decays as $t^{-1/2}$ and
the frequency of oscillation is simply $\omega_{\rm c}$. Note that the attenuation
has nothing to do with any dissipation mechanism. These results, derived by
\citet{terretal05b} are slightly different from those obtained by
\citet{uralov03}. The solution presented here is more appropriate to describe the
physical problem of a flare-induced perturbation and the propagation of the
resulting disturbance in the $x$-direction.

In order to have simple analytical expressions we have considered that the initial
perturbation is a $\delta$-function, but it is possible to consider a source of 
finite width (for simplicity we assume a perturbation with a Gaussian profile of
width $w$). In Figure~\ref{gauss} we have represented the solution for $w=0.1L$
(see continuous line). For a typical flare event the spatial scale of the compact
flare source is much smaller than the mean length of the oscillating loops, hence
values of $w$ below $0.1L$ are close to realistic source widths. For large $t$ the
solution tends to the results of the $\delta$-function perturbation and the
differences are mainly in the wave front but not in the oscillating wake. Hence,
an excitation with a Gaussian shape does not change much (for a reasonable range
of disturbance widths $w$) the displacement profile in comparison with the
$\delta$-function perturbation.

\subsection{Effect of structuring: Leaky and trapped
waves}\label{leakytrappslab}

\begin{figure}[ht]
\center{
  \includegraphics[width=1\textwidth]{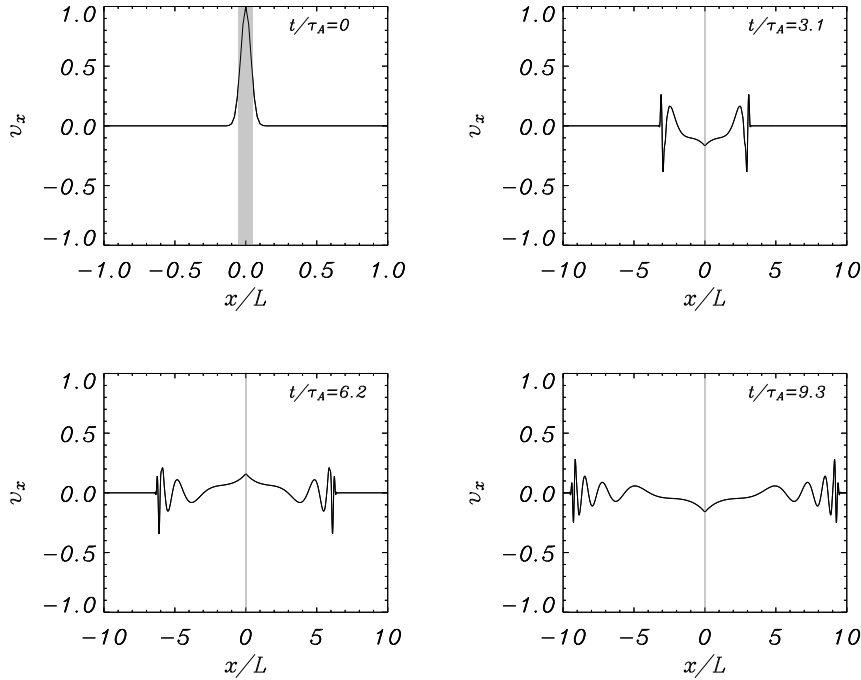}}
\caption{Plot of $v_x$ for different times. The initial perturbation is given by
equation~(\ref{kinkexc}) with $w=a$ and $a=0.05L$. Note the different spatial scale
for $t=0$. The grey area represents the density enhancement
($\rho_{\rm i}/\rho_{\rm e}=3$).}
\label{vxx0}      
\end{figure}

Now we include a density enhancement  (of half-width $a$) representing the loop.
In this slab model, the Alfv\'en speed inside the loop is smaller than the
Alfv\'en speed in the corona ($v_\mathrm {Ai}<v_\mathrm {Ae}$). Since the 
Alfv\'en speed changes in the $x$-direction the analytical solution of the
initial value problem, involving temporal Laplace transforms
\citep[see for example][]{andgoss97}, is much more complex. For this reason we have  solved 
equation~(\ref{wavecut}) numerically using standard techniques
\citep[see][]{terretal05}. To excite the kink mode we assume the following
perturbation 
\begin{eqnarray}\label{kinkexc}
v_x(x,t=0)=v_{x0}\exp\left[{-\left(\frac{x}{w}\right)^2}\right],
\end{eqnarray}

\noindent where we choose $w=a$ (now instead of the displacement, $\xi$, we use
the velocity, $v_x$ ($v_x=\frac{\partial
\xi}{\partial t}$). The numerical solution (see Fig.~\ref{vxx0}) shows that the
induced disturbances propagate away from the slab and contain different
wavelengths due to the dispersive nature of fast waves described before.
However, the shape of the velocity inside the slab and its near surroundings has
a different behaviour due to the presence of the cavity or wave guide. It
resembles the form of the fundamental kink mode eigenfunction: $v_x$ has an
extremum at $x=0$ and decreases exponentially outside the loop. This indicates
that the slab is oscillating with the normal mode. 

\begin{figure}[ht]
\center
\includegraphics[width=0.7\textwidth]{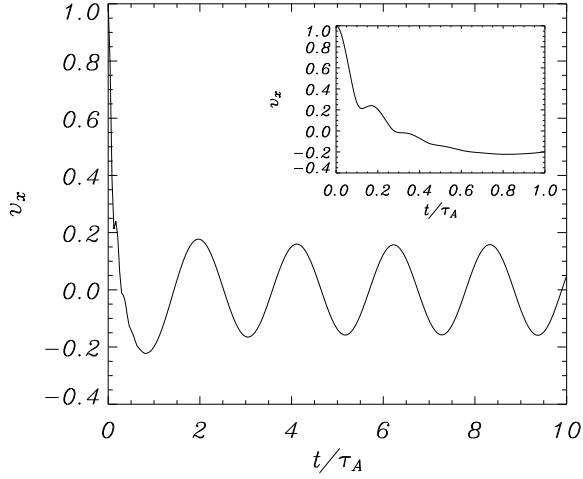}
\caption{\small 
Plot of $v_x$ at the slab centre ($x=0$) for the numerical simulation  of
Figure~\ref{vxx0}. After a short transient phase the loop reaches a stationary
state, oscillating at the frequency predicted by the normal mode analysis. In
the inner plot the short impulsive leaky phase is displayed, showing the  signal for
$t/\tau_\mathrm{A}$ between 0 and 1.} \label{vxx0temp}
\end{figure}

In Figure~\ref{vxx0temp} we have plotted $v_x$ as a function of time at the slab
centre ($x=0$).  The frequency of oscillation for large times is in excellent
agreement with that of the fundamental kink mode calculated by solving the
dispersion relation of the slab (it is basically given by eq.~[\ref{kinkeqslab}]). We also see that there is a short
time interval before the loop reaches the stationary regime. In
Figure~\ref{vxx0temp} we have plotted a detail of this transient, which is
initially dominated by the propagation of the initial disturbance inside the
slab ($0\leq t/\tau_\mathrm{A}<2a/v_\mathrm{Ai}$) and is followed by the
excitation of a leaky mode (for $2a/v_\mathrm{Ai}\leq t/\tau_\mathrm{A}$)
because of the reflection of the disturbance at the slab boundaries. During this
impulsive leaky phase the amplitude shows short period oscillations. Part of the
initially deposited energy in the slab is radiated away through the fundamental
kink leaky mode before the loop oscillates with the corresponding trapped mode.
The period of these oscillations, which are now superimposed to the much larger
period of the trapped kink mode, agrees with the period of the first leaky mode
\citep[see][]{terretal05} given by the following equation (in the limit $a/L\ll1$) \begin{eqnarray} 
\label{pleaky} P&\approx&\frac{2 a}{v_\mathrm{Ai}}\frac{1}{n},
\,\,\,\,\,\,\mbox{$n=1, 2,...$},\label{kinkleakp}\\  \tau_{\rm
d}&\approx&\frac{2 a}{v_\mathrm{Ai}}\left(\ln \left|
\frac{1+v_\mathrm{Ai}/v_\mathrm{Ae}}{1-v_\mathrm{Ai}/v_\mathrm{Ae}}\right|\right)^{-1}.
\label{kinkleaktd}  \end{eqnarray}

\noindent Note that the damping time ($\tau_{\rm d}=\omega_\mathrm{I}^{-1}$) is
independent of $n$, in the limit $a/L\ll1$ all leaky harmonics (i.e. modes with
different $n$) have the same damping time. Moreover, when $v_\mathrm{Ai}\ll
v_\mathrm{Ae}$ the damping time reduces to

\begin{equation}\label{tdleaky}
\tau_{\rm d}\approx a
\frac{v_\mathrm{Ae}}{v_\mathrm{Ai}^2}.
\end{equation}

\noindent The time-signatures in the slab model are now patent, after an initial
disturbance there is a short leaky transient followed by the excitation of the
trapped mode, which oscillates with constant amplitude with time (in ideal MHD,
i.e. without any dissipation mechanism).

It is worth noticing (see Fig.~\ref{vxx0temp}) that the amplitude of oscillation
of the normal mode is constant but smaller than the initial amplitude ($v_{x0}$).
If the perturbation is located in the external medium much less energy is trapped
in the cavity. In this case the motion of the slab is dominated by the wave
passing through it (described in the previous section) rather than by the
eigenmode. In Figure~\ref{gauss} (see circles) we have plotted the evolution of
the slab for a perturbation located in the corona. If we compare this result with
the profile when there is no density enhancement (see continuous line) we find
that the curves are almost the same. Thus, for an external excitation the slab is
unable to trap a significant amount of energy. This issue is discussed later in
detail for the cylindrical loop model (Section~\ref{energtrap}).

\begin{figure*}[ht]
\centering
\mbox{\hspace{-0.cm}{(a)}\hspace{5.6cm}{(b)}\hspace{5.2cm}}\vspace{-0.5cm}
\includegraphics[width=0.4\textwidth,angle=90]{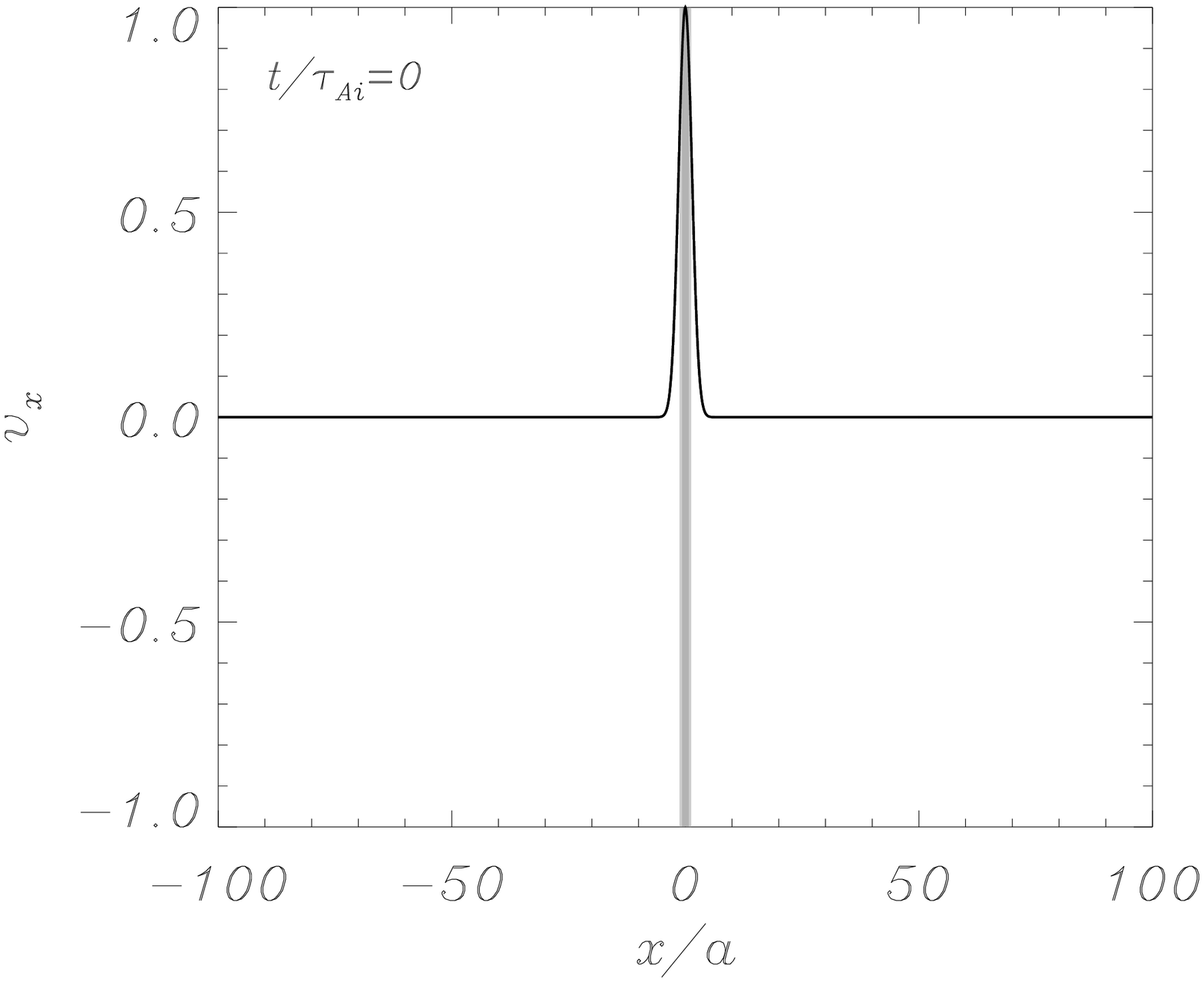}\includegraphics[width=0.4\textwidth,angle=90]{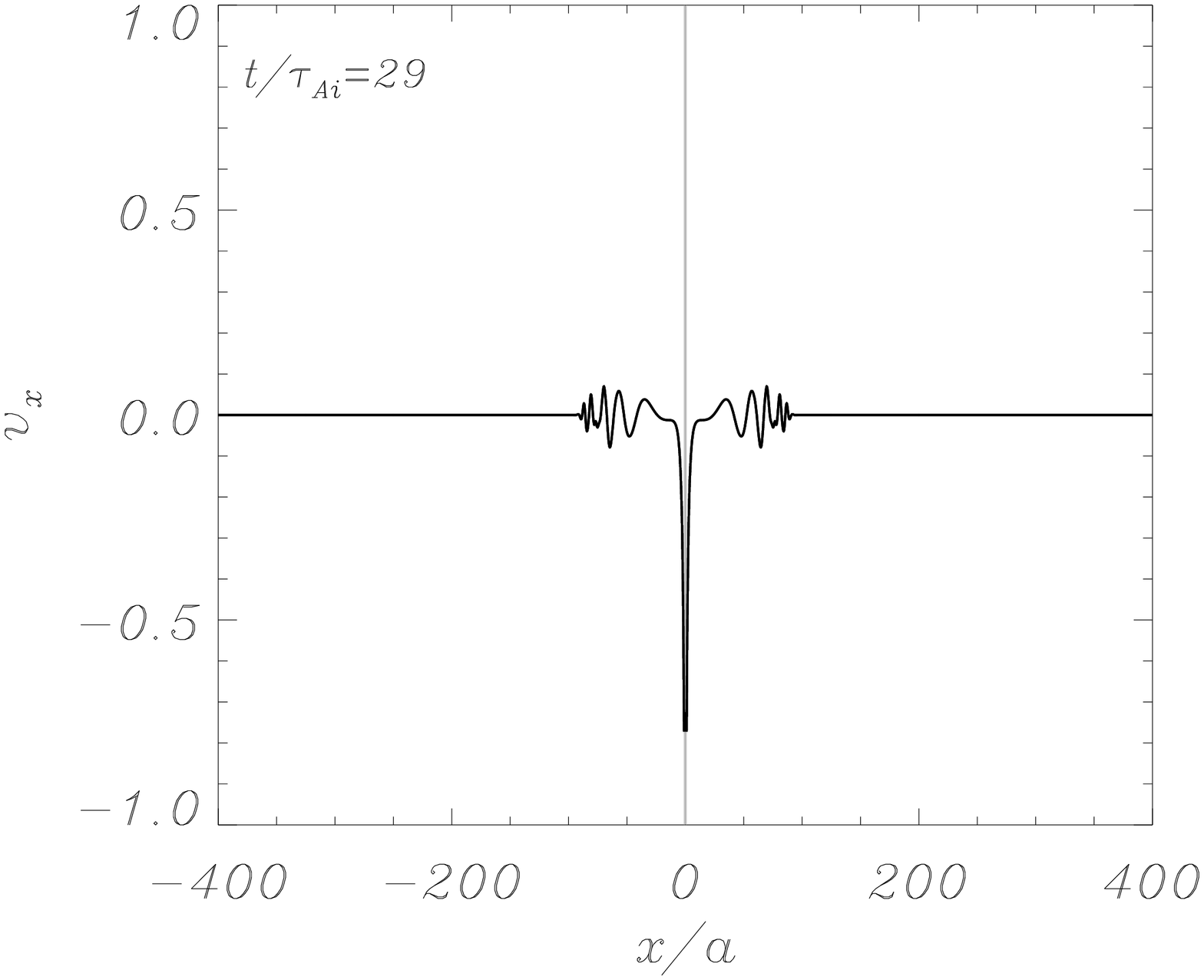}
\mbox{\hspace{-0.cm}{(c)}\hspace{5.6cm}{ (d)}\hspace{5.2cm}}\vspace{-0.5cm}
\includegraphics[width=0.4\textwidth,angle=90]{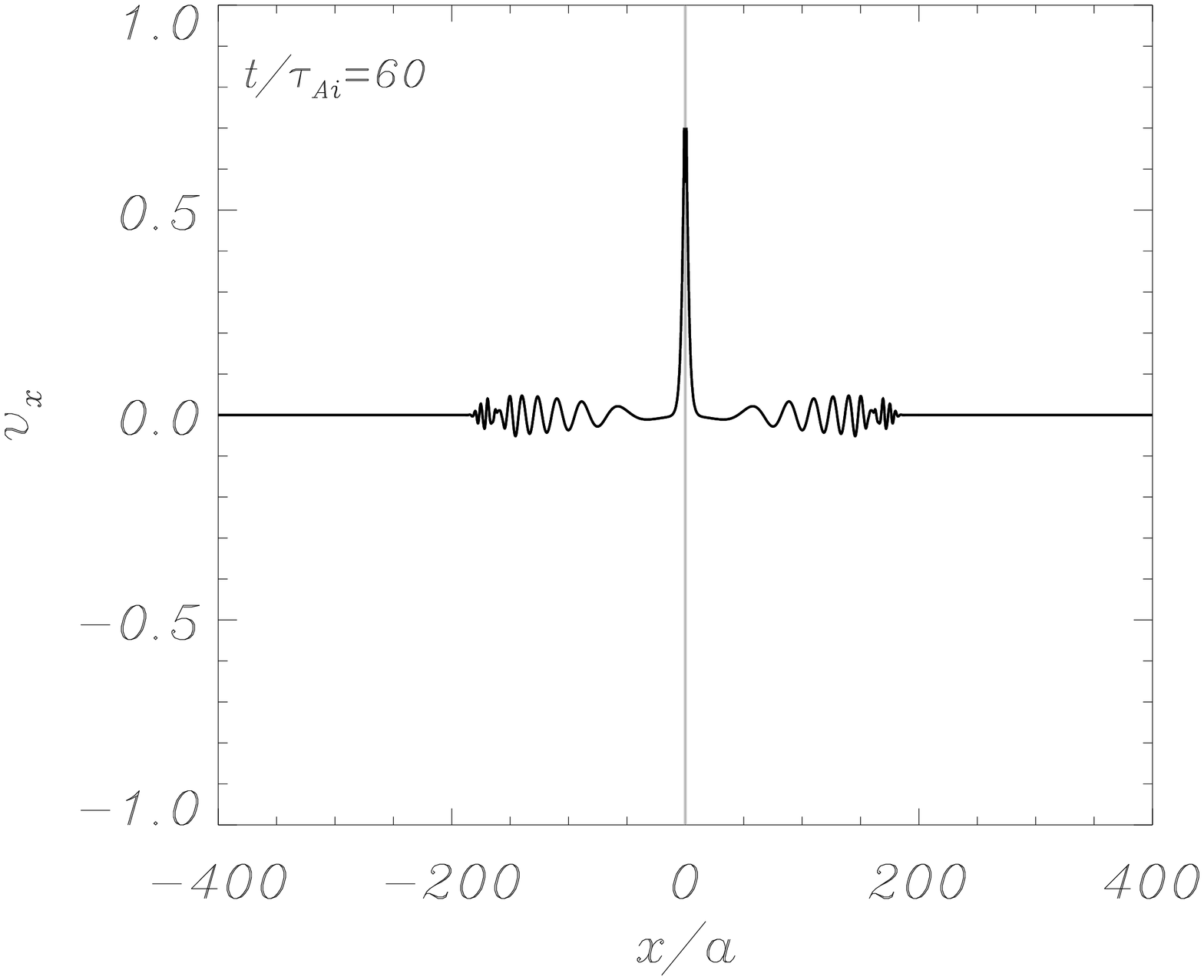}\includegraphics[width=0.4\textwidth,angle=90]{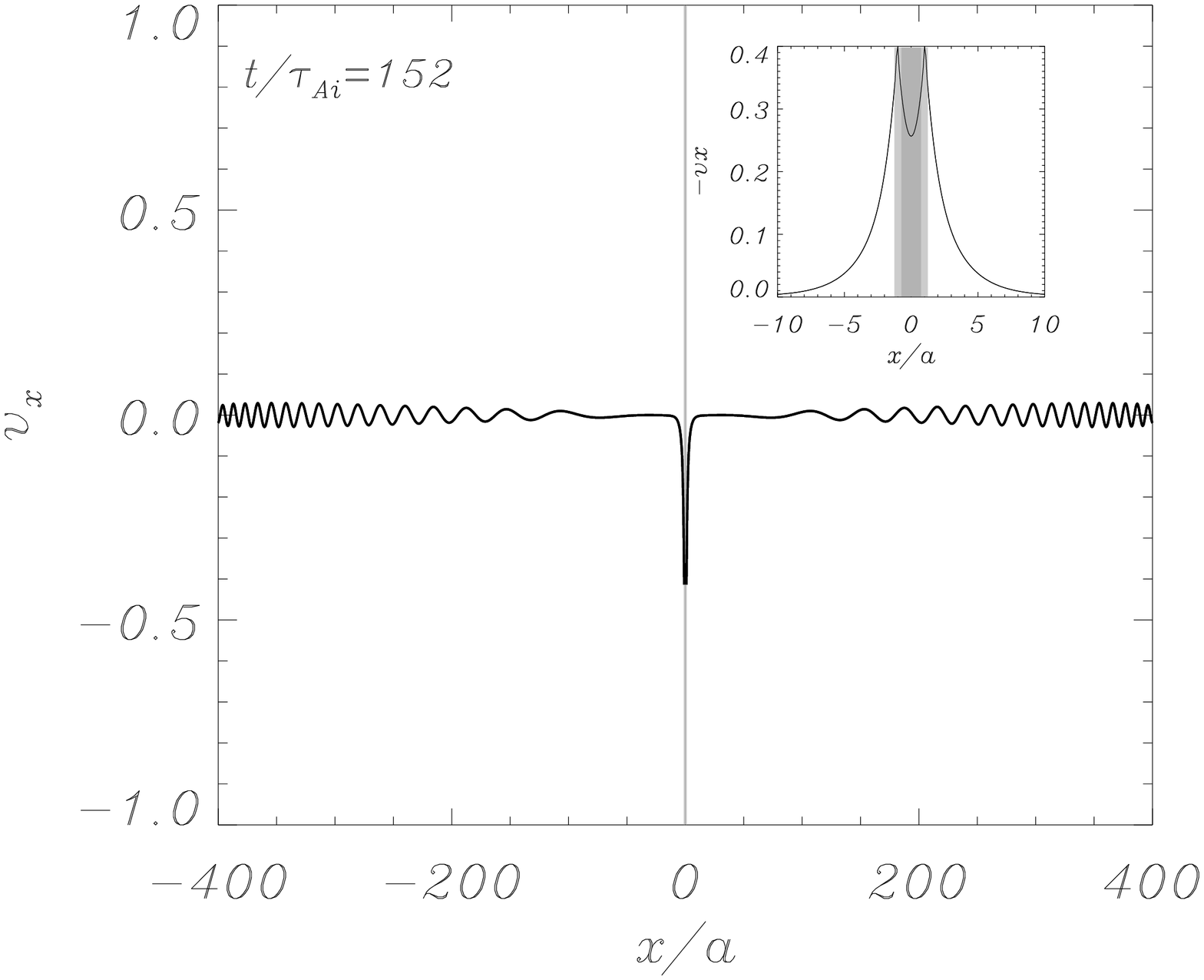}
\caption{Plot of $v_x$ for different times. The initial perturbation is given by
equation~(\ref{kinkexc}) with $w=2a$ and $a=0.02L$. In this simulation
$k_ya=0.5$. Note the different spatial scale
for $t=0$. The grey area represents the density enhancement
($\rho_{\rm i}/\rho_{\rm e}=10$).}
\label{arregui}
\end{figure*}

The effect of oblique propagation ($k_y\neq0$) in the slab, not included in the
previous models, was investigated by \citet{arrterretal07}. The properties of the
eigenmodes in this system are interesting since for large $k_y$ the body kink mode
is converted into a surface kink mode. This is clear in the time-dependent
solution plotted in Figure~\ref{arregui} for $k_ya=0.5$. The system is oscillating
with a trapped mode but the surface nature of the mode is revealed in
Figure~\ref{arregui}d, where we can appreciate that the amplitude of the
oscillation is highly localised around the loop boundaries. Moreover, the
frequency of the oscillations tends to the kink frequency (given by
eq.~[\ref{kinkeq}]). The effect of oblique propagation has been extensively
studied in the past specially in the context of resonant absorption at magnetic
interfaces \citep[see for example][]{leerob86,hollyang88}.

\subsection{Multi-structures}\label{2slabs}

Many loops in the solar corona are not isolated but forming groups or bundles of
loops. For this reason we now consider an idealised system of two parallel loops
modelled as dense plasma slabs of half-width $a$ and length $L$. The distance
between the centres of the slabs is $d$. Such a configuration was studied by
\citet{lunaterretal06}. The normal mode analysis indicates that there are two
kinds of normal modes \citep[see also][]{diazetal05}: solutions symmetric with
respect to $x=0$  (both slabs move in phase) and solutions antisymmetric with
respect to $x=0$  (the slabs move in anti-phase). The symmetric mode respect to
the centre of the structure is the only trapped mode for all distances between the
slabs while the antisymmetric mode is leaky for small slab separations.
Nevertheless, there is a wide range of slab separations for which the fundamental
symmetric and antisymmetric trapped modes are allowed and have very close
frequencies. These modes are excited according to the parity of the initial
perturbation. 

\begin{figure*}[ht]
\centering
\mbox{\hspace{-0.cm}{(a)}\hspace{5.6cm}{(b)}\hspace{5.2cm}}\vspace{-0.5cm}
\includegraphics[width=6cm]{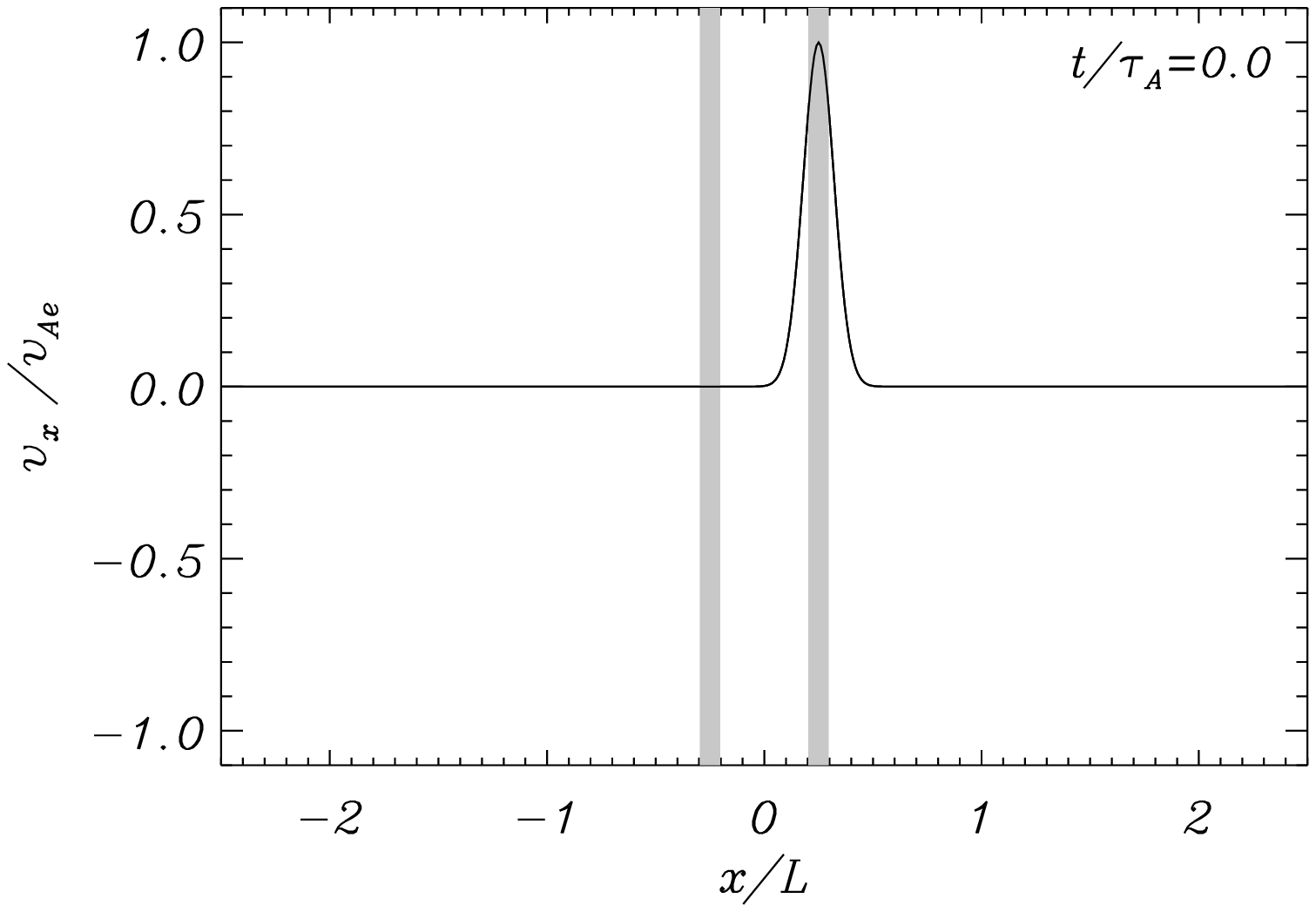}\includegraphics[width=6cm]{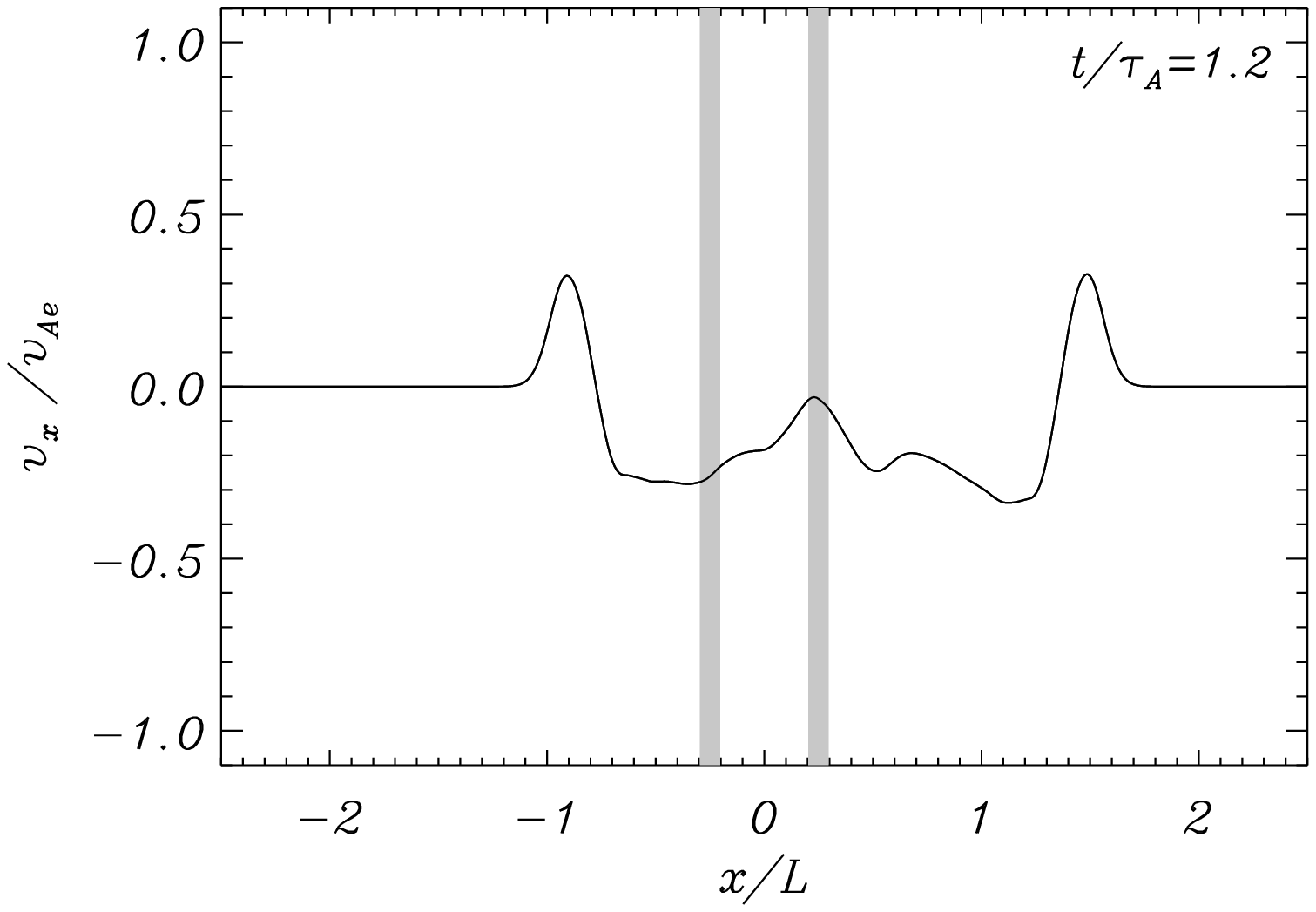}
\mbox{\hspace{-0.cm}{(c)}\hspace{5.6cm}{(d)}\hspace{5.2cm}}\vspace{-0.5cm}
\includegraphics[width=6cm]{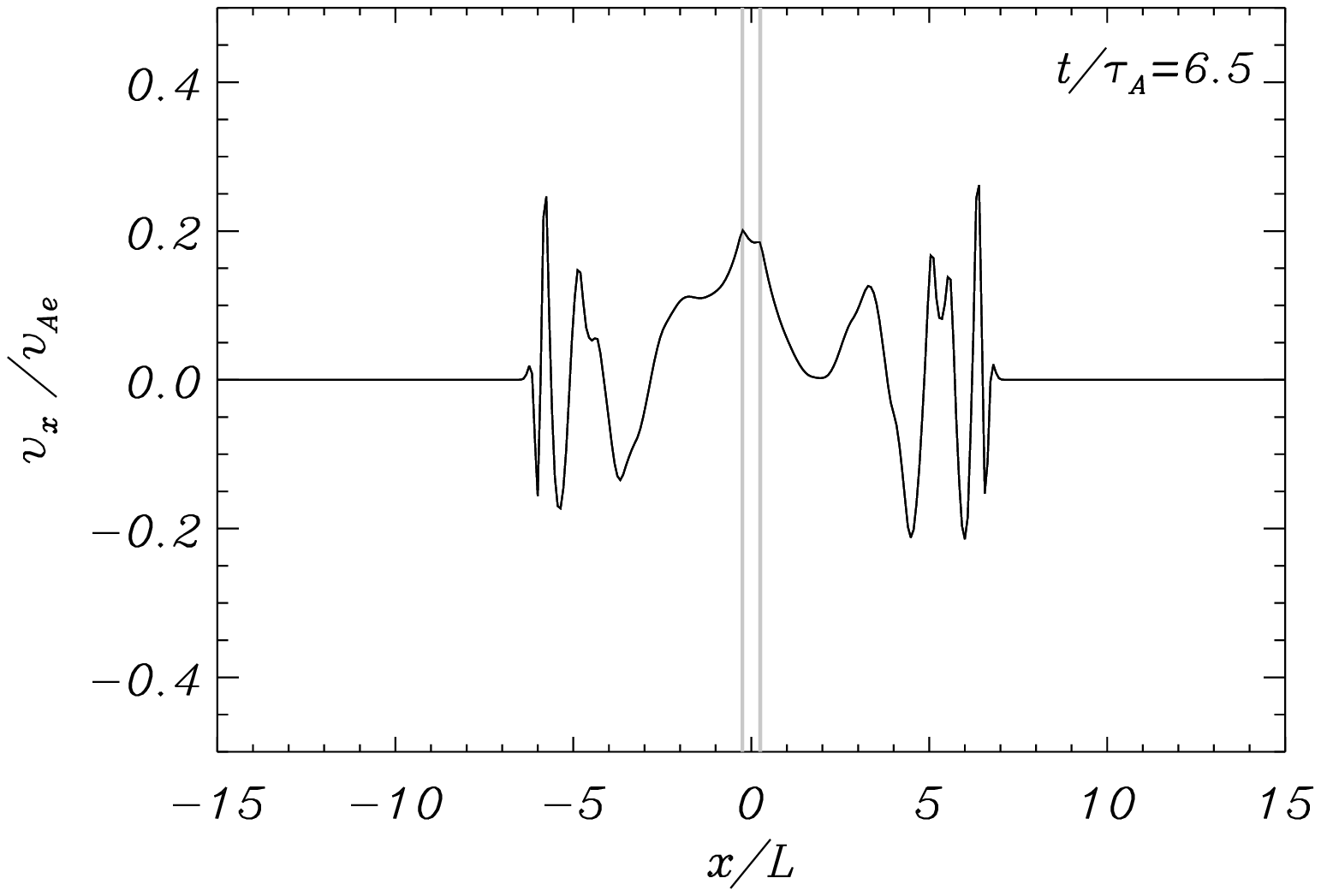}\includegraphics[width=6cm]{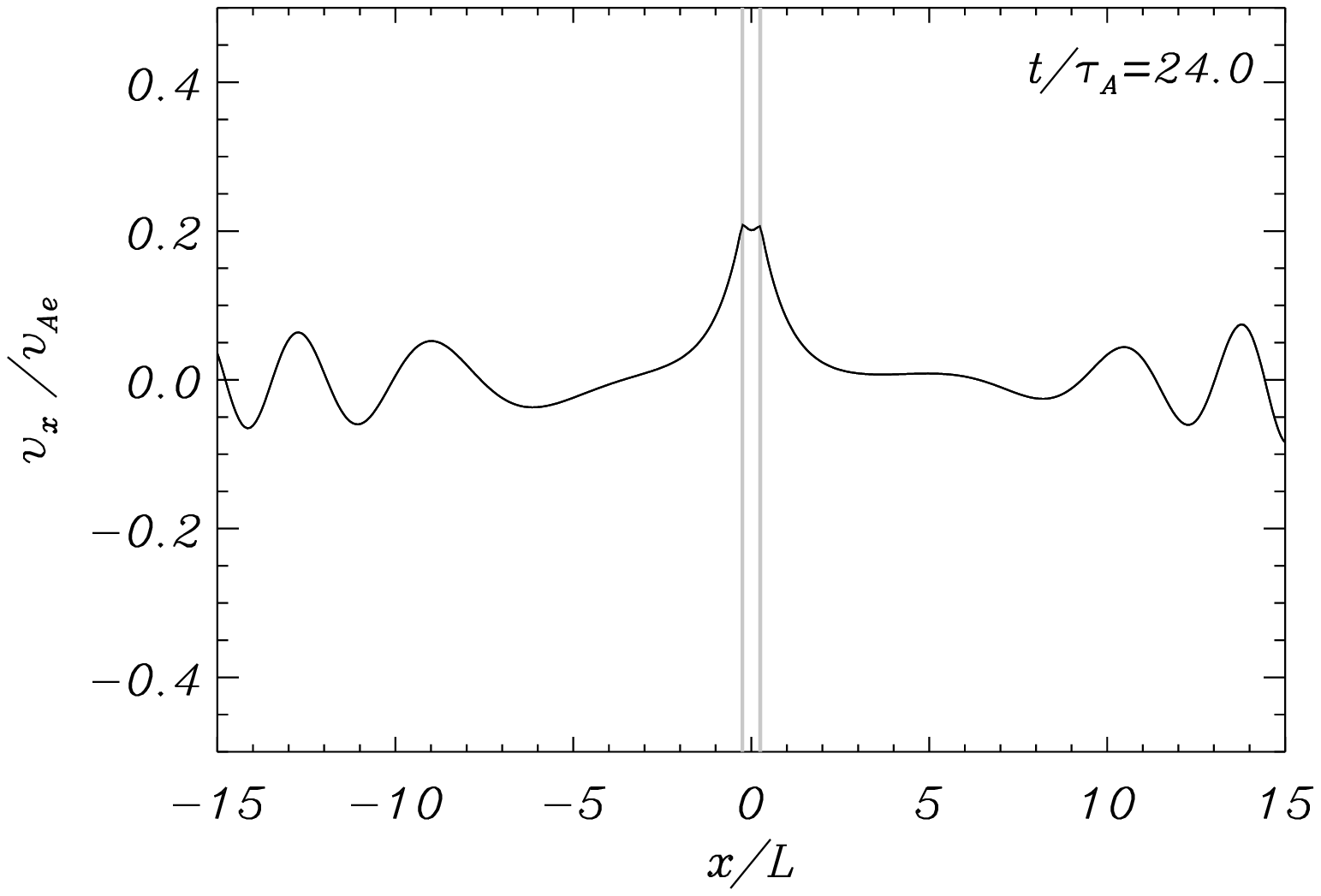}
\caption{Time-evolution of $v_x$ for $d=0.5 L$ and a non-symmetric initial
excitation. The grey areas represent the density enhancements ($\rho_{\rm
i}/\rho_{\rm e}=3$).}
\label{animation_d05}
\end{figure*}

\begin{figure*}
\centering
\mbox{\hspace{-0.cm}{(a)}\hspace{5.6cm}{(b)}\hspace{5.2cm}}\vspace{-0.5cm}
\includegraphics[width=6cm]{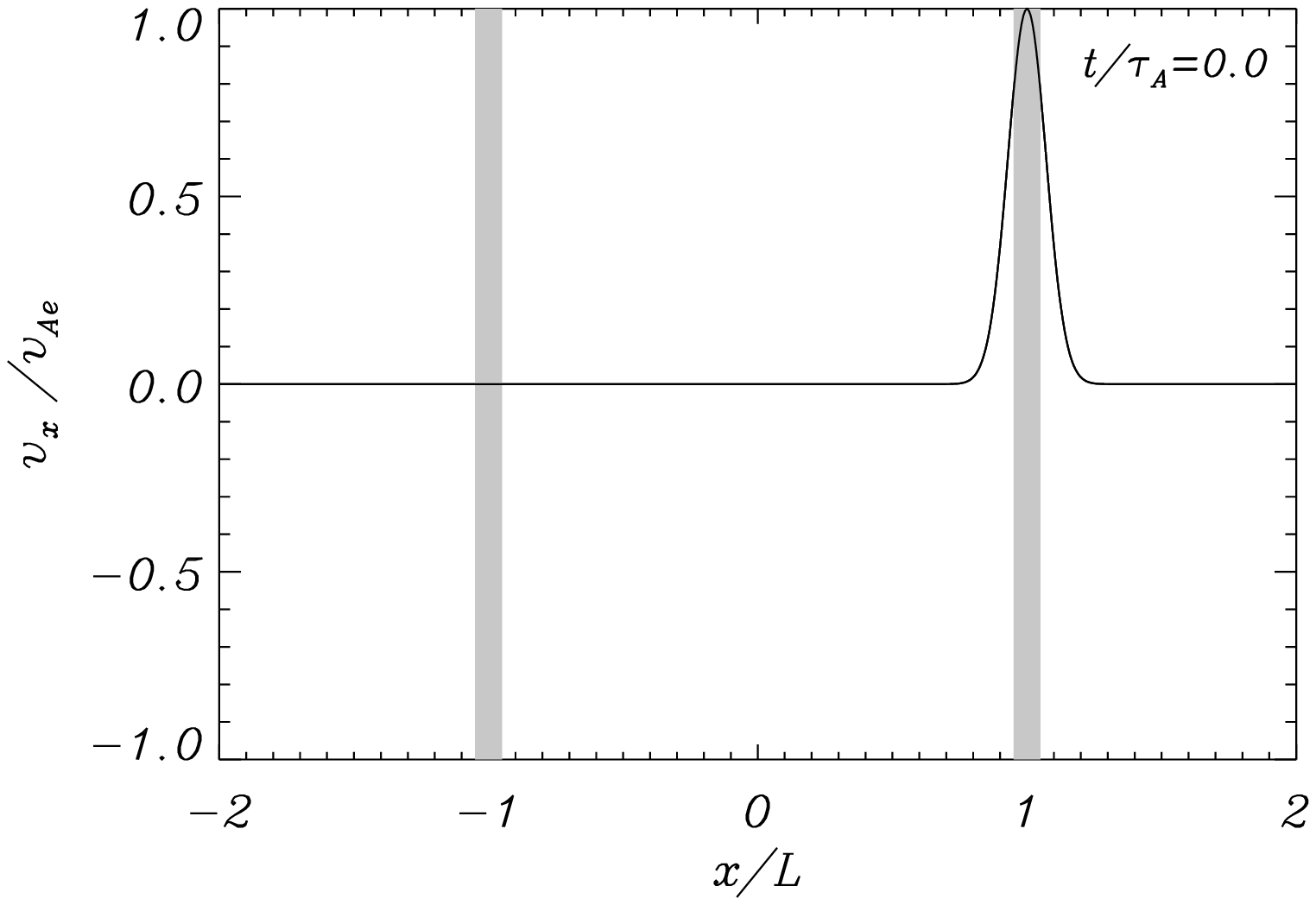}\includegraphics[width=6cm]{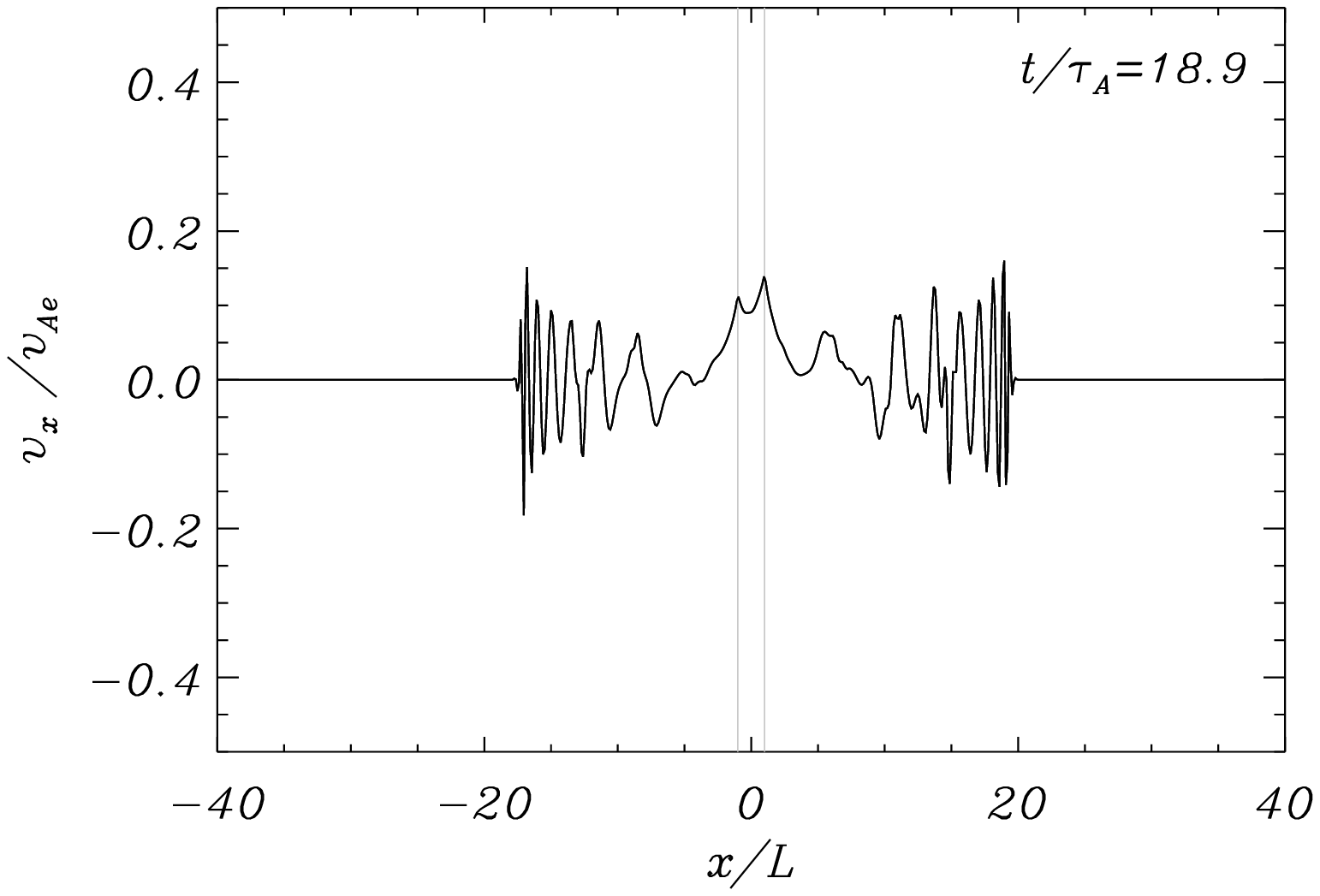}
\mbox{\hspace{-0.cm}{(c)}\hspace{5.6cm}{(d)}\hspace{5.2cm}}\vspace{-0.5cm}
\includegraphics[width=6cm]{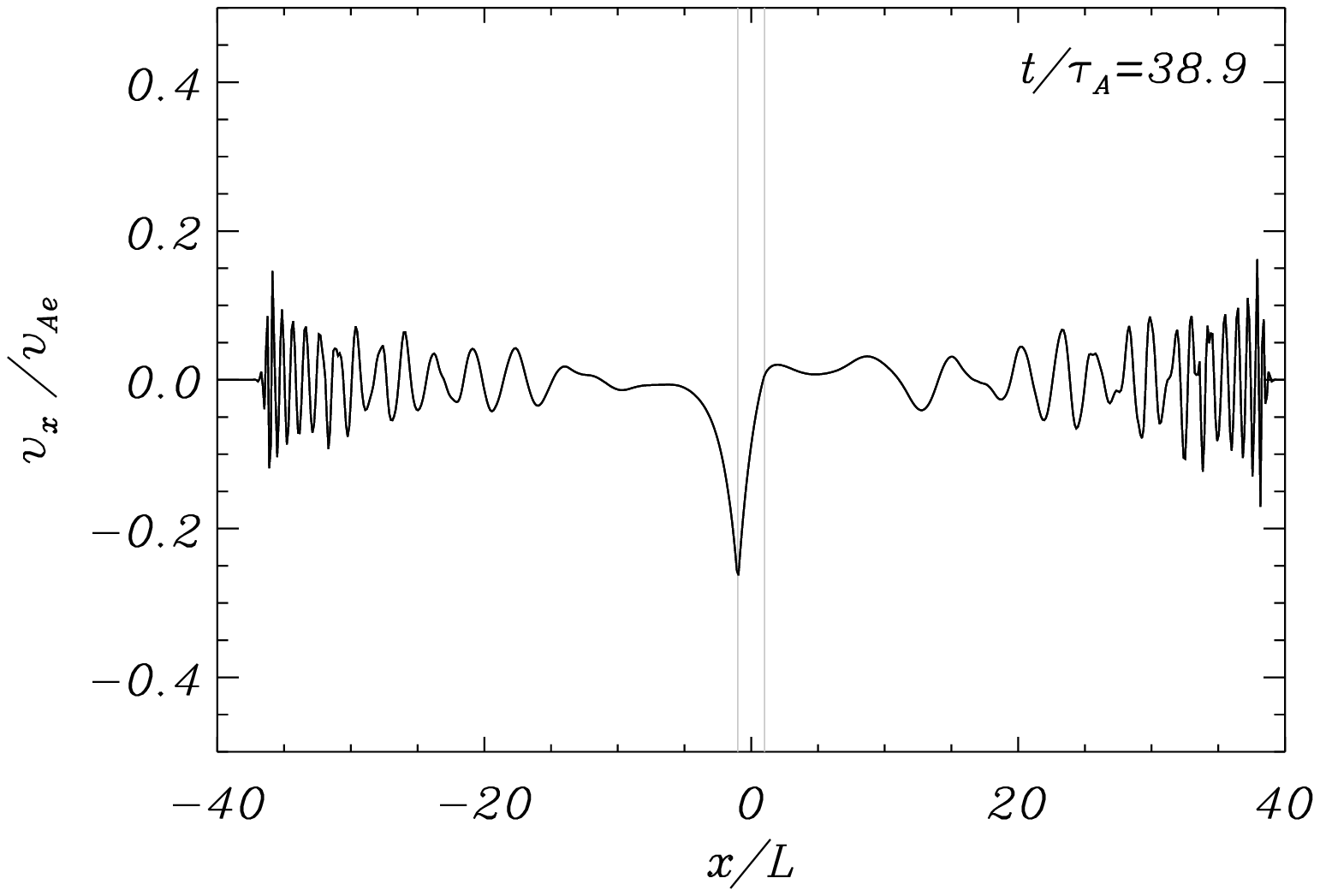}\includegraphics[width=6cm]{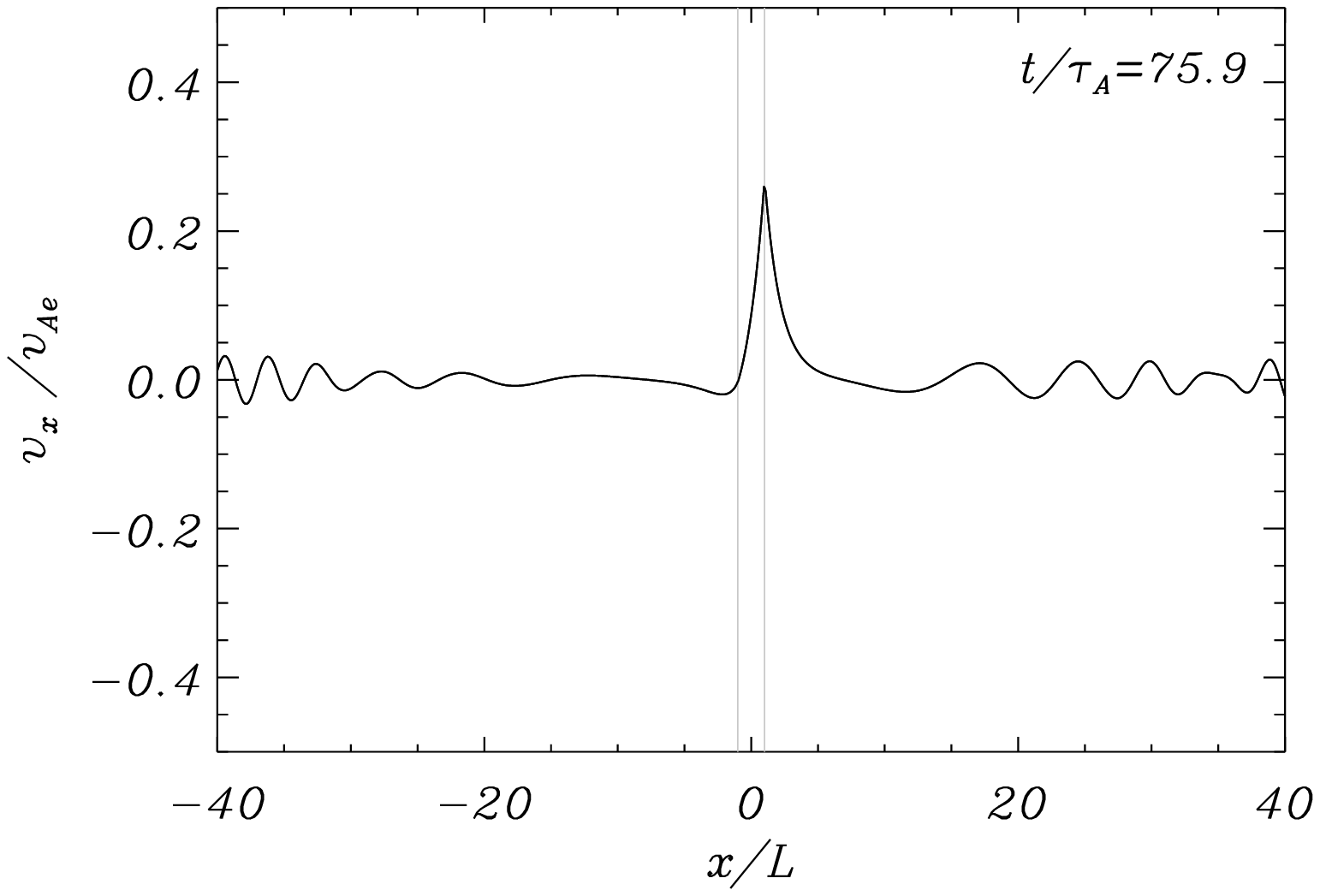}
\caption{
Time evolution of $v_x$ for $d=2L$ and a non-symmetric initial disturbance. Note
the interchange of energy between the two slabs in the last two frames.
}
\label{animation}
\end{figure*}

\begin{figure*}
\center{
\includegraphics[width=6cm]{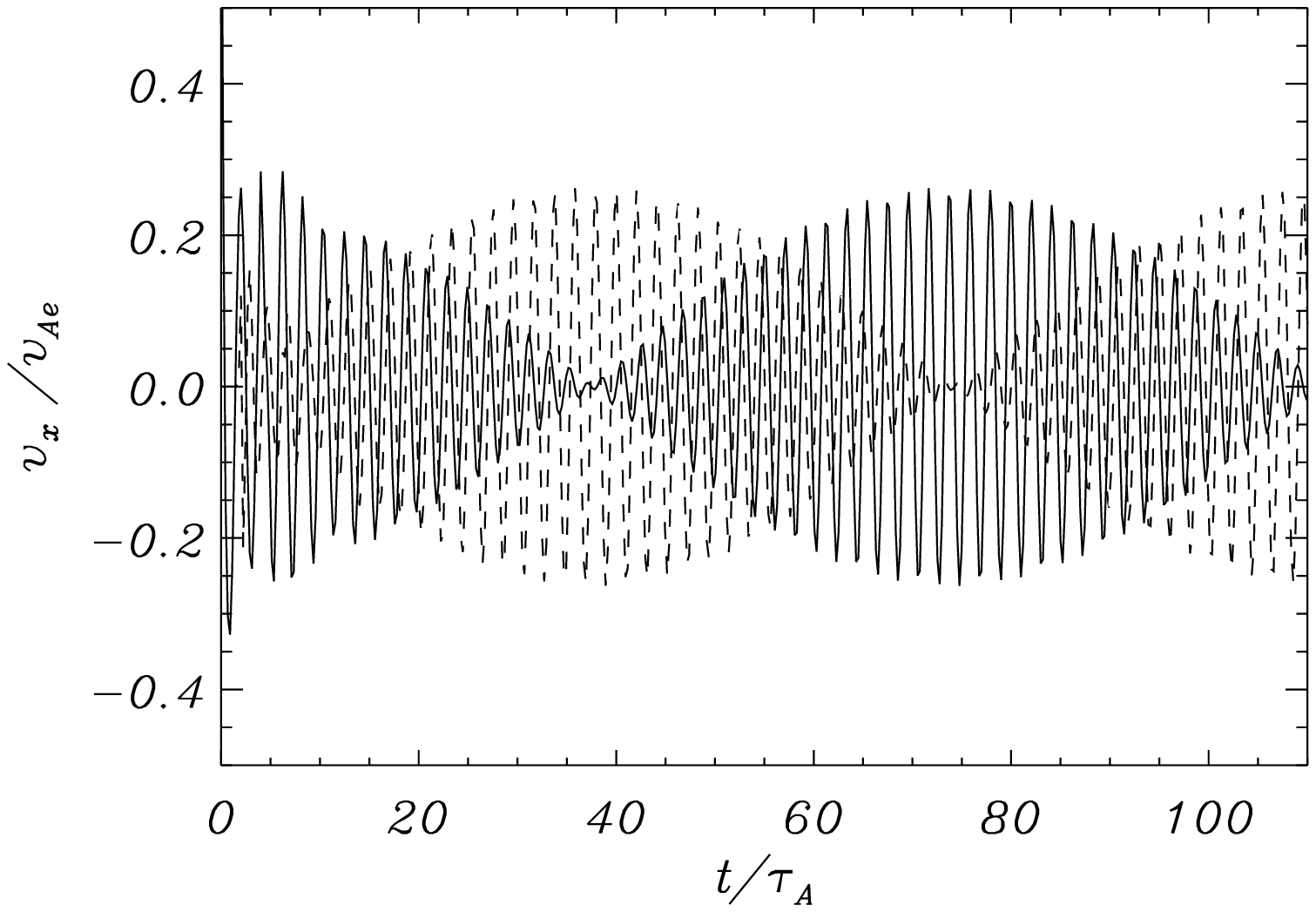}
\includegraphics[width=6cm]{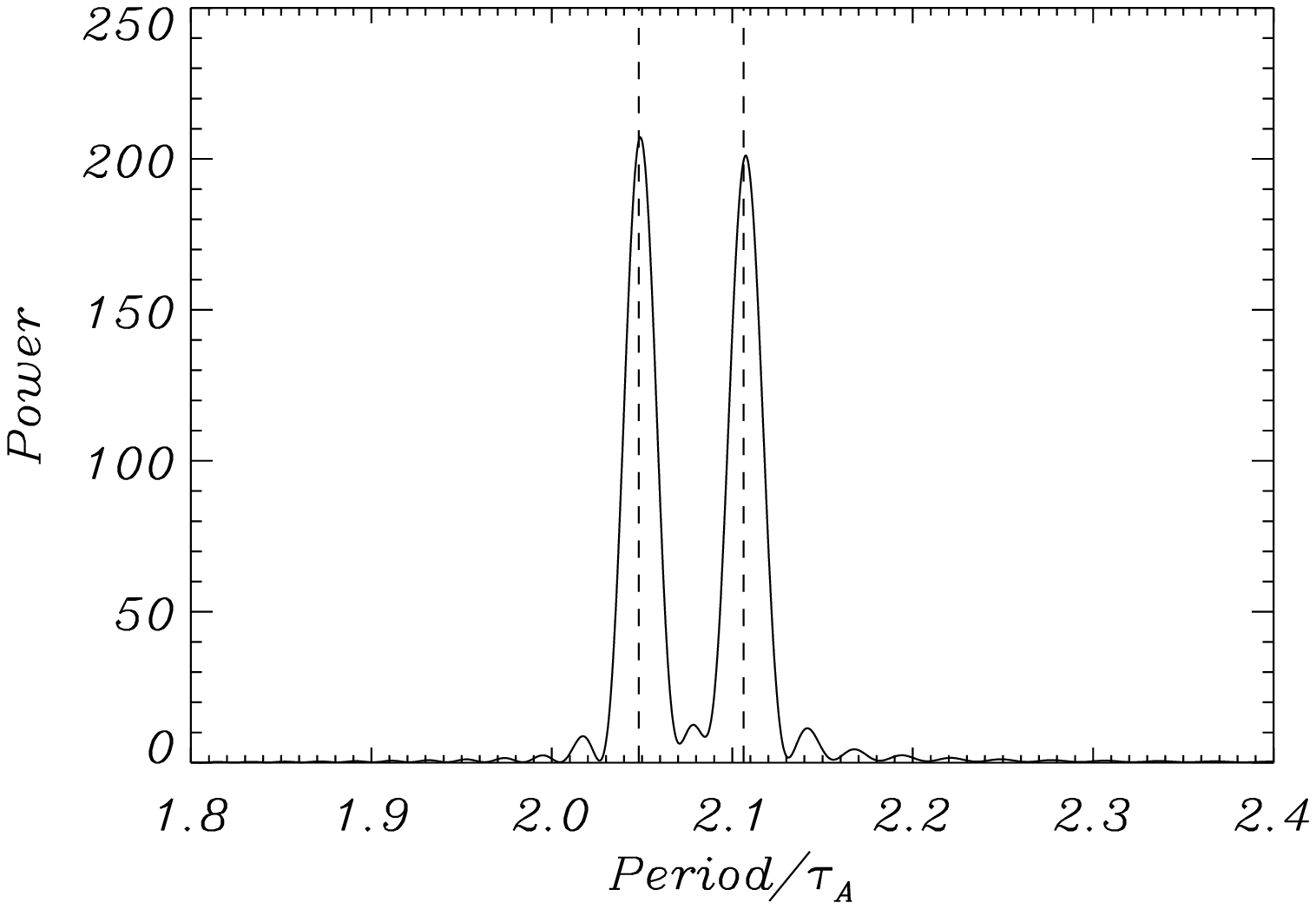}}
\caption{ 
(a) $v_x$ measured at the centre of the slabs for the simulation shown in
Figure~\ref{animation}. The solid and dashed lines correspond to the right and
left slabs, respectively. (b) Power spectrum of the previous signal (solid
line). The two vertical dashed lines indicate the  periods of the fundamental
antisymmetric mode and the fundamental symmetric
mode calculated from the eigenvalue problem.
}
\label{beating_curve}
\end{figure*}

In Figure~\ref{animation_d05} the results of an initial excitation for a loop
separation $d=0.5 L$ are represented. The initial disturbance is half the sum of
the symmetric and antisymmetric initial conditions. During the initial stages of
the temporal evolution (Figs.~\ref{animation_d05}b and c) $v_x$ has no definite
symmetry with respect to $x=0$ because the solution is the sum of the symmetric
and antisymmetric modes. These modes are the fundamental symmetric and the
fundamental antisymmetric, which are, for this particular slab separation trapped
and leaky, respectively. As a consequence, after some time
(Fig.~\ref{animation_d05}d) the antisymmetric mode amplitude is negligible in the
vicinity of the slabs and the system oscillates in a symmetric manner. The
oscillation is collective and different from the motion associated to the
individual eigenmodes of the slabs. Rather different results are found  when the
system is excited with the same initial condition but for a distance between the
slab centres of $d=2 L$. This choice of the slab separation results in the
fundamental antisymmetric mode becoming trapped. The evolution of the system is
plotted for different times in Figure~\ref{animation} and, although after some
time the two slabs seem to move in phase (see Fig.~\ref{animation}b), in a later
stage the right slab has transferred all its energy to the left slab and is
motionless (see Fig.~\ref{animation}c). At an even later time (see
Fig.~\ref{animation}d) the picture is just the opposite, with the left slab fixed
and the right slab in motion,  this is the result of the  continuous exchange of
energy. In Figure~\ref{beating_curve}a we have represented $v_x$ at the centre of
both slabs. Contrary to the behaviour in the stationary regime for symmetric or
antisymmetric initial perturbations, the oscillations do not attain a constant
amplitude, but they instead display a sinusoidal modulation. This is a beating
phenomenon due to the simultaneous excitation of the symmetric and antisymmetric
modes with similar frequencies. These frequencies are recovered from the power
spectrum of the velocity at the centre of right slab (see
Fig.~\ref{beating_curve}b), which shows two power peaks with periods almost
identical to those of the fundamental antisymmetric eigenmode and the  fundamental
symmetric eigenmode.

\subsection{Curved configurations} 

One of the main properties of the straight slab models described before is that
the system always has trapped modes. However, when the configuration is more
complex then the behaviour of the solution can be leaky. For example, in
two-dimensional curved configurations the Alfv\'en frequency changes (in general)
with height, and the phase speed of the body mode might be above the local
Alfv\'en speed at a certain position in the arcade. The character of the wave is
propagating and thus the mode is leaky, being its amplitude attenuated with time
due to the continuous radiation of energy.

\subsubsection{Circular slab}

\begin{figure}[!ht]
\center{
\includegraphics[width=0.3\textwidth,angle=-90]{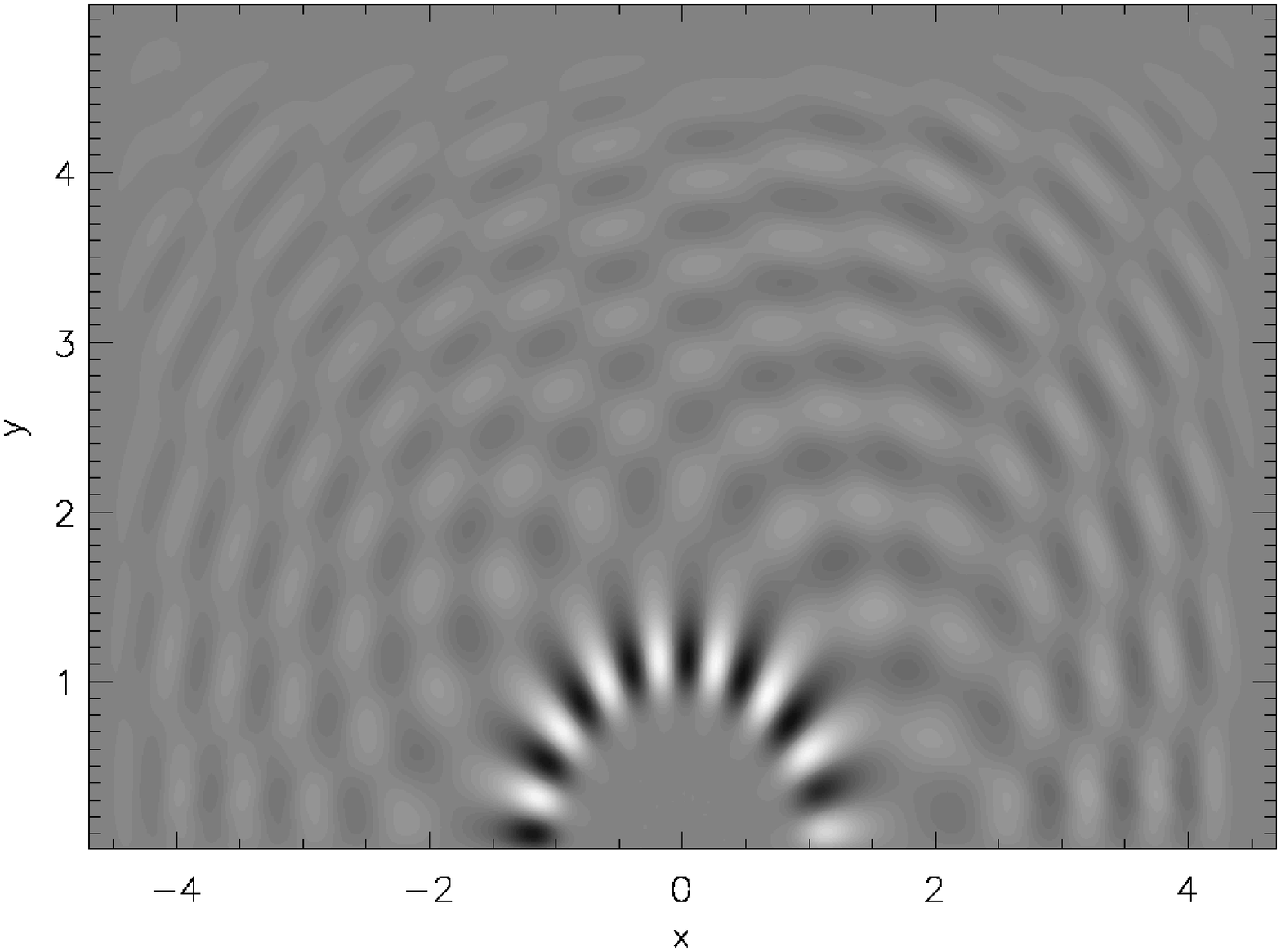}
\hspace{0.25cm}
\includegraphics[width=0.3\textwidth,angle=-90]{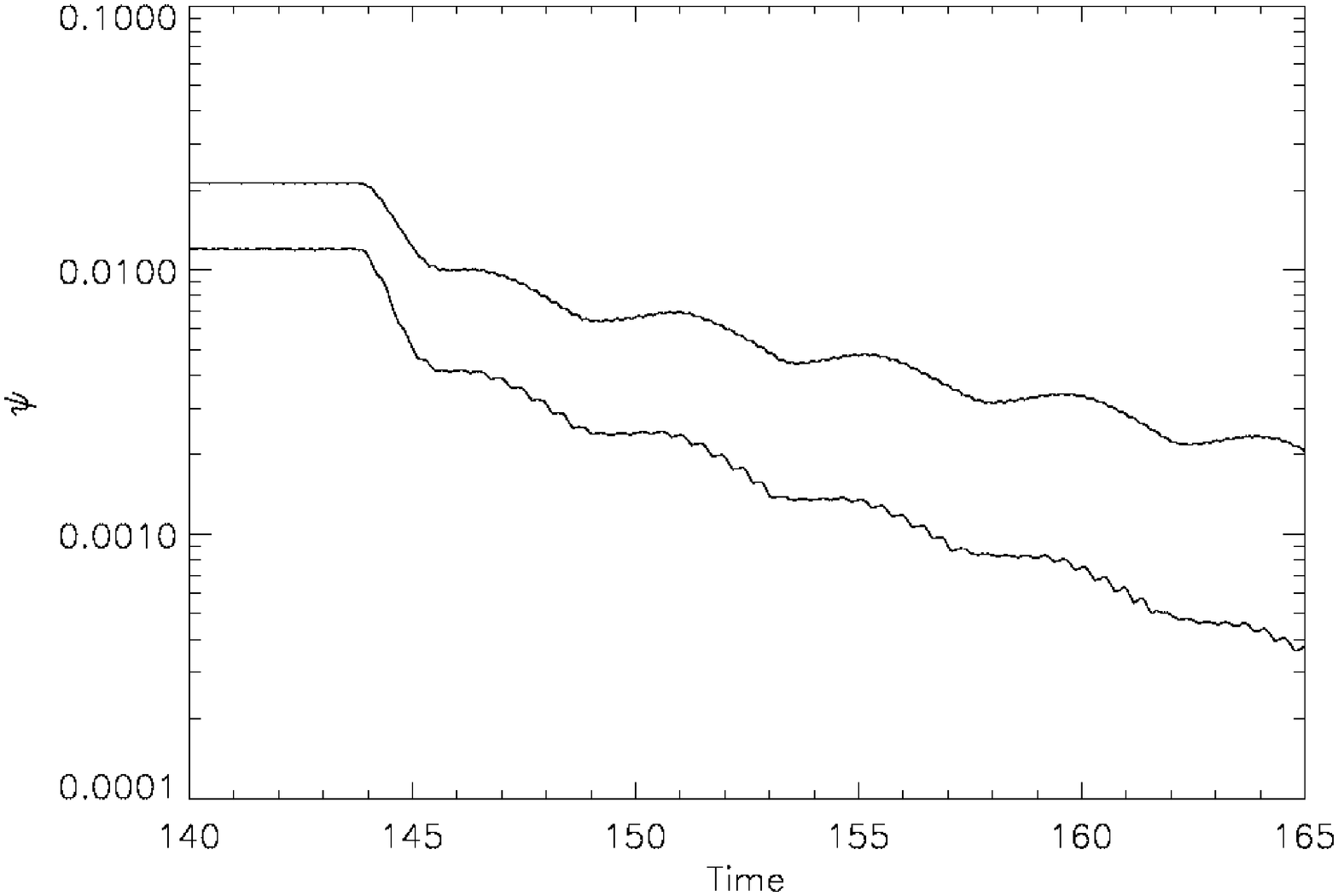}
}
\caption{ \small  (a) Velocity at the end of the simulation after the driver has
stopped. (b) Decay of the velocity inside the slab as a function of time due to
the emission of leaky waves. The system shows three distinct phases, the stable
regions due to the driver, a rapid decay phase when the driver is switched of
and the stable decay of the normal model of the tube. These calculations were
performed by \citet{bradarb05}.}
\label{bradarb05}  \end{figure}

The first time-dependent analysis of the excitation of (vertical) fast waves in a
curved slab (described in Section~\ref{curvedconf}) was done by \citet{bradarb05}.
They considered a driven problem at one footpoint (and line-tying conditions at
the other), and several fast body modes were excited. In Figure~\ref{bradarb05}a
there is an example of the excitation of a particular longitudinal harmonic. We can
appreciate that the energy is not only localised around the slab (with footpoints
around $x=\pm 1$), there are many fronts outside the slab that are a consequence
of the leaky character of the mode. The attenuation of the oscillations with time
due to the wave leakage is evident in Figure~\ref{bradarb05}b. \citet{bradetal06}
extended the previous model and considered in more detail the effect of wave
tunnelling on the vertical fast modes. Later, the eigenmodes of the circular slab
were analytically derived by \citet{veretal06,veretal06a} and \citet{diazetal06}.
\citet{veretal06a} showed that the calculated damping rate of fast magnetoacoustic
waves is consistent with the numerical simulations of \citet{bradarb05}, so a
clear link between the eigenmodes and the time-dependent solution could be
established.

\subsubsection{Slab embedded in a coronal arcade} 

\begin{figure}[!ht]
\center{
\includegraphics[width=0.3\textwidth,angle=-90]{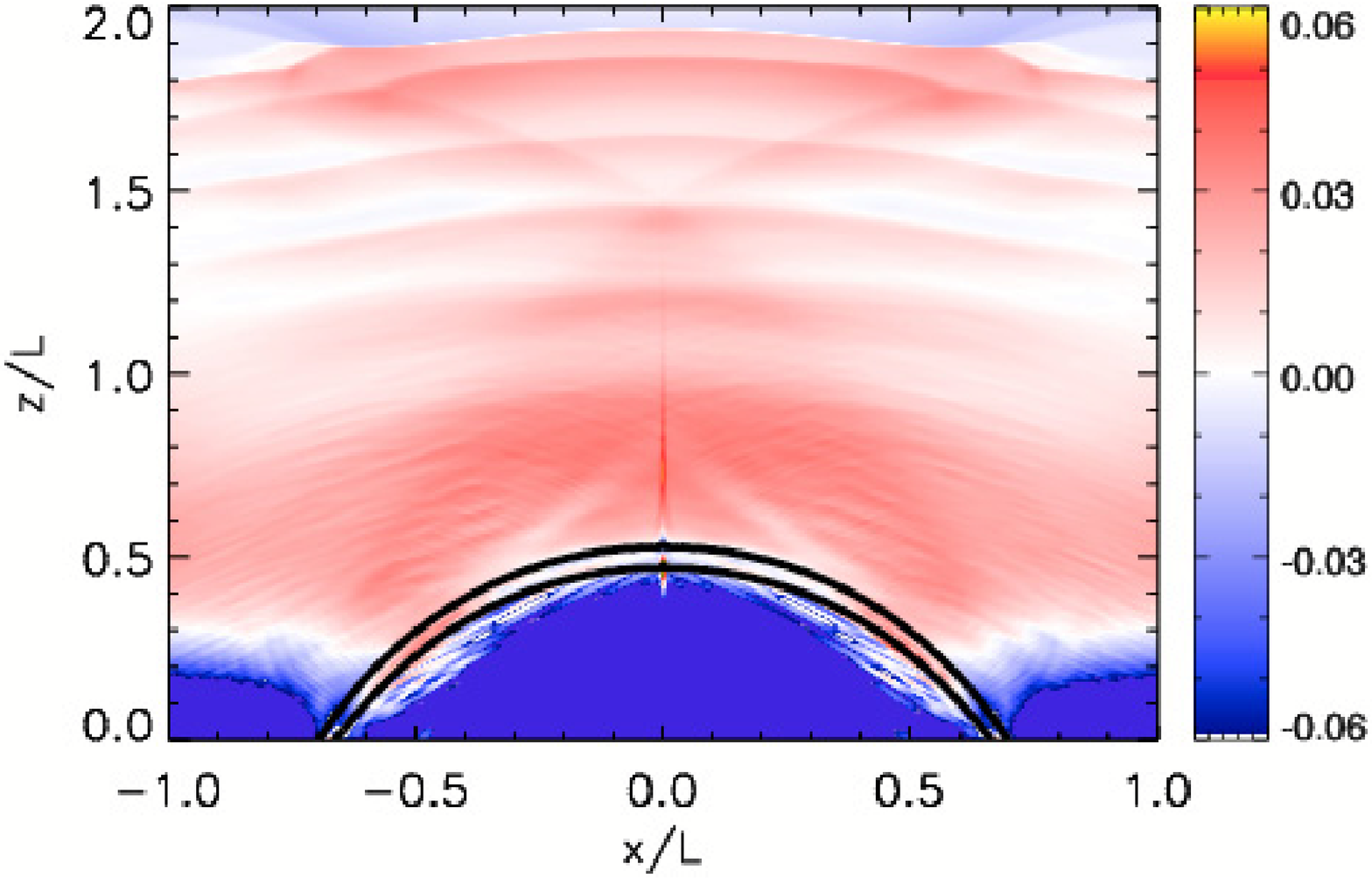}
\hspace{0.25cm}
\includegraphics[width=0.3\textwidth,angle=-90]{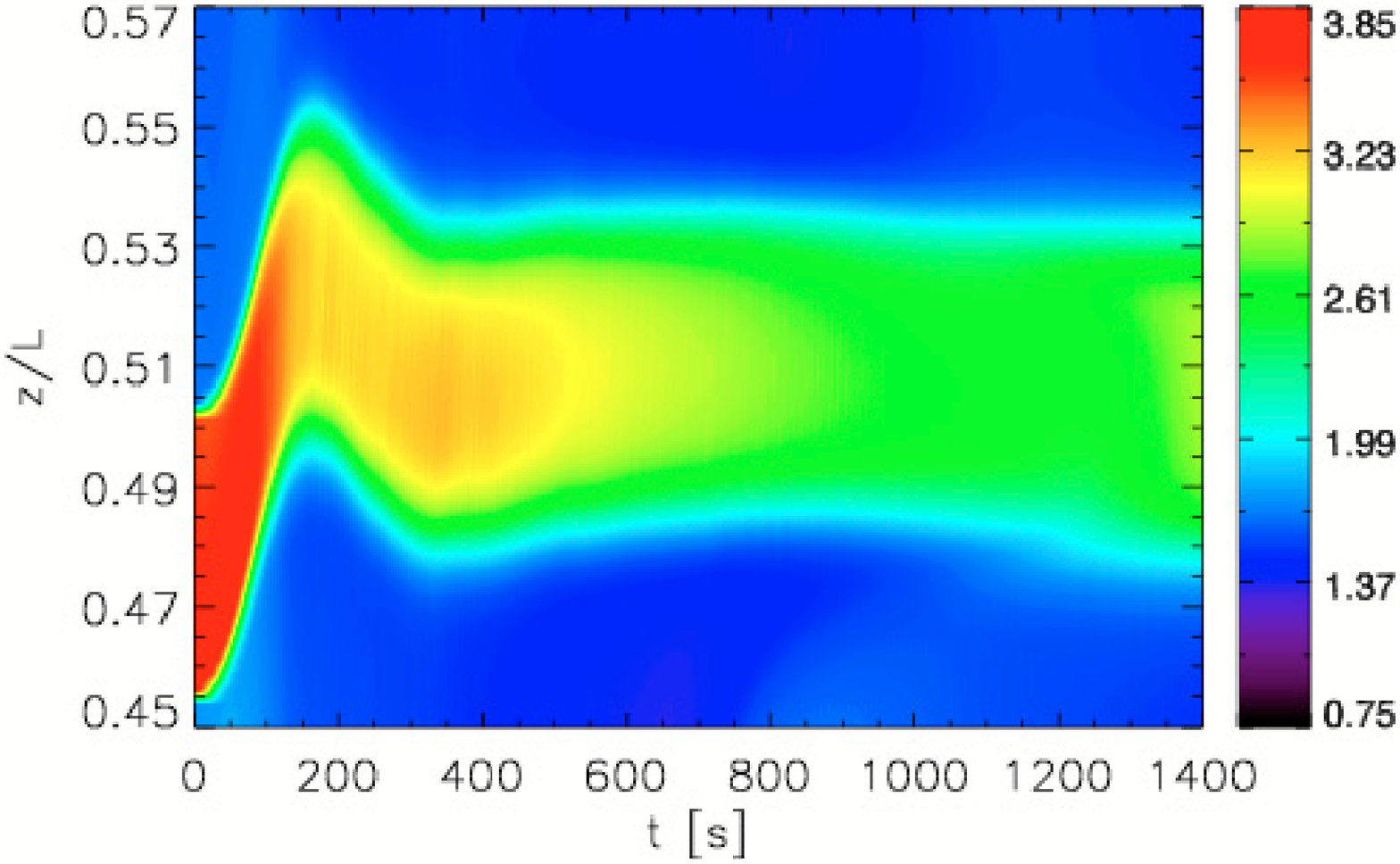}
}
\caption{ \small  (a) Perturbed energy density. The position of the curved loop
is  marked with the black lines. (b) Density evolution at the loop apex. These
results were derived by \citet{selwaetal07}.}
\label{selwa07f}  \end{figure}

The main properties of propagating fast MHD waves in coronal arcades without a
density enhancement have been described by  \citet{olivetal98,terrolietal08}.
Recently, a battery of papers has been devoted to the analysis of the properties
of these waves when a slab is embedded in the arcade. \citet{muretal05} showed
that an impulsively disturbance near the apex of the loop is able to excite
standing kink waves. Since the perturbation was very localised several standing
kink modes were excited at the same time. Similar results were found by
\citet{selwaetal05}, who was able to excite the fundamental standing kink mode
with a perturbation below the loop apex. A more detailed analysis was performed by
\citet{selwaetal06} who studied how the excitation of the kink modes depends on
the position of the pulse  and its width. More recently \citet{selwaetal07} have
extended the previous works and have shown that in the arcade leaky waves are also
excited. In Figure~\ref{selwa07f}a we can appreciate several wave fronts
propagating upwards which are related to the emission of leaky waves (see also
Fig.~\ref{bradarb05}a). It should be noted that these leaky waves are slightly
different from those discussed in Section~\ref{leakytrappslab}. However, the short
period leaky transients should be also present in the curved case, but apparently
there is no clear indication of such oscillations in the previous numerical
studies.

Interestingly, in all the simulations a strong damping of the vertical kink
oscillations is always reported (see Fig.~\ref{selwa07f}b). Although in the arcade
model there is leakage due to the process described in the previous section, it is
not clear that it can account for the strong damping found in the simulations.
Unfortunately, up to know there are no available eigenmode calculations of a slab
in the potential arcade \citep[see][for some preliminary attempts] {smithetal97}
which would help to clarify the origin of the damping. It is also interesting to
mention that in many simulations the loop does not return to its original
position, as can be appreciated in Figure~\ref{selwa07f}b. This issue needs to be
investigated in detail, since in the linear regime it is expected the loop to
oscillate around the initial equilibrium. However, if the amplitude of the initial
oscillation is large enough, the system can relax to a new magnetohydrostatic
equilibrium.

On the other hand, the influence of the dense photosphere on vertical
oscillations has been analysed by \citet{gruetal08}. By including a dense
photosphere-like layer, instead of line-tying conditions, these authors claim
that in this model there is a more efficient excitation and attenuation of the
vertical kink mode due to the leakage through the photosphere.
\citet{grumur08} have analysed the effect of gravity on the oscillations and
they have shown that the reported damping rate decreases as the coronal
scale-height increases. However, this damping rate is still small compared with
the observations. On the other hand, the effect of multi-structure in the arcade
has been modelled by \citet{gruetal06} by including a double stranded loop
and up to five strands.

It is interesting to note that leakage in these two-dimensional structures is
probably overestimated in comparison with three-dimensional loop models. In
three-dimensions the eigenfunction of the kink mode is much more localised near the
tube and the effect of inhomogeneities in the external medium is presumably much
less important \citep[see][]{terretal06a}.

\section{Waves in Cylinders}

We turn now our attention to the analysis of kink body waves in the cylindrical
loop model.

\subsection{Leaky and trapped waves}\label{cylsim}

\begin{figure}[hh] \center{
\resizebox{6.cm}{!}{\includegraphics{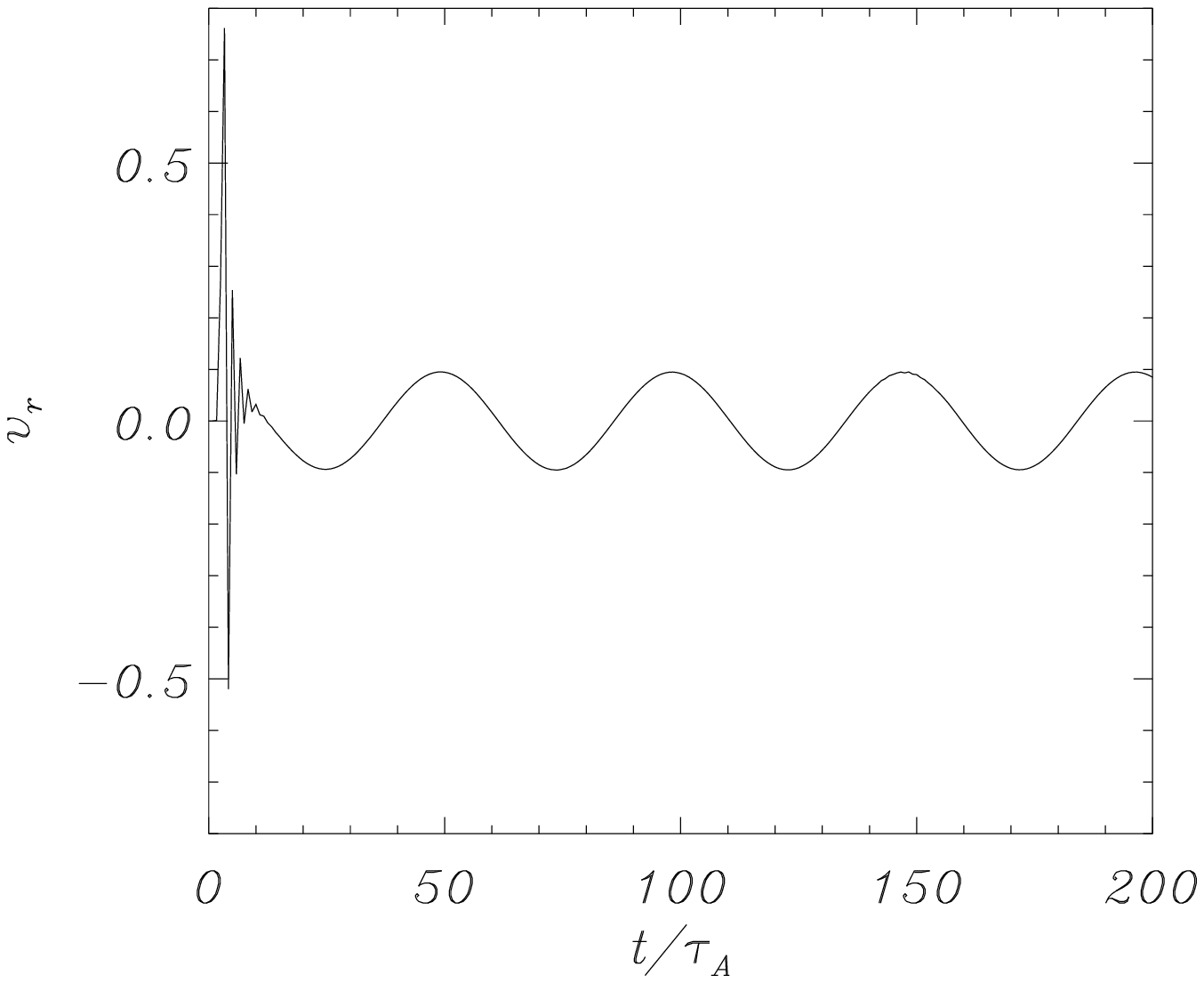}}
\resizebox{6.cm}{!}{\includegraphics{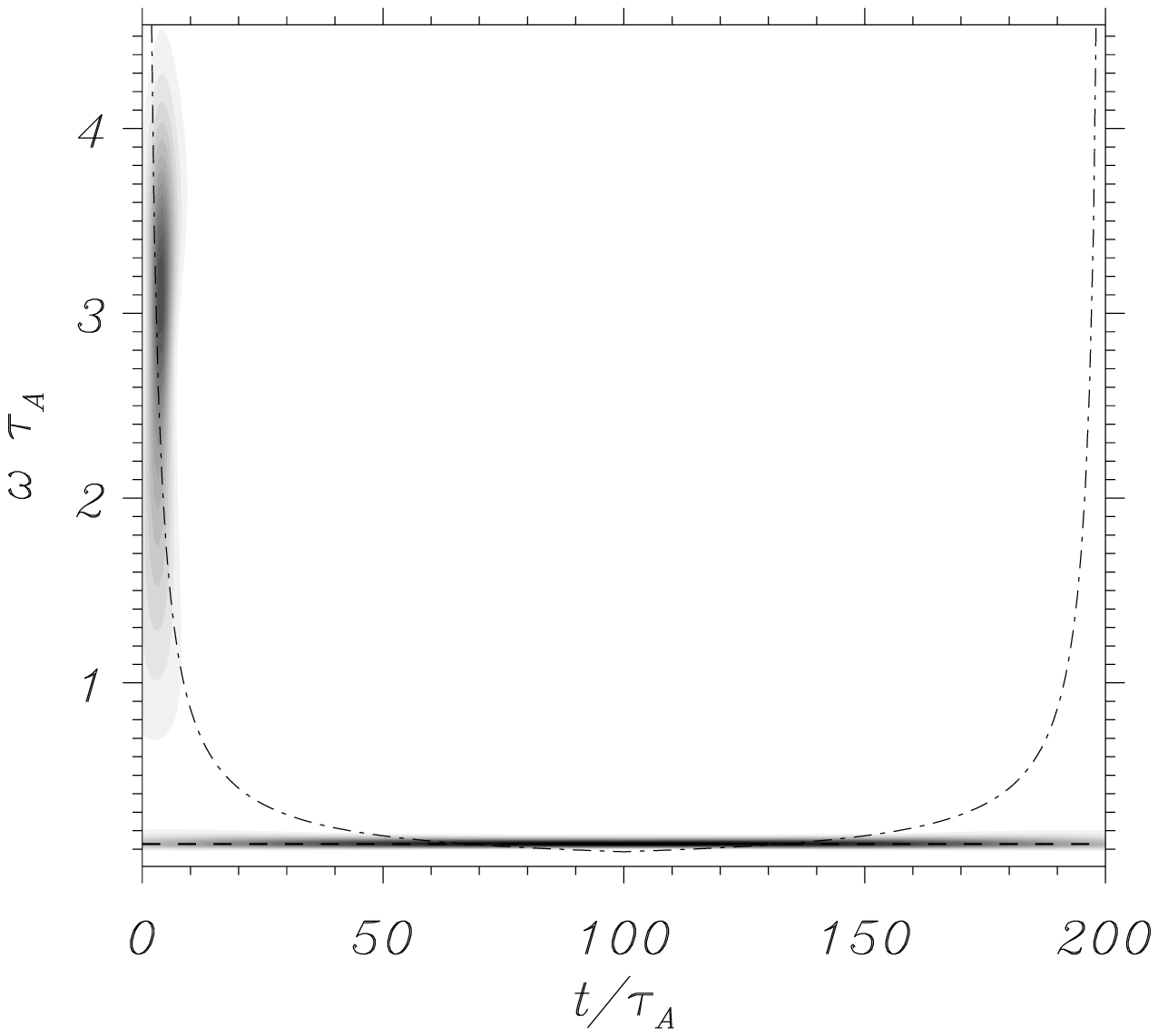}}
} 
\caption{\small {a}) Radial velocity as a function of time at $r=0$. 
{b}) Wavelet
transform of the signal. After a short transient phase, the loop
oscillates with the frequency of the kink mode. The loop length is $L=30\,R$
and the density contrast is $\rho_{\rm i}/\rho_{\rm e}=3$. The initial perturbation if given
by equation~(\ref{kinkexc1D}) with $r_0=5\,R$, $w=R$. The dashed line
represents the frequency of the kink trapped mode calculated from the
dispersion relation.
}\label{kinktemp}
\end{figure}

\begin{figure}[hh] \center{
\resizebox{6.cm}{!}{\includegraphics{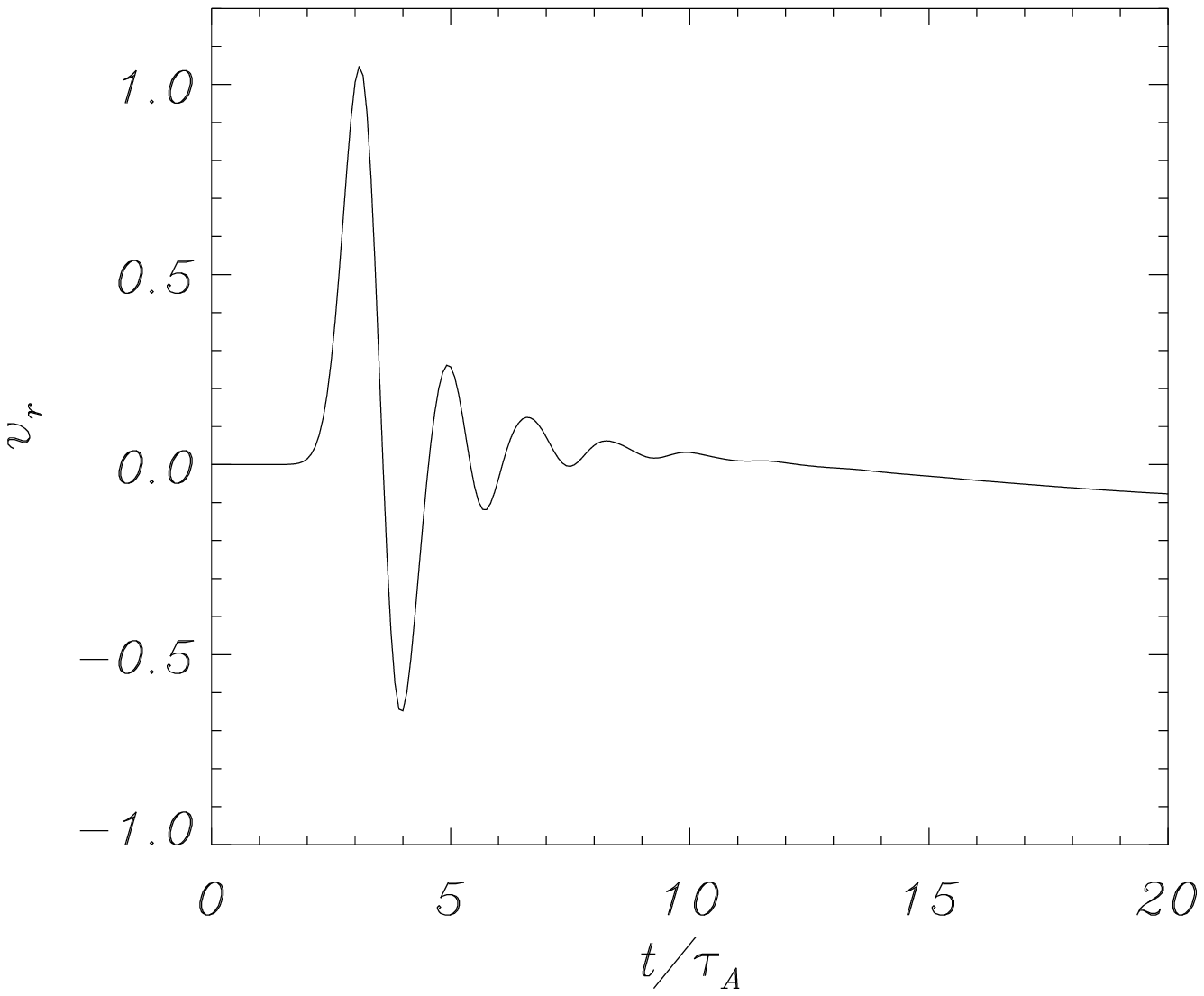}}
\resizebox{6.cm}{!}{\includegraphics{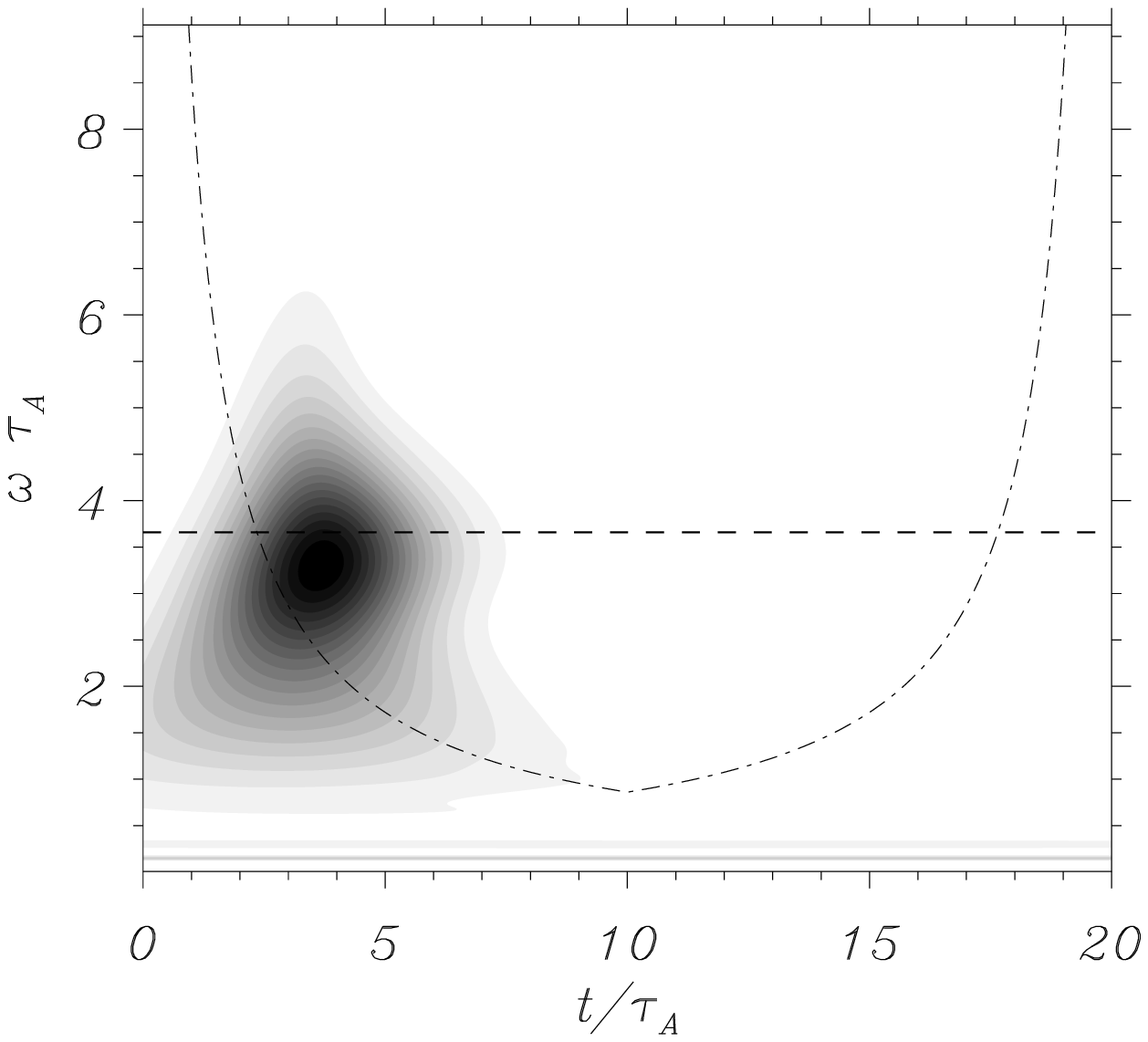}}
} 
\caption{\small {a}) Detail of the radial velocity plotted in Figure~\ref{kinktemp} 
in the range  
$0\leq t/\tau_{\rm A}\leq 20$. {b}) Wavelet transform of the signal. The
leaky mode is clearly identified and there is some indication of the trapped
mode in the power spectrum. 
}\label{kinktemp1}
\end{figure}

The linearised MHD equations are solved in cylindrical coordinates, and since we
are interested in the kink mode, we impose $m=1$ in the equations
\citep[see][]{terrandetal07}. As in the slab model we perturb the system with an
external localised disturbance with the 
following form 
\begin{eqnarray}\label{kinkexc1D}
v_r(r,t=0)=v_{r0}\exp\left[{-\left(\frac{r-r_0}{w}\right)^4}\right],
\end{eqnarray} 

\noindent where $r_0$ is the position of the centre of the disturbance and $w$
is a parameter related to its width. The results 
are qualitatively similar to those found in the slab configuration. The initial
perturbation induces disturbances that propagate towards the loop and also
wavefronts travelling in the opposite direction. Part of the energy of the
initial perturbation is trapped while another part will be radiated through
the excitation of the leaky modes. The radial velocity component at the centre
of the loop is plotted in  Figure~\ref{kinktemp}a. The signal initially shows a
short transient phase ($0\leq t/\tau_{\rm A}\leq 20$) which is followed by a
long-period oscillation. This  oscillation is due to the excitation of the
trapped kink mode. Its frequency is simply the kink frequency of the tube (see
dashed line in Fig.~\ref{kinktemp}b). The amplitude of the oscillation of
the trapped mode remains constant with time. On the other hand, the initial
transitory phase is associated to the leaky modes, represented in more detail in
Figure~\ref{kinktemp1}a. The period of the transient phase is in agreement with
the period of the (fundamental) kink leaky trig mode (see dashed line in
Fig.~\ref{kinktemp1}b). According to \citet{cally86,cally03}, the trig mode has
the following period
\begin{equation}\label{leakper}
P\approx \frac{4 R}{\left(m+2n+\frac{3}{2}\right)v_{\rm Ai}},
\end{equation}
\noindent $n$ being any integer consistent with $P>0$.
Their damping time is
\begin{equation}\label{damp}
\tau_{\rm d}\approx R\, \frac{v_{\rm Ae}}{v_{\rm Ai}^2}.
\end{equation}

\noindent The previous expressions are valid for $v_{\rm Ai}\ll v_{\rm Ae}$ (and
for thin and long loops) and are the modifications due to the cylindrical
geometry of those derived for the slab (see eqns.~[\ref{pleaky}] and
[\ref{tdleaky}]). Again for these leaky modes the period and damping time are
independent of the loop length.  Additionally, in the dispersion diagram there
are other families of trig leaky modes, for example, the ``Type II trig modes"
and the ``orphan mode" \citep[see][]{cally86,cally03}. However, these modes seem
to be very difficult to excite by an initial perturbation. Something similar
happens with the Principal Fundamental Leaky Mode, although it is a solution of
the dispersion diagram (with a frequency very similar to that of the trapped kink
mode for $k_z R\ll1$) it is very difficult to excite by an initial disturbance 
since there is no clear evidence of such modes in the time-dependent solutions
\citep[see][for a detailed investigation about the excitation of this peculiar
mode]{terrandetal07}. In fact, \citet{rudrob06} claim that this mode is spurious.

\subsection{Energy deposition by initial disturbances}\label{energtrap}

\begin{figure}[ht]
\center{
\includegraphics[width=0.45\textwidth]{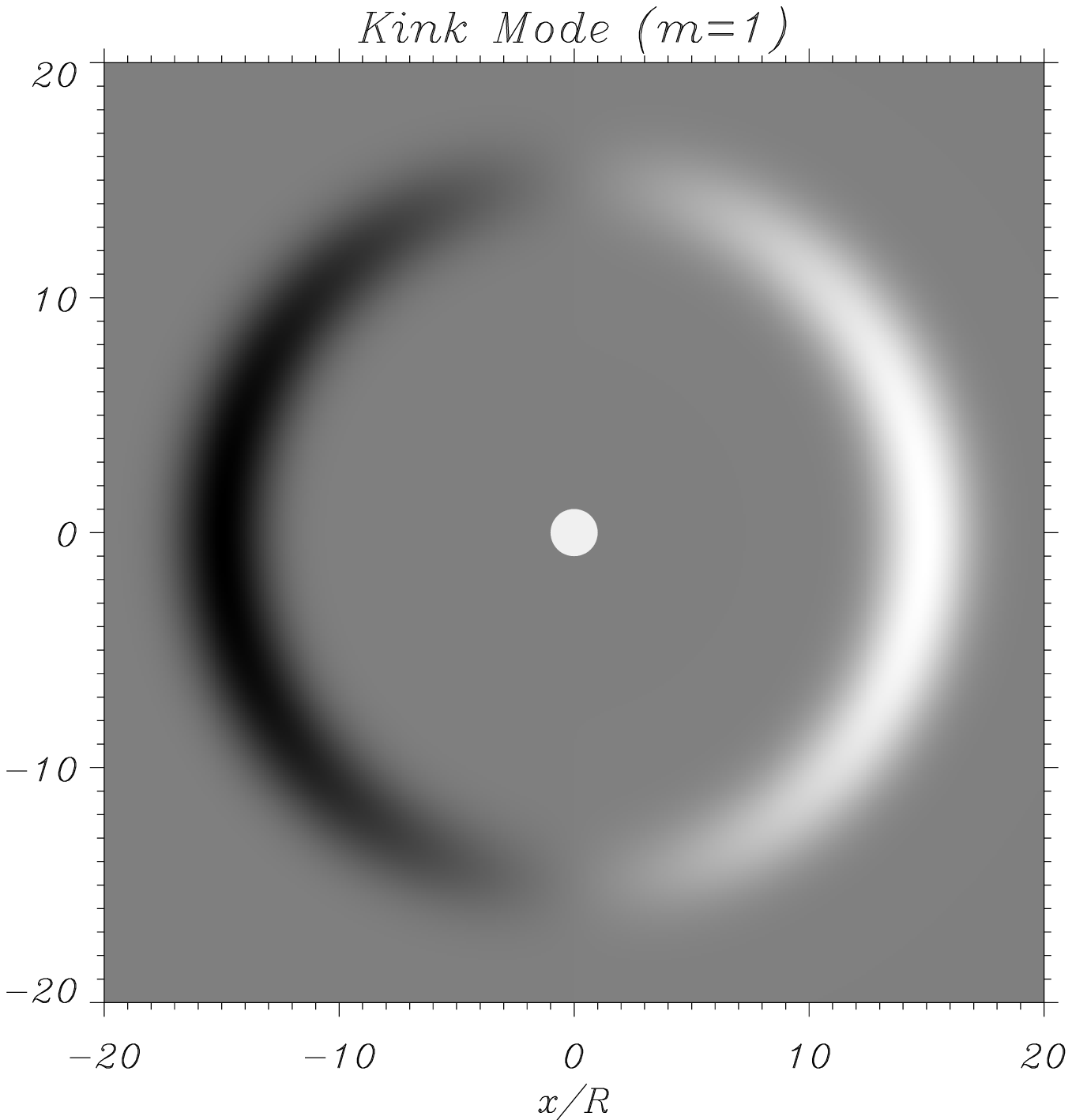}
\includegraphics[width=0.45\textwidth]{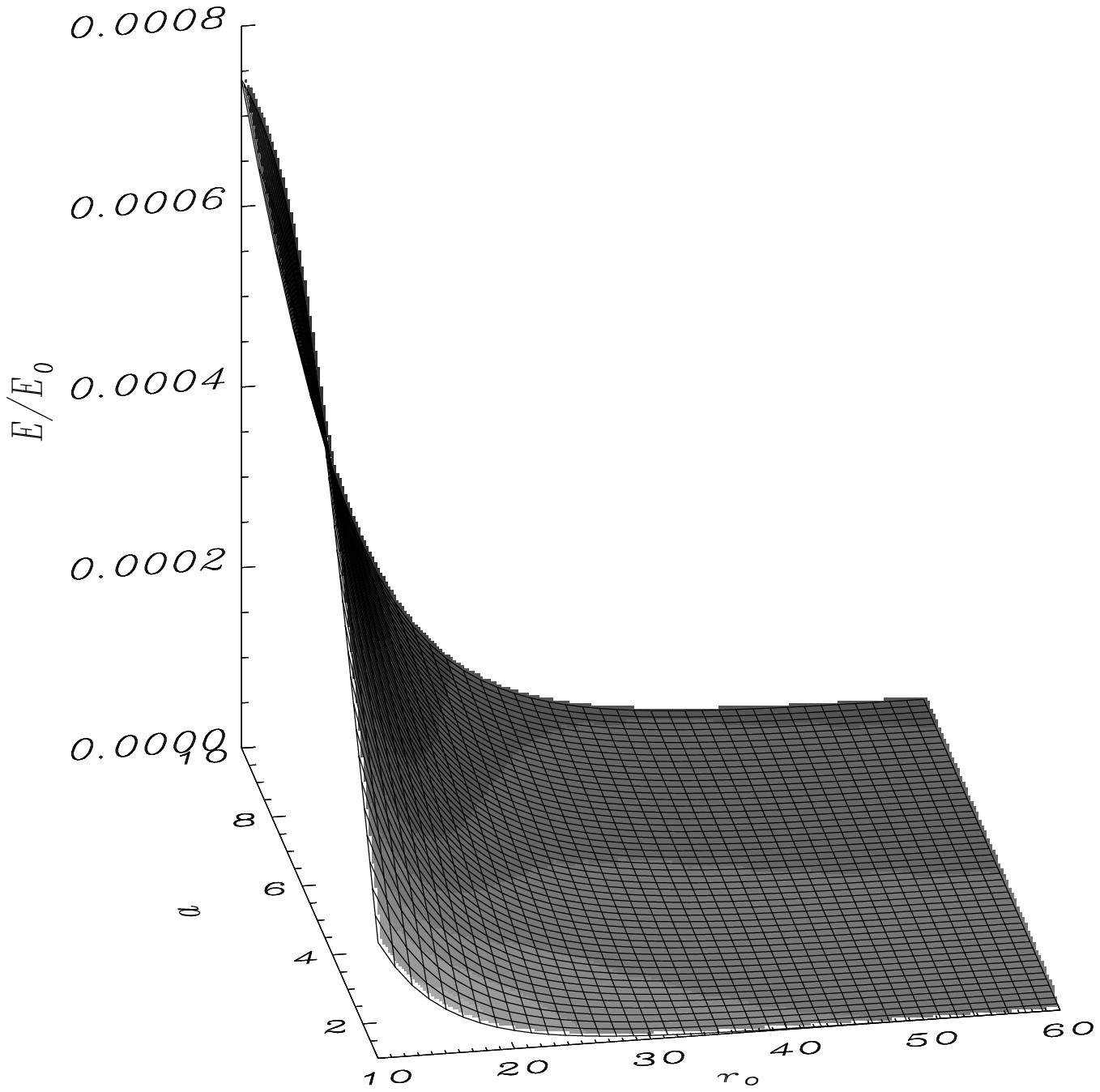}
}
\caption{
\small {({a})} Example of kink perturbation in the
perpendicular plane to the loop axis. For this initial disturbance $r_0=15R$ and
$a=2R$. In the vertical direction the dependence is
sin($kz$). White colour corresponds to
positive magnetic pressure perturbations while black colour represents negative
values. The circular white region
in the centre represents the loop. \small {(b)} Trapped energy as a function of $r_0$ and $a$ for the
$m=1$ mode. For this plot $L=50R$ and $\rho_{\rm i}/\rho_{\rm e}=3$. The behaviour of the
energy with $r_0$ and $a$ is given by equation~(\ref{Etdep}) in the limit $r_0\gg R+a$.}
\label{kinkflutt}
\end{figure}

Once we known the time signatures of the an external disturbance on the loop, we
turn to the theoretical study of how the amplitude of oscillation depends on the
initial disturbance. This problem has been partially addressed in the slab
section, where it was suggested that an external disturbance might deposit a
very small amount of energy in the loop.  To analyse this problem we refer to
the work of \citet{rudrob06} who solved the initial value problem using the
Laplace transform to determine the motion of the tube. These authors have found
that the asymptotic behaviour of the loop for $t\rightarrow \infty$ is given by
the fundamental normal mode of the structure. Since we are mostly interested in
the energy that is trapped in the loop and not in the transients  before the
loop settles in the normal mode (already discussed), instead of solving the full
time-dependent problem numerically and calculating the deposited energy in the
loop we use an approach based on the results of \citet{rudrob06}. These authors
showed that the amplitude of oscillation of the loop is basically proportional
to the convolution between the initial disturbance and the eigenmode of the loop
\citep[see eqn.~(7) in][]{terrandetal07a}. Using this method we can easily
calculated how this amplitude (i.e. the energy of the trapped mode) depends on
the different parameters. In \citet{terrandetal07a} several types of
perturbations (always in the total pressure) were studied. Since we are
interested on the kink modes the simplest perturbation is an $m=1$ disturbance
localised in the external medium.  \begin{eqnarray}\label{pert} P_0=P_N\,
e^{-{\left(\frac{r-r_0}{a}\right)}^2}, \end{eqnarray}

\noindent where $P_N$ is a normalisation constant, $r_0$ is the location of the
Gaussian and $a$ is the width at half height (do not confuse with the loop
radius $R$). This allows us to study the behaviour of the loop for different
locations of the disturbance and different widths of the initial pulse. The
profile of the initial perturbation for $m=1$  using the radial dependence given
by equation~(\ref{pert}) is displayed in Figure~\ref{kinkflutt}a. It is not a good
representation of  disturbance produced by for example, a flare or eruption,
however its analysis is useful to understand more realistic excitations. In
Figure~\ref{kinkflutt}b the trapped energy for the $m=1$ mode is plotted. We see
that it decreases quite rapidly with $r_0$ while it smoothly increases with the
width of the initial perturbation $a$. It is possible to find the asymptotic
behaviour of the energy for $a\ll r_0$, \begin{eqnarray}\label{Etdep}  E\sim
\frac{e^{-2 |\Lambda_e| r_0}}{r_0^2}\,a. \end{eqnarray}

\noindent Due to the strong dependence on $r_0$ in general only a small amount
of the energy of the initial perturbation is trapped in the loop. For example,
for a perturbation located at a distance similar to the loop length $r_0=40R$
($L=50R$ and $a=R$), the trapped energy is $9.4\times10^{-7}E_0$ ($E_0$ being
the energy of the initial disturbance). This results clearly indicates that the
loop is able to trap a very small part of the energy of the initial disturbance.

The method to calculate the energy can be easily extended to azimuthally and
vertically localised perturbations. For example, we can study an azimuthally 
localised perturbation of the form 
\begin{eqnarray}\label{perta}
P_0=P_N\,
e^{-{\left(\frac{r-r_0}{a}\right)}^2}
e^{-{\left(\frac{\varphi}{\delta}\right)}^2}
\sin(\varphi).
\end{eqnarray}

\begin{figure}[ht]
\center{
\resizebox{6cm}{!}{\includegraphics{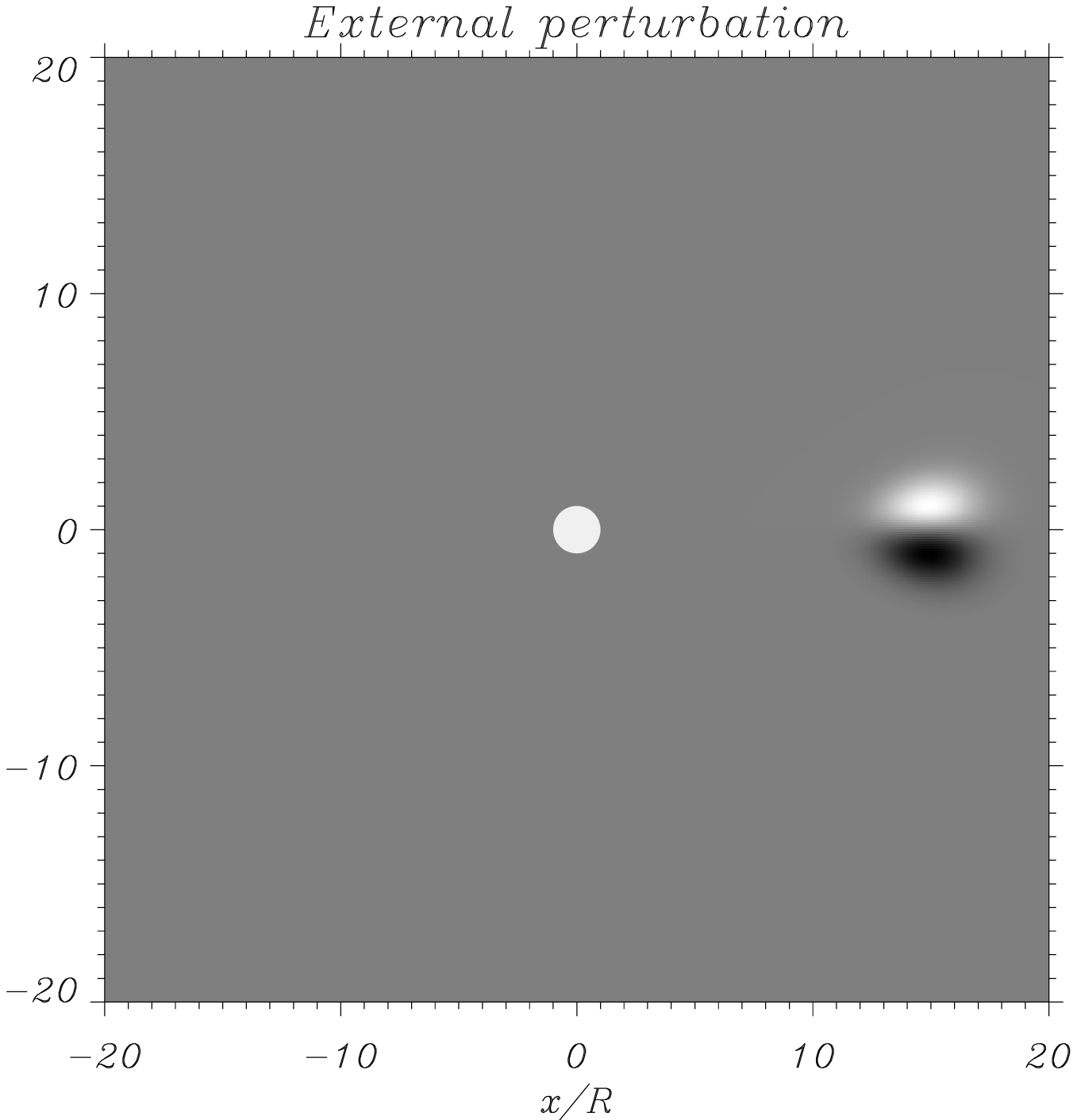}}
\resizebox{6cm}{!}{\includegraphics{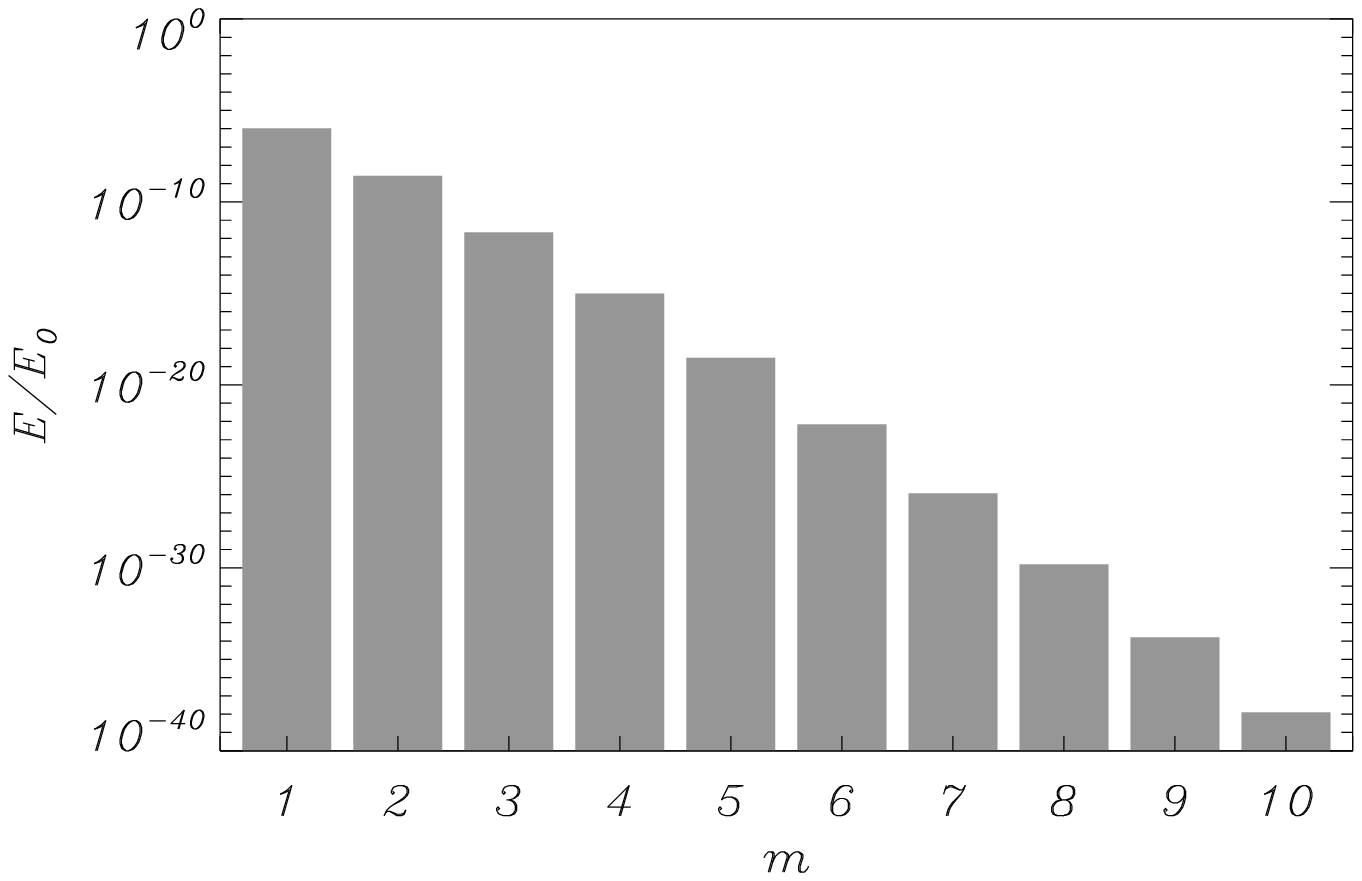}}}
\caption{
\small {(a)} Example of an azimuthally localised perturbation. The circular white region
represents the loop. {(b)} Trapped energy of each eigenmode. Note that the
vertical axis is in logarithmic scale. The difference in energy between
consecutive modes increases with $m$. For example, the trapped energy for the $m=1$ is
$4\times10^2$ larger than the energy of the $m=2$. For this
plot $r_0=40R$, $a=4R$ and $\delta=0.5$ ($L=50R$).}
\label{azimuth}
\end{figure}

\noindent The shape of the perturbation and the energy distribution for the
different $m$'s are plotted in Figure~\ref{azimuth}. The trapped energy
associated with each $m$ is displayed in Figure~\ref{azimuth}b. It is clear that
the energy decreases with $m$ and that the kink mode ($m=1$) has the largest
energy. The energy trapped by the kink mode is between two and three orders of
magnitude larger than the energy of the first fluting mode ($m=2$). Even for
$m=1$ the trapped energy is quite small and, as expected, is lower than that for
an excitation with a single $m$ since in that case all the initial energy was in
one single mode. Qualitatively similar results are found when the width of the
perturbation in the angular direction, $\delta$, is changed. Obviously, the
energy of the modes in the initial perturbation is different, but again the
trapped energy distribution is dominated by the kink mode. 

\begin{figure}[ht]
\center{
\resizebox{6cm}{!}{\includegraphics{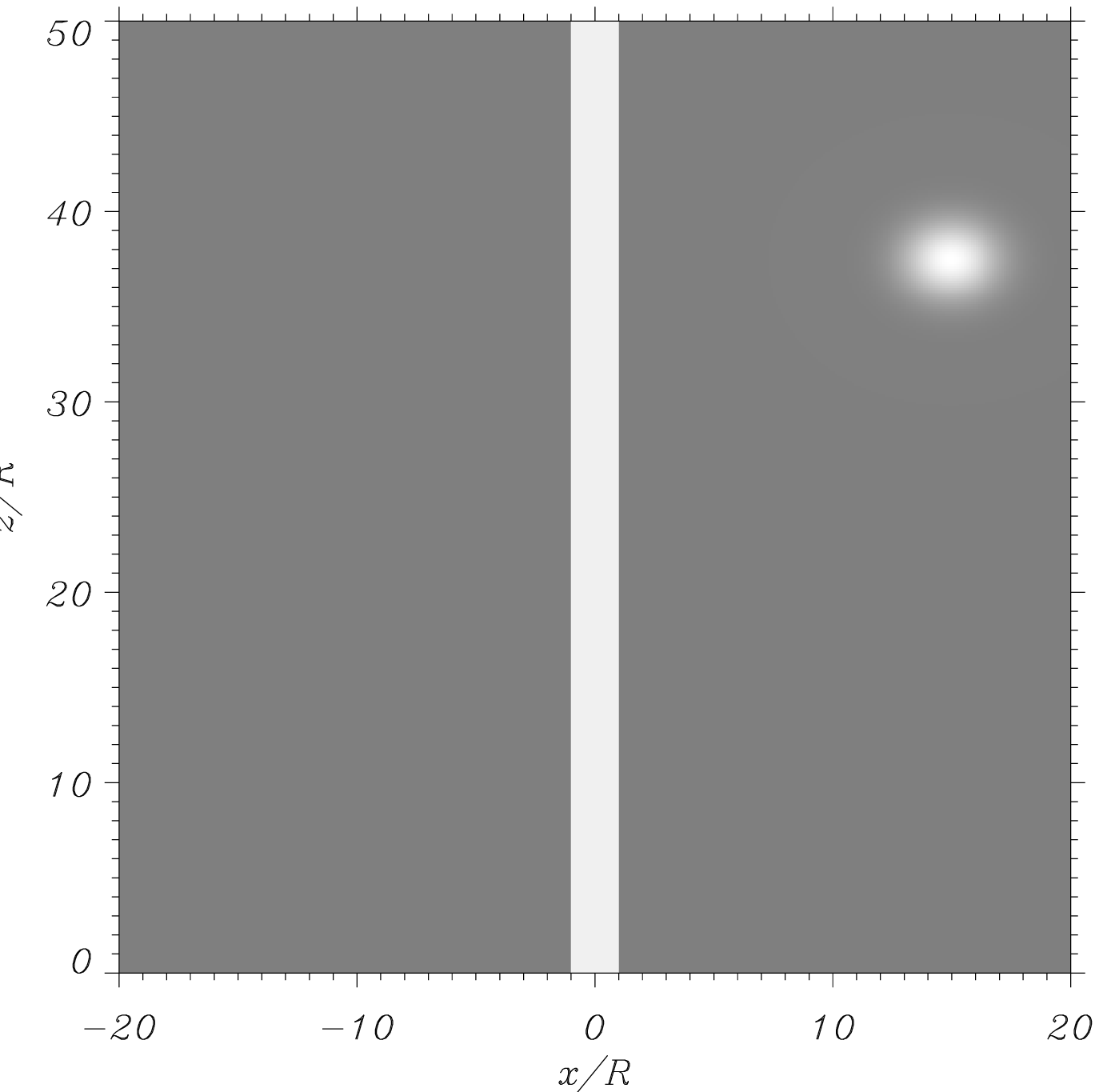}}
\resizebox{6cm}{!}{\includegraphics{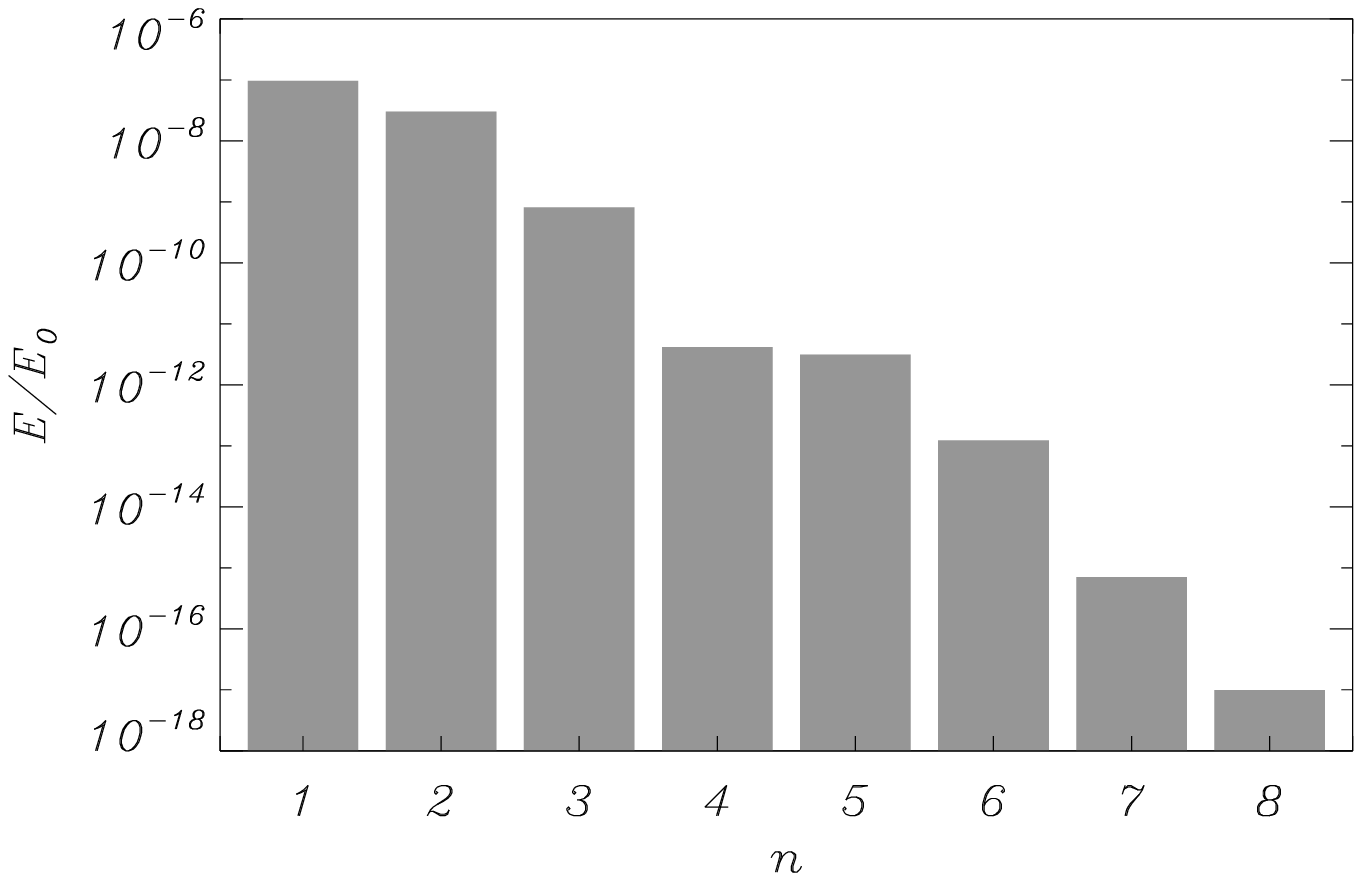}}
}
\caption{
\small {(a)} Example of longitudinally localised perturbation. The white area represents the
loop. {(b)} Trapped energy of each eigenmode (the vertical axis is in logarithmic scale). Note
that the fundamental mode has the largest energy. For this case $r_0=40R$, $a=4R$ ($L=50R$), $z_0=L/4$ and
$\Delta=L/10$.}
\label{verst}
\end{figure}

For a longitudinally localised perturbation we assume the following dependence
\begin{eqnarray}\label{pertv}
P_0=P_N\,
e^{-{\left(\frac{r-r_0}{a}\right)}^2}e^{-{\left(\frac{z-z_0}{\Delta}\right)}^2}
\sin(z \pi/L),
\end{eqnarray}

\noindent where $z_0$ is the longitudinal location of the Gaussian and $\Delta$ is
the width of the perturbation in the longitudinal direction. The sinusoidal
dependence on $z$ has been included in order to satisfy the line-tying condition.
This perturbation is represented in Figure~\ref{verst}a and it corresponds to the
case $z_0=L/4$, i.e. a perturbation located at a quarter of the loop length. Again
the kink mode has the largest energy (see Fig.~\ref{verst}b). These results
suggests that longitudinal harmonics are in principle more easily excited than
azimuthal harmonics (which have energy differences of the order $10^3$).

Although there are estimations of the various types of energy released in real
flares, it is not clear what amount of energy is associated with MHD waves. From
the information of the previous model and the information of the coronal loop
oscillations, this energy can be estimated. For example from the amplitude of
the oscillating loops it is easy to estimate that they typically  have an 
energy of  $10^{19}\, \rm J$. If the perturbation is located at the loop length,
then the energy of the initial disturbance should be $10^6$ times larger, i.e.
of the order $10^{25}\, \rm J$. This quantity is, in order of magnitude, similar
to the estimations of the other types of energy release. Thus, the energy of the
oscillating loops can be potentially used as a seismological tool to determine
the amount of energy released in real flares (related to fast MHD waves). It is
interesting to note that \citet{ballaietal08} based on a model of forced
oscillations (driven by EIT waves) have estimated energies of the same order of
magnitude. 

\subsection{Role of an inhomogeneous layer at the loop boundary}\label{resabs}

The effect of a non-uniform layer on the kink oscillation has been extensively
studied in the literature. The reader is referred to the review by
\citet{goossens08} (and references therein) about the process of resonant
absorption and to its application to explain the reported damping of loop
oscillations.

\begin{figure}\center{
\includegraphics[width=0.7\textwidth]{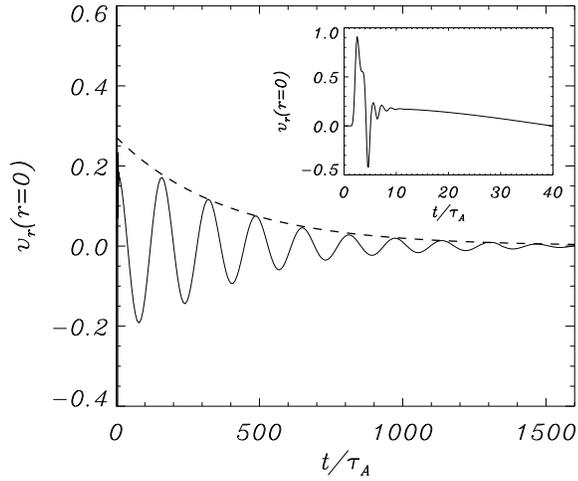}}
\caption{
Plot of $v_r$ at the centre of the loop as a function of time for the numerical
simulation for the initial perturbation given by equation~(\ref{kinkexc1D}). After a short leaky transient, detailed in the inner plot, the loop oscillates with the
corresponding quasi-mode. The dashed line is a fit of the form
$\exp(-t/\tau_{\rm d})$. In this simulation, $r_0=5R$, $w=2R$, $l=0.5R$, $\rho_{\rm
i}/\rho_{\rm e}=3$ and $R/L=0.01$.}
\label{tempr_0}
\end{figure}

Here, we just want to illustrate by means of a simple time-dependent solution the
process of resonant absorption that takes place at the loop boundary. For this
reason, as in the homogeneous cylinder, we set up an initial disturbance given by
equation~(\ref{kinkexc1D}) and we analyse the evolution of the system.
Figure~\ref{tempr_0} shows the radial velocity component, $v_r$, at the loop
centre ($r=0$) as a function of time for a loop with an inhomogeneous layer of 
$l=0.5R$. We can clearly identify several phases. There are initially two large
extrema followed by short-period oscillations which are quickly attenuated (see
inner plot of Fig.~\ref{tempr_0}). After this short leaky transient the loop
oscillates with a much longer period. Now the amplitude of this mode is attenuated
with time (compare with Fig.~\ref{kinktemp}a for the homogeneous loop, i.e.
$l=0$). This damping is due to the conversion of energy from the global kink mode
to the localised Alfv\'enic modes. By fitting an exponential function of the form
$\exp(-t/\tau_{\rm d})$ to the envelope of the signal we have numerically
determined the damping time, $\tau_{\rm d}$. There is quite a good agreement
between the time-dependent results and the quasi-mode (which is basically the
resistive eigenmode in the limit of small dissipation). The calculation of the
quasi-mode provides with valuable information about the behaviour of the system,
as was also shown by \citet{rudrob02} (they solved the time-dependent problem
analytically using Laplace transforms).

\begin{figure}\center{
\includegraphics[width=1\textwidth]{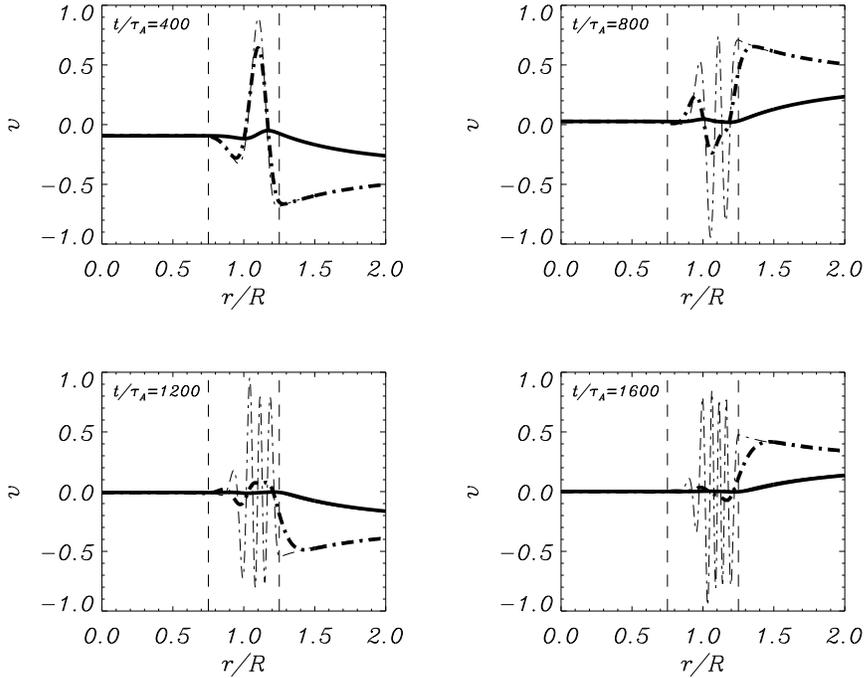}}
\caption{
Plot of $v_r$ (continuous line) and $v_{\theta}$ (dot-dashed line) around the
resonant position for different times for the numerical simulation of
Figure~\ref{tempr_0}. Thin lines correspond to $R_m\rightarrow\infty$ and thick
lines to $R_m=10^5$. Although the behaviour of $v_{\theta}$ in the inhomogeneous
layer is quite different for the two values of $R_m$, the damping times are the
same. The vertical dashed lines mark the boundaries of
the inhomogeneous layer.}
\label{temp1}
\end{figure}

To show in detail the behaviour of the solutions in the inhomogeneous layer we
have plotted the time evolution of the radial and the azimuthal velocity
components in the range $0<r/R<2$ (Fig.~\ref{temp1}). For $R_m\rightarrow\infty$
we see that small scales are generated in time in the region $R-l/2<r<R+l/2$, the
amplitude of the azimuthal component being much larger than that of the radial
component due to the energy conversion. The spatial scales in the inhomogeneous
layer decrease due to the process of phase mixing \citep[see for
example][]{heypri83} since the local Alfv\'en frequency changes with position.
Figure~\ref{temp1} also shows the time evolution for the same situation but for
$R_m=10^5$ (thick lines). This is an unrealistically small value but it is
illustrative since it clearly shows the effect of dissipation. Although the
damping time is almost the same as for the ideal case, now the spatial scales are
much larger since the short scales have been smoothed out by the effect of
diffusion. However, the damping time is independent of the value of the
resistivity for $R_m$ large. In this limit, resistivity only determines the
behaviour in the inhomogeneous layer but it does not affect the conversion of
energy and in consequence it does not change the damping time. It is clear that
resistivity plays a secondary role and that the calculations can be done even in
ideal MHD ($R_m\rightarrow\infty$). This property of resonant absorption is
particularly attractive sine no anomalous Reynolds numbers are required to explain
the damping of the oscillations.

\subsection{Nonlinear effects on the kink oscillation} The previous results for
the cylindrical loop were derived using the linearised MHD equations.
Nevertheless, it is also interesting to study the nonlinear regime. One of the
consequences of a large amplitude (nonlinear) standing kink oscillation is the
generation of flows along the tube due to the ponderomotive force. This problem
was studied by \citet{terrofm04} in the context of coronal loops oscillations
\citep[see also][]{hollweg71,rankinetal94}. Due to the nonlinear terms, an initial
disturbance produces a force imbalance along the loop which results in an up-flow
from its legs. The accumulation of mass at the top of the loop (for the
fundamental standing kink mode) can produce strong density enhancements, which
secularly grow with time as
\begin{eqnarray}\label{vz}
\frac{\rho_2}{\rho}\simeq\frac{1}{4}\left(\frac{V_0}{V_A}\right)^2(\omega
t)^2\cos(2k_z z),\label{rho1}  \end{eqnarray}
where $V_0$ is the amplitude of the
initial disturbance. However, if the plasma beta is different from zero the gas
pressure gradient becomes dominant and inhibits the concentration of mass. The
density enhancements are then of the form  \begin{equation}\label{rho1s}
\frac{\rho_2}{\rho}\simeq\frac{1}{2}\left(\frac{V_0}{V_A}\right)^2\left(\frac{\omega}{\omega_s}\right)^2[1-\cos(\omega_st)]\cos(2k_z
z), \end{equation} where $\omega_s=2k_z\,c_s$, being $c_s$ the sound speed.
This expression can be potentially used as an indirect way to determine the sound
speed in the solar corona if such enhancements are reported in the observations
\citep[see][for an example]{terrofm04}. Thus, although the zero-$\beta$
approximation is in general justified under coronal conditions, the role of the
gas pressure is important in nonlinear studies of standing kink oscillations (and
also in propagating kink body modes) since it avoids the eventual destruction of
the tube due to the ponderomotive force.

\begin{figure*}[!ht]
\center{
\includegraphics[width=0.45\textwidth]{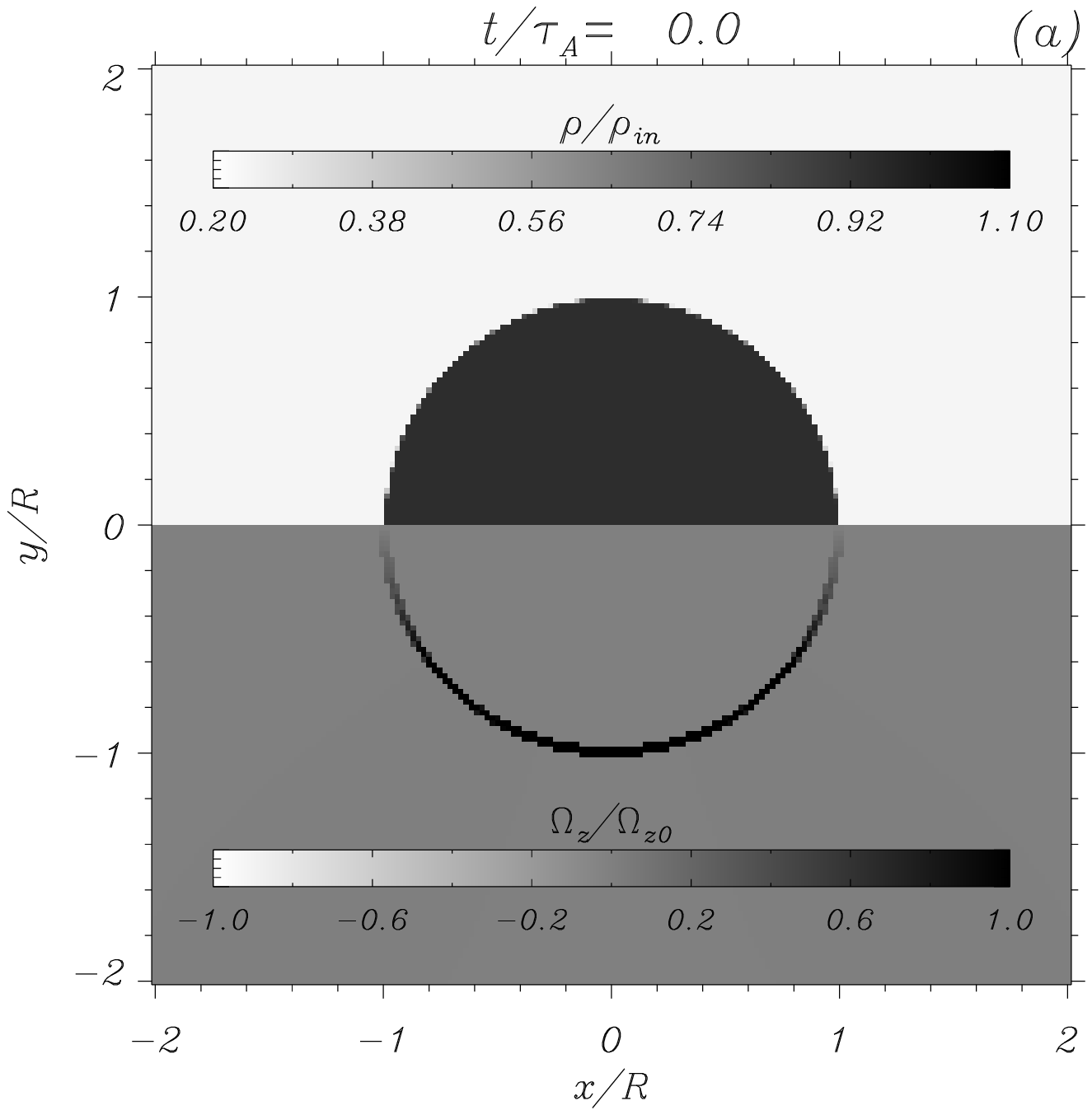}
\includegraphics[width=0.45\textwidth]{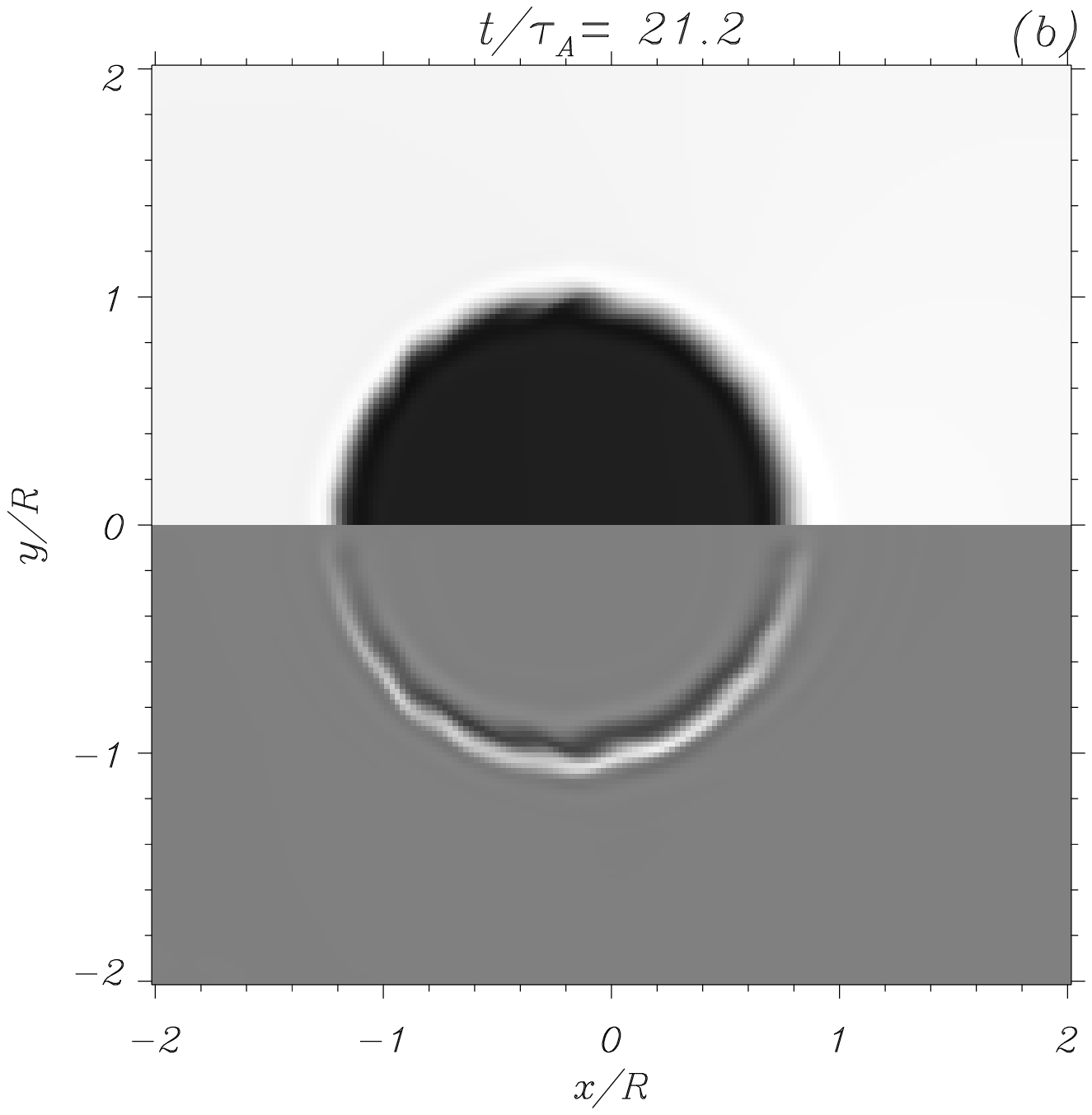}
\includegraphics[width=0.45\textwidth]{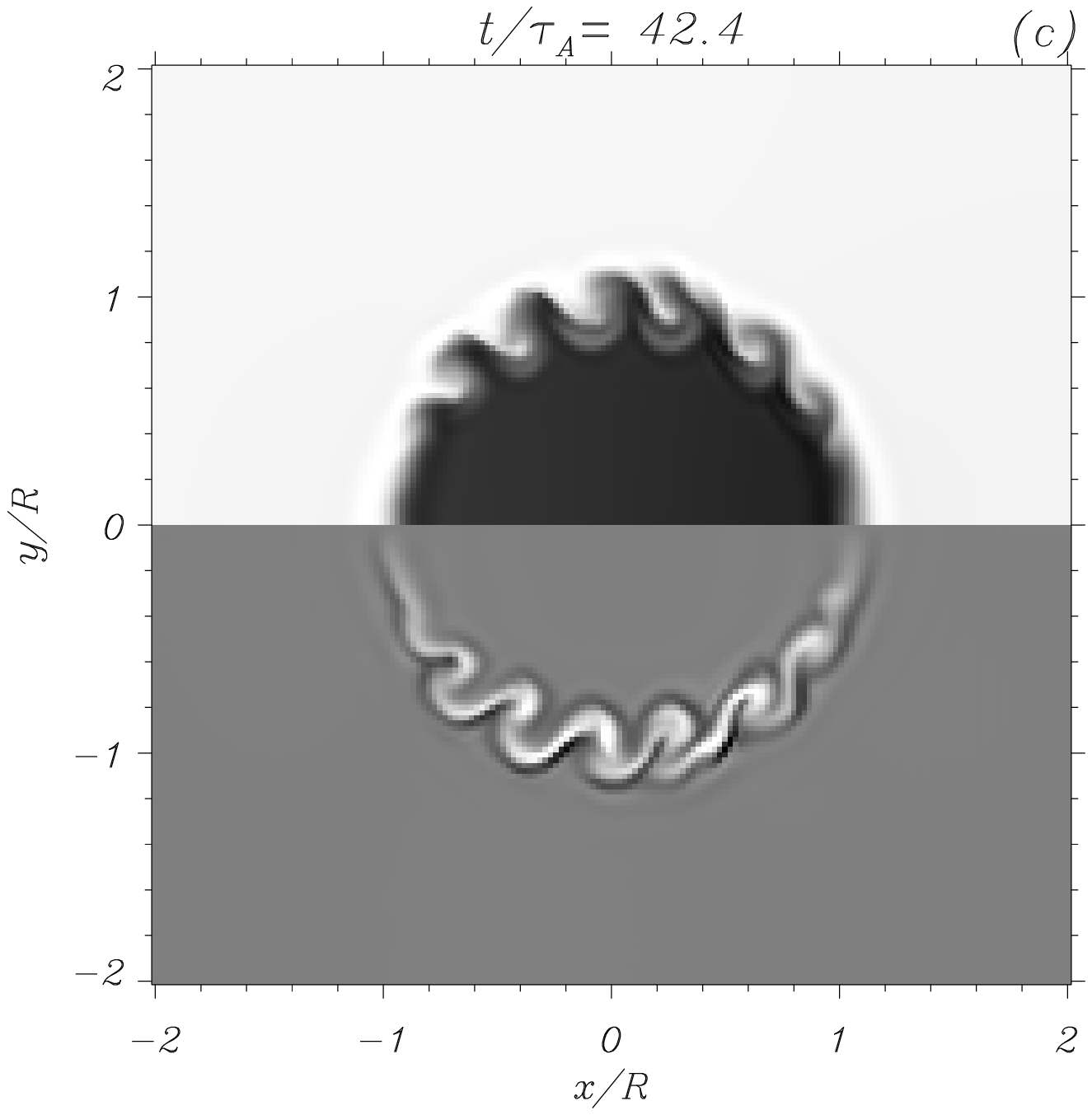}
\includegraphics[width=0.45\textwidth]{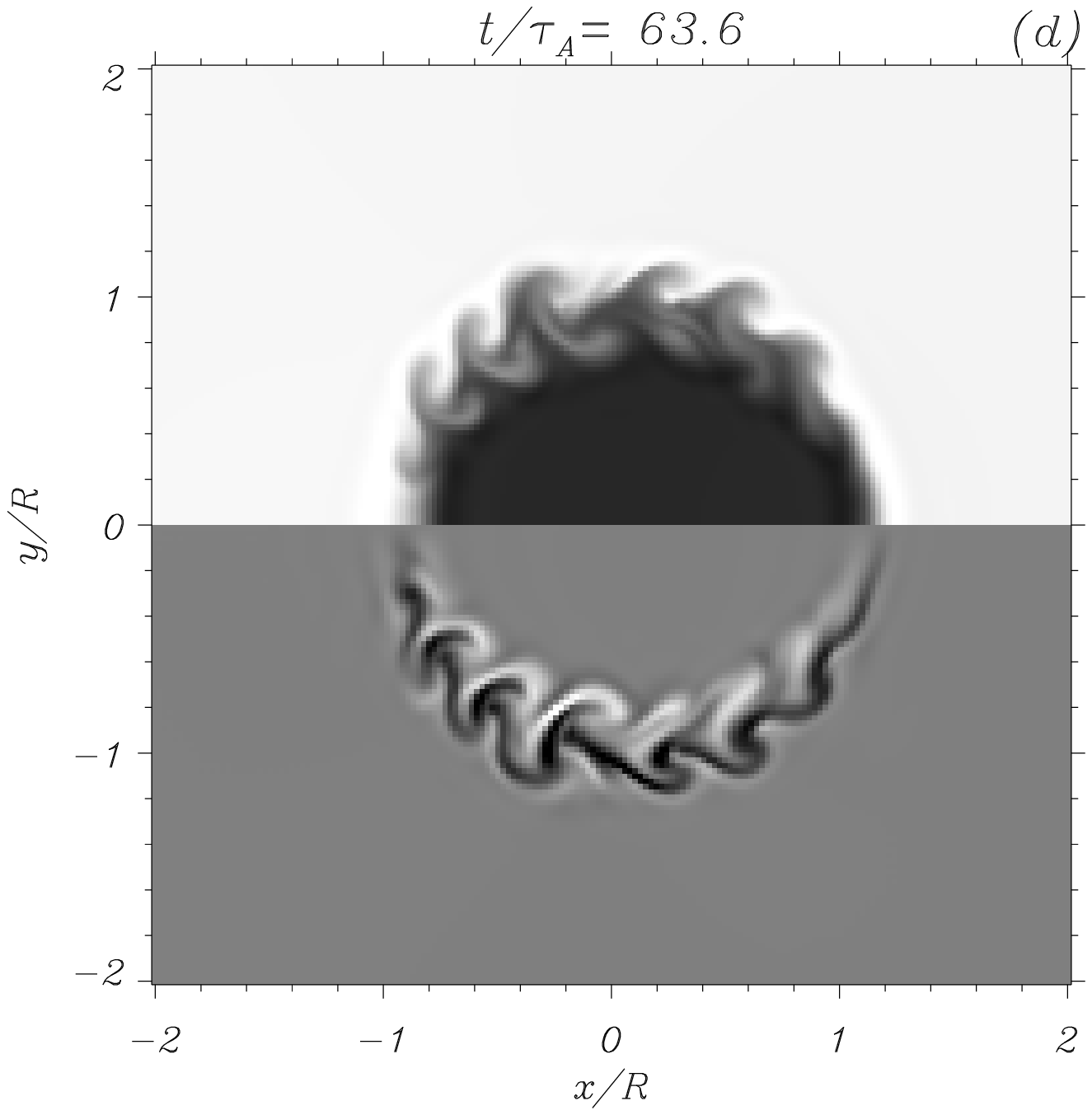}
}
\caption{ 
\small  The top half of each panel displays the density (symmetric respect to
$y=0$) and the bottom half the vorticity (antisymmetric respect to $y=0$) at
$z=L/2$ (half the loop length) for
different times. In this simulation the following parameters have been used: 
$L=10R$, $\rho_{\rm i}/\rho_{\rm e}$=3, $v_0=0.1v_{\rm Ai}$, and a full domain of
$[-6R,-6R]\times[-6R,6R]\times[0,10R]$. The kink oscillation develops a
Kelvin-Helmholtz instability  at
the loop boundary.}  \label{density}
\end{figure*}

On the other hand, nonlinearity can significantly change the motion of the plasma
fluid at certain locations in the tube. It is important to remark that the kink
oscillation of a cylindrical loop with a sharp or smooth transition involves
strong shear motions at the tube boundary (see for example Fig.~\ref{temp1}).
Such motions might be liable to develop a Kelvin-Helmholtz instability. The most
likely place for the KHI to occur is where the velocity is largest and the
magnetic field is perpendicular to the velocity \citep[see for
example][]{ranketal93}. For the fundamental standing kink mode this is precisely
the antinode of the velocity, located at half the loop length from the
photosphere. In order to understand this problem we show the results of solving
the full nonlinear MHD equations in three-dimensions
\citep[see][]{terradasetal08}. This is the extension of the linear problem
studied in Sections \ref{cylsim} and \ref{resabs}. Now the initial perturbation
is located inside the tube. The evolution of the density (for a sharp transition
model) and the longitudinal component of the vorticity of a representative case
in a weak nonlinear regime is shown in Figure~\ref{density} at different time
intervals. The initial perturbation  produces a lateral displacement of the tube
in the $x-$direction. As in the linear regime the tube starts to oscillate around
the initial position, but small length scales quickly develop at the boundary
(see Fig.~\ref{density}b). These small scale structures grow with time (see
Fig.~\ref{density}c) and several rolls form at the loop edge. Around $x=\pm R$
and $y=0$ the density is almost undisturbed because there are no shear motions at
this position. At later stages of the evolution (Fig.~\ref{density}d) the small
spatial scales are still localised at the boundary. As a result of the
instability the shape of the tube at the boundary has been considerably changed.

On the other hand, if instead of a sharp transition a thick inhomogeneous layer
between the tube and the environment is included then the motions are
characterised by strong phase-mixed scales (see Fig.~\ref{temp1} for the linear
results). In this situation the instability is still present, but it develops
more slowly than in the sharp transition case. This is due to the fact that the
thicker the layer the later the generation of small length scales due to  phase
mixing, and thus the eventual onset of the instability. This process seems to be
related to the development of the KHI for torsional Alfv\'en waves described by
\citet{brownpri84}. Interestingly, for thick layers the attenuation of the
central part of the tube due to resonant absorption (i.e. the damping rate) is
not significantly altered by the changes at the boundary due to the shear
instability. 

Finally, it must be noted that up to now, there is no evidence of such
instability from the observation of oscillating loops. It is known that magnetic
twist, not included in the model, might decrease or even suppress the instability
since the presence of a magnetic field component along the flow stabilises the
KHI. Nevertheless, a tube with very large azimuthal magnetic field  is subject to
the instabilities of the linear pinch. Therefore, the azimuthal component of the
magnetic field of a stable flux tube in the solar corona is probably constrained
to be in a specific range. The absence of observational evidences of the
instability might be also due to a lack of spatial resolution of the telescopes.
This issue about this instability is very promising and needs to be investigated
in more detail in the future from both the theoretical (including twist in the
tube models) and the observational point of view.

\subsection{Multi-structure}

Now we leave the single tube model to study the collective motion of several tubes
again in the linear regime. We start with two parallel homogeneous straight
cylinders of radius $R$, length $L$, and separation between centres $d$. This is
the equivalent to the slab configuration analysed in Section~\ref{2slabs}. In this
two-tube problem \citet{lunaetal08} showed that there are four collective
fundamental kink-like trapped modes \citep[see also][]{vandetal08}. The velocity
field is more or less uniform in the interior of the loops, and so they move
basically as a solid body, while the external velocity field has a more complex
structure. The four velocity field solutions have a well defined symmetry with
respect to the $y$-axis. The symmetric mode has the velocity field inside the
tubes lying in the $x$-direction and is symmetric with respect to the $y$-axis.
This is the $S_x$ mode, where $S$ refers to the symmetry of the velocity field and
the subscript $x$ refers to the direction of the velocity inside the tube. The
same nomenclature is used for the other modes. The following mode has the velocity
inside the cylinders mainly in the $x$-direction and  it is antisymmetric with
respect to the $y$-axis, we call this mode $A_x$. Similarly, we have the modes
$S_y$ and $A_y$. The pressure field of the $A_x$ and $S_y$ modes is symmetric 
with respect to the $y$-axis, while that of the $S_x$ and $A_y$ modes is
antisymmetric.

\begin{figure}[!ht]
\center{
\includegraphics[width=0.4\textwidth]{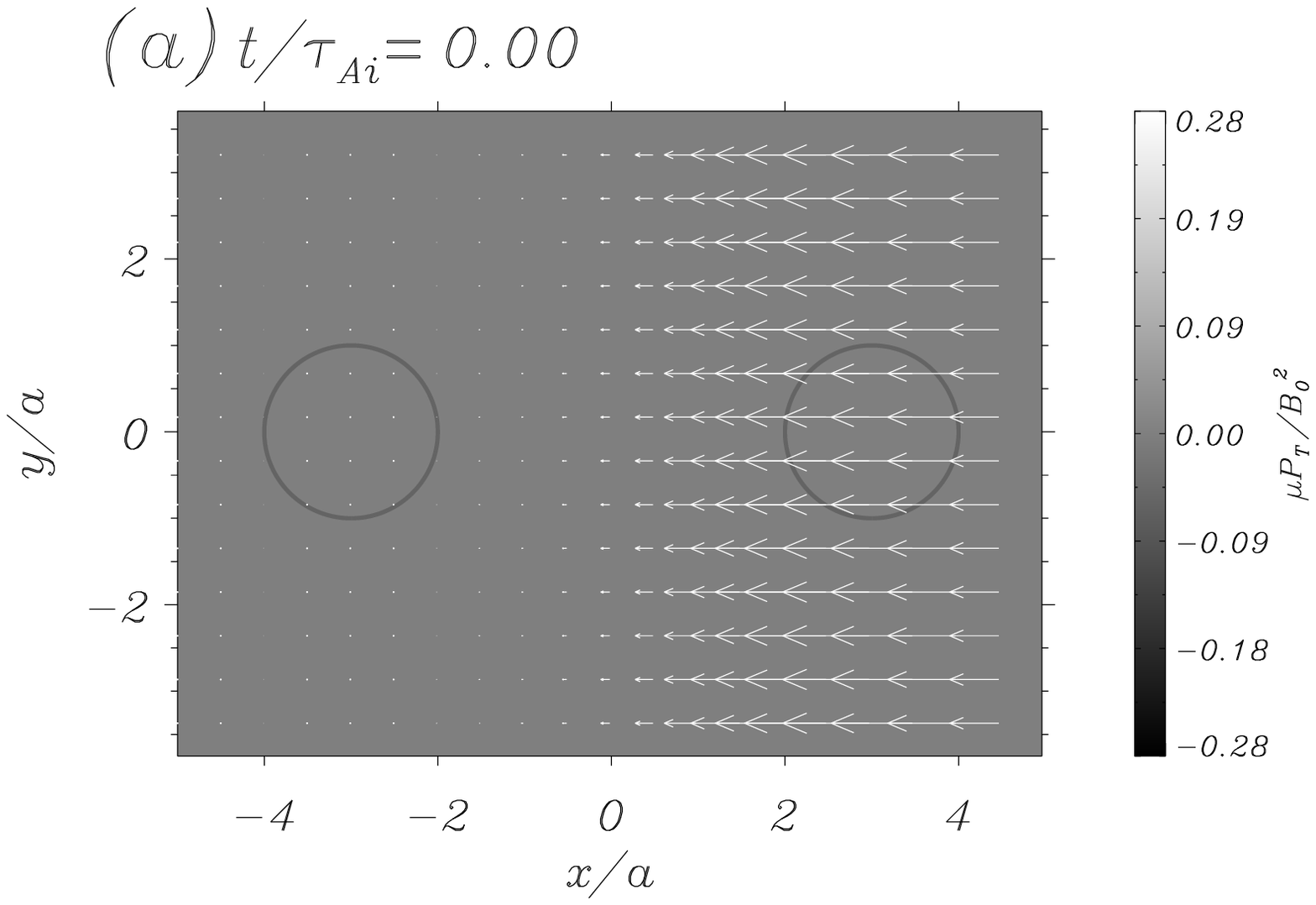}
\hspace{0.25cm}
\includegraphics[width=0.4\textwidth]{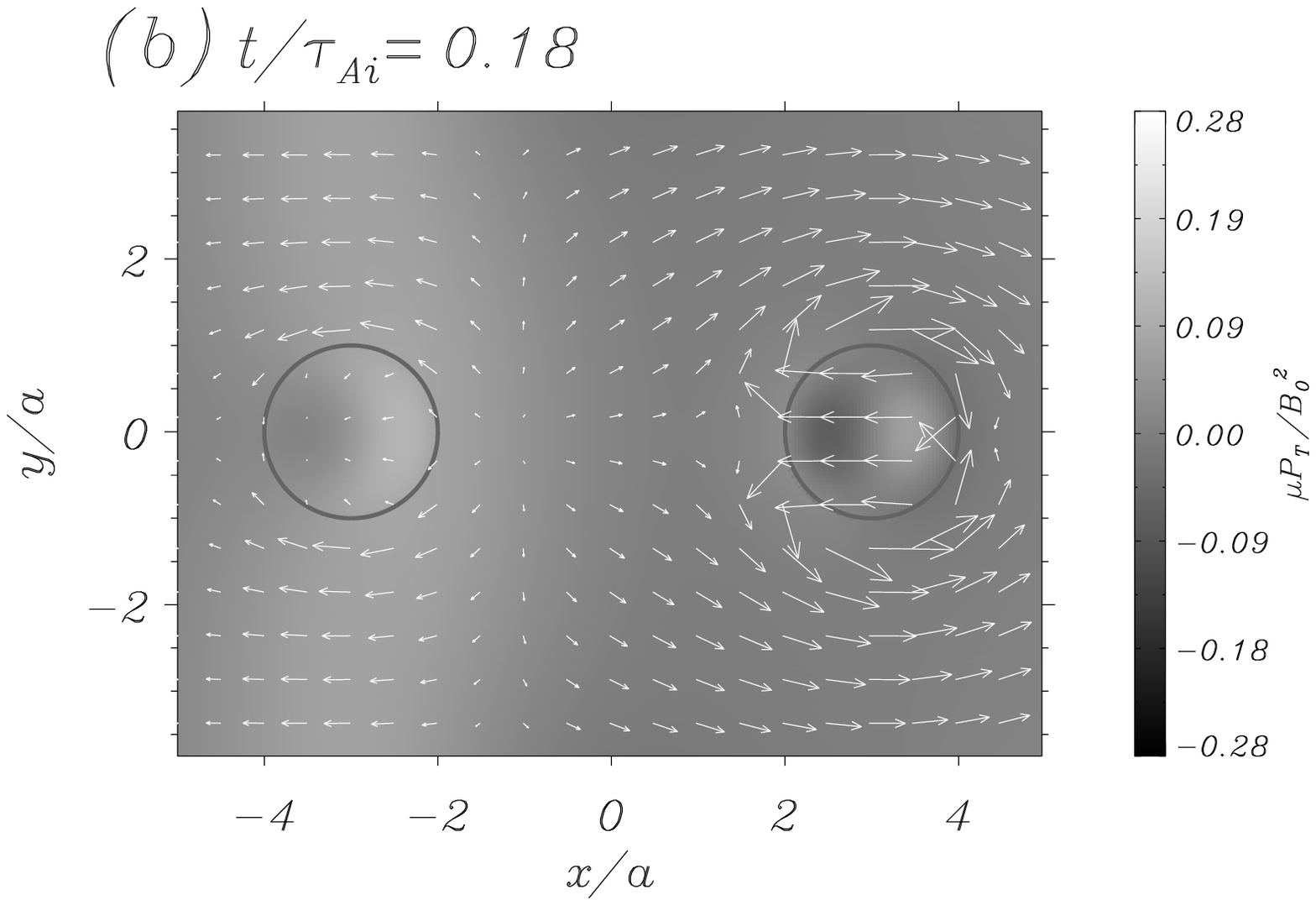}
\includegraphics[width=0.4\textwidth]{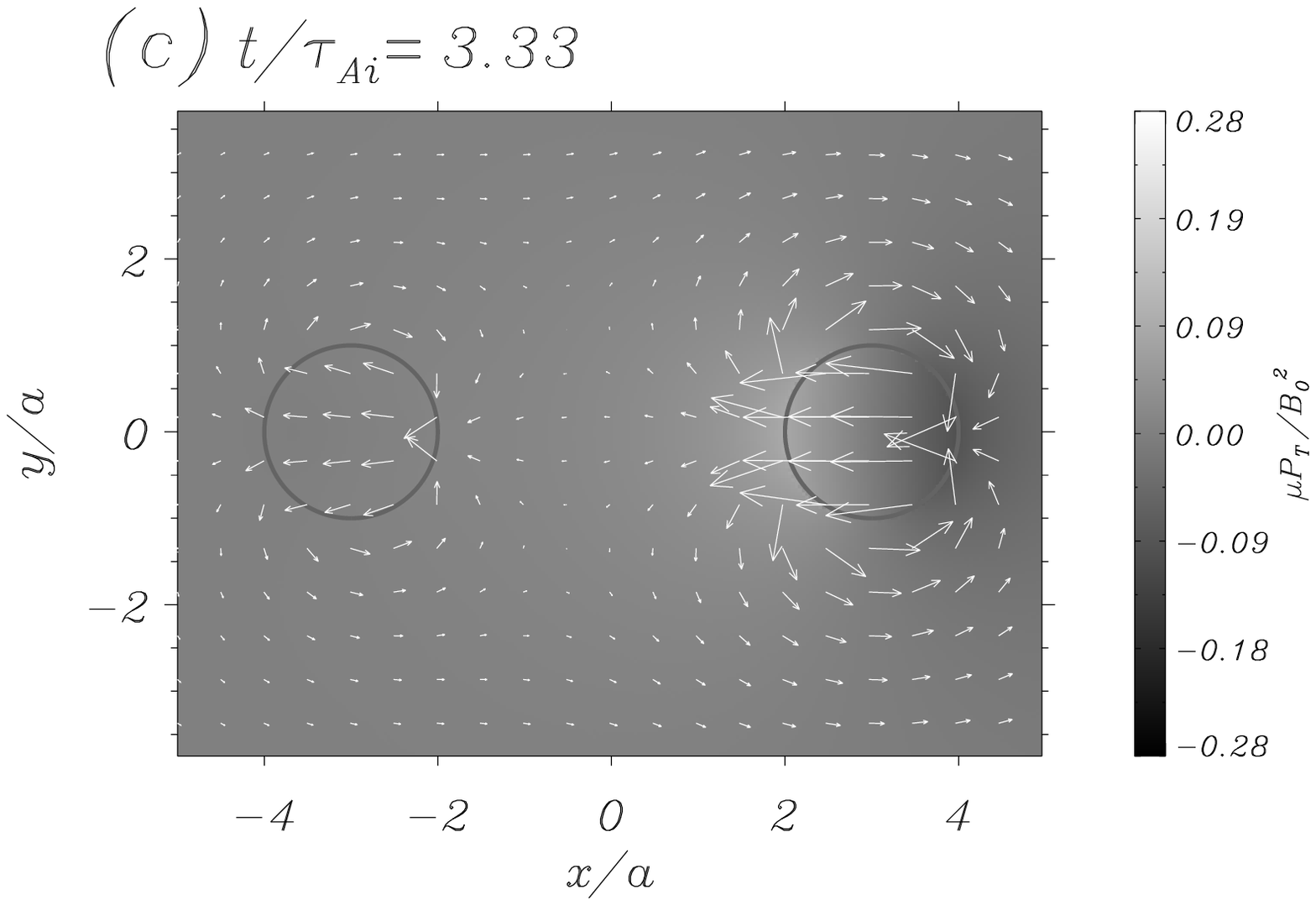}
\hspace{0.25cm}
\includegraphics[width=0.4\textwidth]{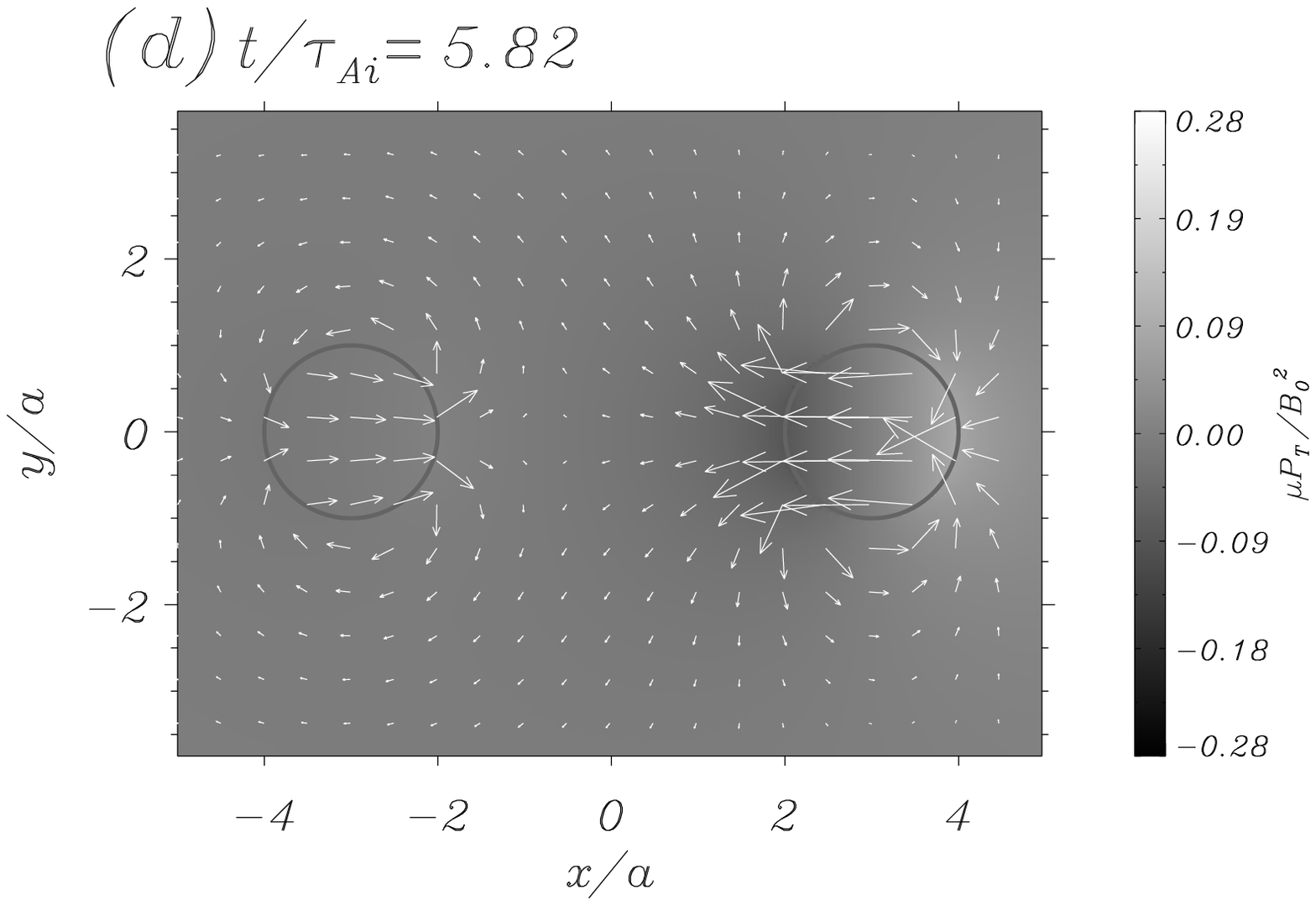}}
\caption{
Time-evolution of the velocity field (arrows) and total pressure field (
contours), for a separation between loops $d=6 R$ (in this model $a=R$). The two circles mark the positions of the loops at
$t=0$. The panels show different evolution times. In (a) the initial condition
over the velocity field is represented. In panel (b) the tubes are in
the transient phase. In panels (c) and (d) the system oscillates in the
stationary phase with a superposition of
the $S_x$ and $A_x$ normal modes.
}
\label{t_evol_0}
\end{figure}

In Figure~\ref{t_evol_0}, the time evolution is shown for a pulse in the $v_x$
component initially centred on the right loop (see Fig.~\ref{t_evol_0}a). In
Figure~\ref{t_evol_0}b, the pulse reaches the left tube and passes through it,
the system still being in the transient phase. In Figures \ref{t_evol_0}c and
\ref{t_evol_0}d the system oscillates in the stationary phase. The oscillatory
amplitude in the left loop grows with time in the stationary phase,  while the
amplitude in the right loop decreases in the time interval shown in
Figures~\ref{t_evol_0}c and \ref{t_evol_0}d. The left tube begins its motion
through the interaction with the right loop, i.e. by a transfer of energy from
the right loop to the left loop. This process is reversed and repeated
periodically: once the left loop has gained most of the energy retained by the
system, so that the right loop is almost at rest, the left tube starts giving
away its energy to the right cylinder, and so on. We find again the beating
phenomenon (see Section~\ref{2slabs}), since the initial disturbance excites the
$S_x$ and $A_x$ modes with the same amplitude.

\begin{figure}[!ht]
\center{
\includegraphics[width=0.45\textwidth]{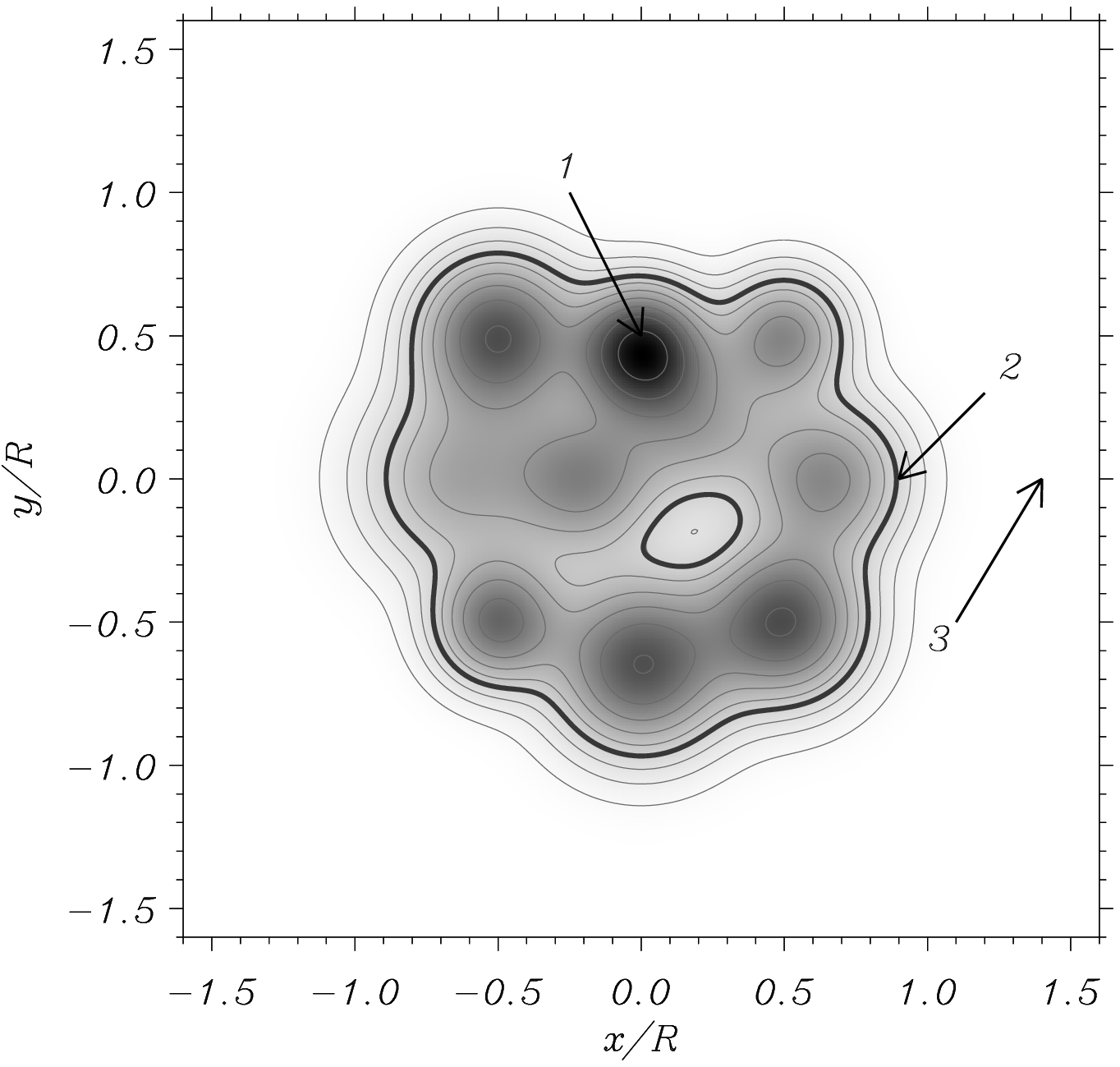}
\includegraphics[width=0.45\textwidth]{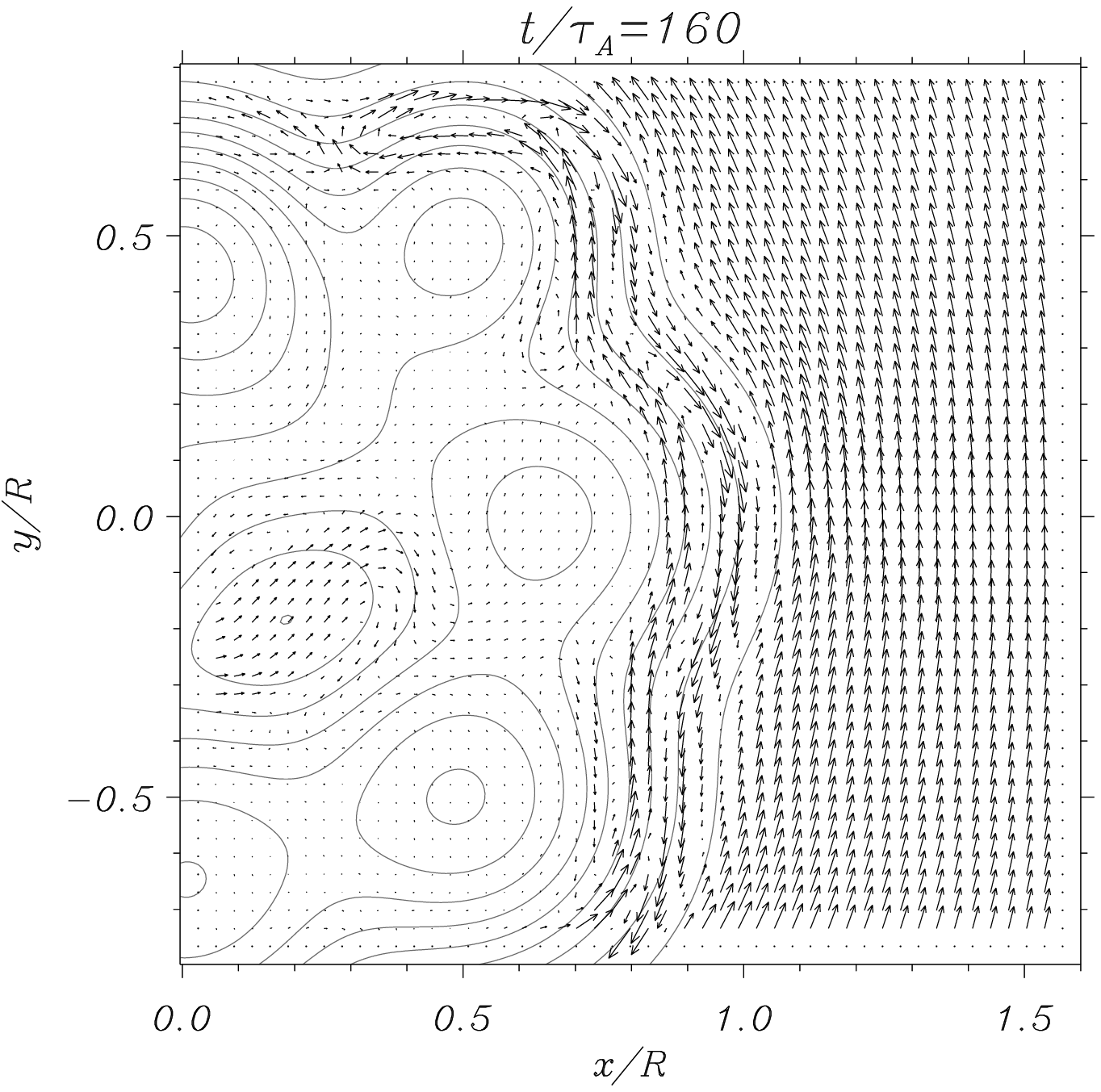}
}
\caption{ \small 
(a) Cross section of the density of the multi-stranded loop model. The
contour lines represent curves of constant Alfv\'en frequency, (they also represent contours of constant density).
The thick line corresponds to the Alfv\'en frequency that matches the frequency
of the global mode. (b) Detail of the velocity field at
$t=160\,\tau_{\rm A}$.}  \label{density1}
\end{figure}

\begin{figure}[!ht]
\center{
\includegraphics[width=0.4\textwidth]{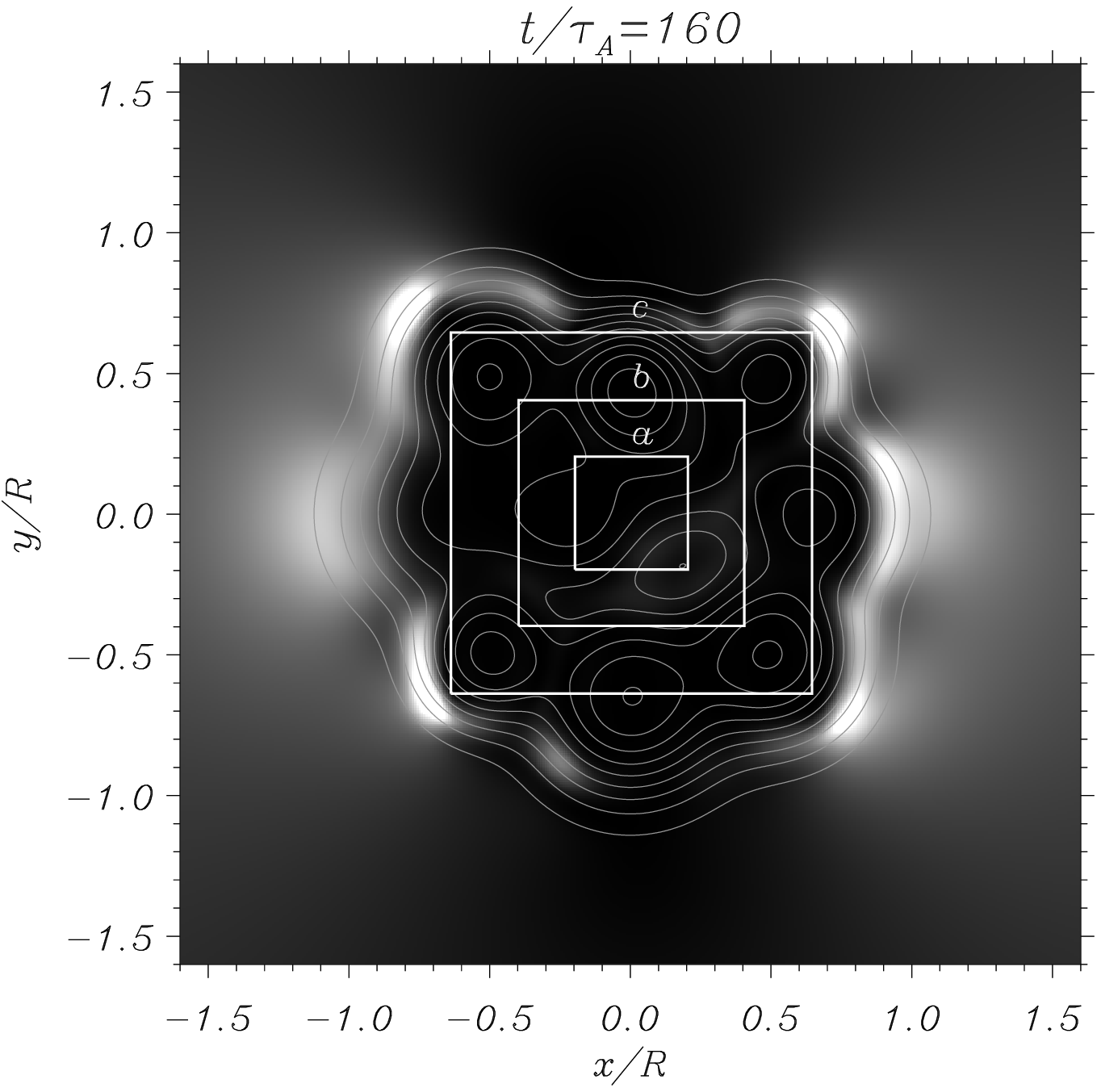}
\includegraphics[width=0.525\textwidth]{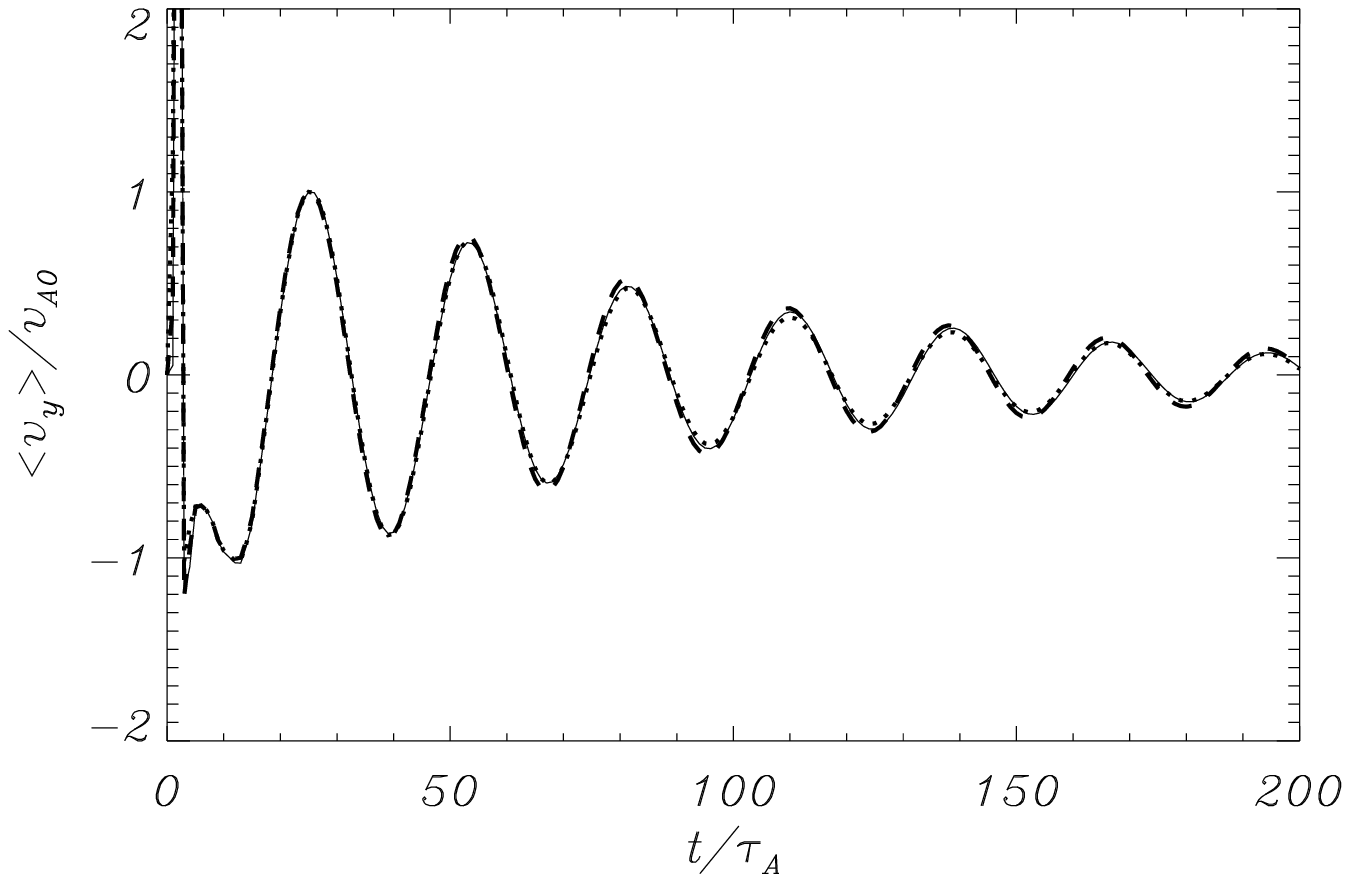}
}
\caption{ \small  (a) Energy distribution at a given instant (when most of the
energy has been transferred to the inhomogeneous layers). The three
square boxes labelled with $a$, $b$, and $c$ mark the regions
where the velocity averages
have been computed to determine the damping time. (b) Time evolution of the spatially averaged $v_y$ component as a 
function of time. The continuous, dot-dashed, and dotted lines
correspond to the (normalised) averages in the square regions marked with labels
$a$, $b$, and $c$ in (a), respectively. The profile of
the three signals is almost the same since the loop is oscillating with a
global mode. The damping per period in this configuration is $\tau_{\rm d}/P=2.3$. }
\label{resonanceenerg}  \end{figure}

The extension of the previous model to multi-stranded configuration but with
nonuniform layers has been carried out by \citet{terrarretal08}. In this work a
smooth density transitions between the tubes is allowed. In
Figure~\ref{density1}a the two-dimensional distribution of the density (the cross
section of the composite loop) is plotted for a particular configuration. The
loop density has an inhomogeneous distribution with quite an irregular cross
section and boundary. In this so-called ``spaghetti model" the distance between
the strands is quite small and a strong dynamical interaction between them takes
place. When an initial perturbation is launched in the system the structure
basically oscillates in the direction of the perturbation. This kink-like motion
of the bundle of loop results in a strong mode conversion at the inhomogeneous
layers. We can see the shear motions due to the resonant coupling in
Figure~\ref{density1}b (see the phase-mixed scales at the boundary). As in the
cylindrical model described in Section~\ref{resabs} the energy of the kink-like
mode is transferred to the layer, this is clear from the energy distribution at a
particular given time displayed in Figure~\ref{resonanceenerg}a. In
Figure~\ref{resonanceenerg}b the averages of the velocity in three different
regions (marked with the labels $a$, $b$ and $c$) are represented. We see that
the signals are almost the same, they have the same dominant period and the same
damping time. This is due to the global nature of the motion, and the attenuation
with time is the consequence of the energy conversion due to resonant absorption
(this result is completely equivalent to that found in the single cylindrical
inhomogeneous loop model, see Fig.~\ref{tempr_0}). These results clearly
demonstrate that the mode conversion takes place in quite irregular geometries.
Regular magnetic surfaces are not necessary for this mechanism to work
efficiently as was already anticipated by \citet[][]{ruderman03} using an
elliptical cross section loop model. This indicates that resonant absorption is
quite a robust damping mechanism and although we have analysed a particular
system, the behaviour found in the present equilibrium is expected to be quite
generic of inhomogeneous plasma configurations. The resonant coupling between
fast and Alfv\'en modes can hardly be avoided in a real situation since
inhomogeneity is certainly present in coronal loops,  for these reasons, resonant
absorption seems to be quite a natural damping mechanism. 

\subsection{Curved configurations}

A curved cylindrical model involves a three-dimensional model. Kink oscillations
in three-dimensional configurations are discussed in this volume by L. Ofman, so
the reader is referred to this review and references therein. Here we just want
to comment a few points about these works. One of the striking findings of the
three-dimensional studies is that, as in some two-dimensional works, the damping
rate found in the simulations is much stronger that the reported by the
observations. This damping is usually attributed to the wave leakage due to 
curvature. However, up to now there are no comprehensive numerical studies about
this process in 3D. Moreover, if there is leakage in three-dimensions, because
the frequency of the kink mode is above the local Alfv\'en frequency, then it is
unavoidable to have an external resonance (basically where the frequency of the
kink mode is equal to the local Alfv\'en frequency). This external resonance
\citep[see][]{terretal06a} will also contribute to the attenuation of the kink
oscillation due to resonant absorption. The implications of this mechanism have
not been yet addressed in three-dimensional studies. 

\section{Conclusions and Discussion}

We have described the main efforts that have been done so far to understand
standing kink oscillations in coronal loops. We have reviewed the  time-dependent
studies in simple slabs and cylindrical loop models. The effect of curvature and
inhomogeneity has been also described in some detail, and  preliminary studies
about the possible effects of nonlinearity have been explained. However, the
time-dependent modelling of kink oscillations is still in a primitive stage and
there are several issues that need to be addressed, being the attenuation of the
oscillations one of the most interesting. 

In particular, the analysis of the damping by wave leakage in two and
three-dimensional models needs to be investigated in detail, this might help to
provide clues about the problem of the strong damping found in most of the
numerical simulations. Nevertheless, here it is worth to draw the attention about
the overlooked problem of numerical dissipation. Very often, in many numerical
studies the effect on the results of the intrinsic numerical viscosity of the
numerical scheme is not studied or quantified. Moreover, in some numerical studies
dissipation is explicitly added to stabilise the code. The issue of the
dissipation is very relevant, specially if we want to extract quantitative
information about the damping rates with the goal of identifying the possible
damping mechanism that produce the attenuation. Too much artificial dissipation
can simple mask the physics behind the damped oscillations and lead to the wrong
interpretation of the results.

Another remark about the numerical simulations is that in general in most of the
works the full set of nonlinear MHD equations are numerically solved.
Nevertheless, in the majority of the published papers about kink oscillations, the
linear regime is considered (the amplitude of the waves are very small in
comparison with the local Alfv\'en or sound speed). Before any problem is
attempted to be studied we should first ask whether it is worth to solve the full
nonlinear equations instead of the simple linearised system of equations. One of
the advantages of the linear regime is that (for simple equilibrium
configurations) we can make Fourier analysis in some direction to simplify the
problem (since that direction is incorporated in the equations through a certain
wavenumber). This is not possible when the full nonlinear MHD equations are
solved. Obviously, if we are interested in nonlinear processes, such as shocks or
some types of instabilities then it is required to solve the nonlinear MHD
equations.

Finally, we would like to point out that in other branches of sciences, for
example in computational fluid dynamics, it is a standard practice to attempt to
repeat experiments soon after they are published. This is a necessary aspect of
the scientific method  and it can viewed as an excellent learning tool. We think
that it might be interesting to apply the idea of reproducible research \citep[see
the paper about this topic of][]{leveque06} in numerical MHD. This would mean to
make available the codes used to create the figures and tables in a paper in such
a way that the reader can download the codes and run them to reproduce the results
shown in the paper.



%

%

\begin{acknowledgements}
This paper was inspired by a ISSI workshop held in Bern, Switzerland, August
2008.
\end{acknowledgements}


\end{document}